\documentclass[apjs, numberedappendix]{emulateapj}

\usepackage{natbib}
\usepackage{epsfig,amsmath} 
\usepackage{graphicx}
\usepackage{txfonts}
\usepackage{amssymb}
\slugcomment{The Astrophysical Journal Supplement Series, in press}

\newcommand{\be}{\begin{equation}}
\newcommand{\ee}{\end{equation}}
\newcommand{\beq}{\begin{eqnarray}}
\newcommand{\eeq}{\end{eqnarray}} 
\newcommand{\bez}{\begin{eqnarray*}}
\newcommand{\eez}{\end{eqnarray*}}

\newcommand{\ovl}{\overline}
\newcommand{\bc}{\begin{center}}
\newcommand{\ec}{\end{center}}

\newcommand{\ve}{\mbox{\boldmath $e$}}
\newcommand{\vecx}{\mbox{\boldmath $x$}}
\newcommand{\vecp}{\mbox{\boldmath $p$}}
\newcommand{\vecQ}{\mbox{\boldmath $Q$}}
\newcommand{\vnabla}{\mbox{\boldmath $\nabla$}}
\newcommand{\vomega}{\mbox{\boldmath $\omega$}}
\newcommand{\vOmega}{\mbox{\boldmath $\Omega$}}

\newcommand{\fourx}{\underline{x}}
\newcommand{\fourp}{\underline{p}}
\newcommand{\unb}{\underline{\nabla}}

\newcommand{\noccx}{n}

\newcommand{\vl}{\mbox{\boldmath $l$}}
\newcommand{\vn}{\mbox{\boldmath $n$}}  

\newcommand{\betab}{\beta_{\rm b}}
\newcommand{\Gammab}{\Gamma_{\rm b}}
\newcommand{\Thetacmb}{\Theta_{\rm cmb}}
\newcommand{\Tcmb}{T_{\rm cmb}}
\newcommand{\rmd}{{\rm d}}
\newcommand{\fe}{f_{\rm e}}
\newcommand{\me}{m_{\rm e}}
\newcommand{\Te}{T_{\rm e}}

\newcommand{\Ne}{N_{\rm e}}
\newcommand{\Nph}{N_{\rm ph}}
\newcommand{\etae}{\eta_{\rm e}}
\newcommand{\re}{r_{\rm e}}
\newcommand{\sigmat}{\sigma_{\rm T}}
\newcommand{\taut}{\tau_{\rm T}}

\newcommand{\chijzero}{\chi_{\!j\,0}}

\newcommand{\chijn}{\chi_{\!j\,n}}
\newcommand{\chitwon}{\chi_{\!2n}}
\newcommand{\chionen}{\chi_{\!1n}}
\newcommand{\chizeron}{\chi_{\!0n}}

\newcommand{\Deltajn}{\Delta_{jn}}
\newcommand{\psiij}{\psi_{ij}}
\newcommand{\PPsiij}{\Psi_{ij}}
\newcommand{\gxi}{g({\xi})} 

\begin{document}
%

\title{Theory of Compton scattering by anisotropic electrons}

\shorttitle{Theory of Compton scattering by anisotropic electrons}
\shortauthors{Poutanen \& Vurm}


\author{Juri Poutanen and Indrek Vurm\altaffilmark{1}}

\affil{Astronomy Division, Department of Physics, P.O.Box 3000, 
90014 University of Oulu, Finland; juri.poutanen@oulu.fi,  indrek.vurm@oulu.fi }
 \altaffiltext{1}{Also at Tartu Observatory, 61602 T\~{o}ravere, Tartumaa, Estonia}

%




	    	    

\begin{abstract}
Compton scattering plays an important role in various astrophysical objects such as accreting black holes and neutron stars, pulsars, and relativistic jets, clusters of galaxies as well as the early Universe. In most of the calculations it is assumed that the electrons have isotropic angular distribution in some frame. However, there are situations where the anisotropy may be significant due to the bulk motions, or anisotropic cooling by synchrotron radiation, 
or anisotropic source of seed soft photons. 
We develop here an analytical theory of  Compton scattering by anisotropic distribution of electrons that can simplify significantly the calculations. 
Assuming that the electron angular distribution can be represented by a second order polynomial over cosine of some angle (dipole and quadrupole anisotropy), we integrate the exact Klein-Nishina cross-section over the angles.
Exact analytical and approximate formulae  valid for any photon and electron energies are derived for the redistribution functions describing Compton scattering of photons with arbitrary angular distribution by anisotropic electrons. 
The analytical expressions for the corresponding photon scattering cross-section on such electrons as well as the mean energy of scattered photons, its dispersion and radiation pressure force are also derived. 
We applied the developed formalism to the accurate calculations of the thermal and kinematic Sunyaev-Zeldovich effects for arbitrary electron distributions. 
\end{abstract}
   

%

\section{Introduction}

Compton scattering is one of the most important radiative process that shapes the spectra of various sources: black holes and neutron stars in X-ray binaries, pulsars and pulsar wind nebulae, jets from active galactic nuclei, and the early Universe. Compton scattering kernel takes a simple form if electrons are ultra-relativistic with the Lorentz factor $\gamma\gg 1$ \citep{BG70}. In a general case, when no restrictions are made on the energies of photon and electrons,  \citet{Jones68} derived the kernel for isotropic electrons and photons. The formulae there  contain a few misprints, but even by correcting those (see e.g. \citealt{PW05}) they cannot be used for calculations because of a number of numerical cancellations \citep[see e.g.][]{Bel09}. 
An alternative derivation to that kernel was given  by \citet{Bri84} and \citet[][ NP94 hereafter]{NP94}, who showed how 
to extend numerical scheme to cover all photon and electron energies of interest in astrophysical sources. 
 
In real astrophysical environments, the radiation field does not need to be isotropic and a more general redistribution function is required to describe angle-dependent Compton scattering. NP94 extended previous results to the situation when the photon distribution   can be represented as a linear function of some polar angle cosine (Eddington approximation), deriving an analytical formula for the first moment of the kernel. \citet{AA81} were first to derive a redistribution function for arbitrary photon angular dependence (see also \citealt{PKB86}). \citet{KPB86} and \citet{K87} have developed numerical methods to compute the kernel efficiently and with a high accuracy. All these works neglect the effect of photon polarization. 

\citet{NP93} have derived a general Compton scattering redistribution 
matrix for Stokes parameters assuming an isotropic electron distribution. 
A general relativistic kinetic equation incorporating the effects of induced scattering and polarization of photons as well as electron polarization and degeneracy has been derived by \citet{NP01}.   

In this paper, we propose a method to extend previous results to the case where the electron distribution is no longer isotropic, but can have weak anisotropies which can be represented as a second order polynomial of the cosine of some polar angle. 
The proposed  formalism can find its application in a number of astrophysical problems. 
The distortion of the cosmic microwave background (CMB) caused by the hot electron gas in clusters of galaxies (i.e.  kinematic and thermal Sunyaev-Zeldovich effect) is an obvious application. The electron distribution, isotropic in the cluster frame, can be Lorentz transformed to the CMB frame, 
resulting in a dipole term linear in cluster velocity and a small quadrupole correction. Compton scattering then can be directly computed in the CMB frame. 
Another possible application concerns the models of outflowing accretion disk-coronae or jets \citep{B99PE,MBP01}. If the outflow velocities 
are non-relativistic, the radiation transport can be considered directly in the accretion disk frame following recipe by \citet{PS96}, 
with the Lorentz transformed electron distribution.  

Weak anisotropies in the electron distribution can arise in high-energy sources with ordered magnetic field, 
because of the pitch angle-dependence of the synchrotron cooling rate and/or anisotropic injection of high-energy electrons \citep[e.g.][]{Bjornsson85,Roland85, Crusius98,Schopper98}. 
An anisotropic electron distribution is also a very natural outcome of the photon breeding operation 
in relativistic jets with the toroidal magnetic field \citep{StP06,SP08}, 
because the electron-positron pairs born inside the jet by the external high-energy photons move perpendicularly to the field. 

Our method is also extendable to the polarized radiation using the techniques developed in \citet{NP93}. It is also in principle possible to calculate the scattering redistribution function in the case when the electron distribution is expressible as an arbitrary order expansion over the polar angle cosine. Unfortunately, in the latter case the analytical expressions become extremely cumbersome and the advantage over direct numerical integration becomes small.

Although here we consider only photon scattering it is also possible to apply the same method for electrons interacting with the photons in the case when photon angular distribution is expressible as an expansion of powers of the polar angle cosine. This can be of interest only in the deep Klein-Nishina regime where the electron can lose or gain a large fraction of its initial energy in one scattering and continuous energy loss approach is not applicable.

Our paper is organized as follows. In Section \ref{sec:rke} we introduce the relativistic kinetic equation for Compton  scattering and define the redistribution function and total cross-section. The expressions for the total cross-section, the mean energy and dispersion of scattered photons, and the radiation pressure force are given in Section \ref{sec:cross}. The exact analytical  formulae for the redistribution function for mono-energetic anisotropic electrons as well as approximate formulae valid in Thomson regime are derived in Section \ref{sec:redistr}. We present the numerical examples of redistribution functions in Section \ref{sec:numer}, where we also develop the relativistic theory of Sunyaev-Zeldovich effect.  We summarize our findings in Section \ref{sec:concl}.

\section{Relativistic kinetic equation}
\label{sec:rke} 

Let us define the dimensionless photon four-momentum as $\fourx =\{ x, \vecx \}= x \{ 1,\vomega\}$, where $\vomega$ is the unit vector in the photon propagation direction and $x \equiv h\nu/\me c^2$. The photon distribution  will be described by the occupation number $\noccx$.  The dimensionless electron four-momentum is $\fourp  = \{ \gamma, \vecp\}= \{ \gamma, p\vOmega\} = 
 \gamma \{ 1 , \beta\vOmega\}$, where $\vOmega$ is the unit vector along the electron momentum, $\gamma$ and $p=\sqrt{\gamma^2-1}$  are the electron Lorentz factor and its momentum in units of $\me c$, and $\beta=p/\gamma$ is the electron velocity in units of speed of light. 
The momentum distribution of electrons is described by the relativistically invariant distribution function $\fe(\vecp)$ (see \citealt{BB56}; NP94).
 
The interaction between photons and electrons via Compton scattering (in linear approximation, i.e. ignoring induced scattering and electron degeneracy) can be described by the explicitly covariant relativistic kinetic equation for photons (\citealt{Pom73,NP93}; NP94):
\beq \label{eq:rke}
\fourx \cdot \unb \noccx(\vecx) &=&  \frac{\re^2}{2}  
\int \frac{\rmd^3 p}{\gamma} \frac{\rmd^3 p_1}{\gamma_1}  \frac{\rmd^3 x_1}{x_1} 
\: \delta^4(\fourp_1 + \fourx_1 - \fourp - \fourx)   \: F 
\nonumber \\ 
&\times & 
\left[ \noccx(\vecx_1)  \fe(\vecp_1)-  \noccx(\vecx) \fe(\vecp) \right] ,
\eeq
where $\unb=\{\partial/c \partial t, \vnabla\}$ is the four-gradient, $\re$ is the classical electron radius, $F$ is the Klein-Nishina reaction rate \citep{LLVol4}
\begin{equation} \label{eq:kn}
F  = \left( \frac{1}{\xi} - \frac{1}{\xi_1}\right)^2 + 2 \; \left( \frac{1}{\xi} - \frac{1}{\xi_1}\right)
+ \frac{\xi}{\xi_1} + \frac{\xi_1}{\xi},
\end{equation}
and 
\begin{equation} \label{eq:xixi1}
\xi = \fourp_1\cdot\fourx_1= \fourp\cdot\fourx, \qquad \xi_1 =  \fourp_1\cdot\fourx =  \fourp\cdot\fourx_1
\end{equation}
are the four-products of corresponding momenta. The second equalities in both equations (\ref{eq:xixi1}) arise  from the four-momentum conservation law. 
The invariant scalar product of the photon four-momenta can be written in the laboratory frame as well as in the frame related to a specific electron 
\be \label{eq:q_kk1}
q \equiv \fourx \cdot \fourx_1 =xx_1 (1-\mu) = \xi\xi_1(1-\mu_0) = \xi- \xi_1, 
\ee 
where $ \mu= \vomega\cdot \vomega_1$ is the cosine of the photon scattering angle in some frame and $\mu_0$ is the corresponding cosine in the electron rest frame.  

In  any frame, the kinetic equation can be also put in the usual form of the radiative transfer equation \citep{NP93}:
\beq \label{eq:rte}
\lefteqn{ 
\left(  \frac{1}{c} \frac{\partial}{\partial t} + \vomega \cdot\vnabla \right) \noccx(\vecx) =
-   \sigmat \: \Ne\ \overline{s}_0(\vecx)\ \noccx(\vecx)  
} \nonumber \\ &+ & \sigmat \Ne\ \frac{1}{x}
\int_{0}^{\infty} x_1 \rmd x_1 \int \rmd ^2 \vomega_1 \: R(\vecx_1 \rightarrow \vecx) \noccx(\vecx_1) ,
\eeq
where $\Ne$ is the electron density in that frame. Here we have defined the photon redistribution function 
\be \label{eq:rk1k}
R(\vecx_1 \rightarrow \vecx) = \!
\frac{3}{16\pi} \frac{1}{\Ne} \!\! \int \!\!
\frac{\rmd^{3}p}{\gamma} \frac{\rmd ^{3}p_1}{\gamma_1} \fe(\vecp_1) F \delta^{4} (\fourp_1 + \fourx_1 - \fourp - \fourx)  
 \ee
and the total scattering cross-section (in units of Thomson cross-section $\sigmat$) 
\be \label{eq:totcross}
\overline{s}_0(\vecx)  =\! 
\frac{3}{16\pi}  \frac{1}{\Ne} \frac{1}{x} \! \int \!\! \frac{\rmd ^3 p}{\gamma} 
\frac{\rmd ^3 p_1}{\gamma_{1}} \frac{\rmd ^3 x_1}{x_{1}}
 \fe(\vecp) \  F \ \delta^{4} (\fourp_1 + \fourx_1 - \fourp - \fourx)  .
\ee

\subsection{Electron distribution and scattering geometry}

Let us now consider a specific frame $E$. Our basic assumption  is that the anisotropy of the electron distribution in this frame can be expressed as a second order polynomial expansion in the cosine of the polar angle in some coordinate system $(\vl_1,\vl_2,\vl_3)$: 
\begin{equation}   \label{eq:fe}
\frac{1}{\Ne}\fe(\vecp) = \fe(\gamma,\etae) = \sum_{k=0}^{2} f_k(\gamma) P_k(\etae) , 
\ee
where $\Ne$ is the electron density in frame $E$, $\etae = \vOmega\cdot\vl_3$ is the cosine of the polar angle of the electron momentum, $P_k(\etae)$ are the Legendre polynomials, and we now switched to the  dimensionless distribution function $\fe(\gamma,\etae)$ normalized to unity: 
\begin{equation} 
\int \rmd^2 \Omega \int \fe(\gamma,\etae) \; p^2 \rmd p =1. 
\end{equation}
The moments $f_0$, $f_1$ and $f_2$ determine the energy spectrum of electrons and the relative magnitudes of the isotropic and anisotropic components.
The distribution function  for mono-energetic electrons of energy $\gamma_0$ can be described by 
\be \label{eq:fe_mono}
\fe(\gamma, \etae)= \frac{1}{4\pi\, p\gamma}\delta(\gamma-\gamma_0) \left[ 1 + \frac{f_1}{f_0} \etae + \frac{f_2}{f_0} P_2(\etae) \right] ,
\ee
where the ratios $f_1/f_0$ and $f_2/f_0$ are constants. 

\begin{figure}
\centerline{\epsfig{file=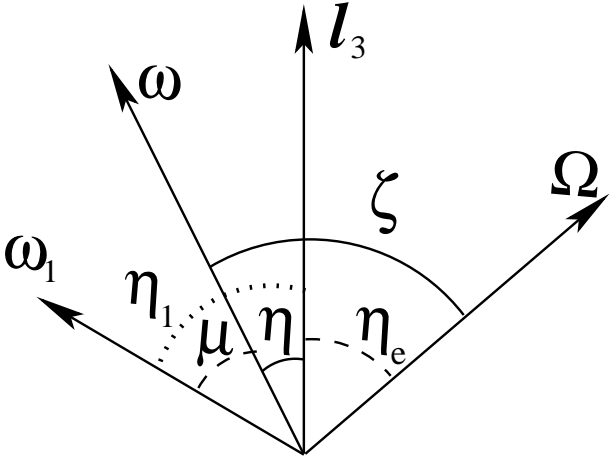,width=8.2cm} }
\caption{The scattering particles' momentum vectors and the vector $\vl_3$.
Note that the shown angular variables are the cosines of the respective angles.}
\label{fig:geom1}
\end{figure}

The directions of photons in this coordinate system (see Fig.~\ref{fig:geom1}) are given by 
\beq 
\vomega & =&   \sqrt{1-\eta^2}\cos\phi\ \vl_1 +  \sqrt{1-\eta^2}\sin\phi\ \vl_2 + \eta\ \vl_3 , \\
\vomega_1 & =&   \sqrt{1-\eta_1^2}\cos\phi_1\ \vl_1 +  \sqrt{1-\eta_1^2}\sin\phi_1\ \vl_2 + \eta_1\ \vl_3 , 
\eeq
so that the cosine of the scattering angle is 
\be \label{eq:mu_scat}
\mu=\vomega\cdot\vomega_1=\eta\eta_1 + \sqrt{1-\eta^2} \sqrt{1-\eta^2_1} \cos(\phi-\phi_1) . 
\ee

\section{Total cross section and mean powers of energies}
\label{sec:cross}

\subsection{Total cross-section}

Let us  simplify the  expression for the total cross-section.  We follow here the approach described in NP94. 
We rewrite the cross-section as 
\be \label{eq:ovlsigma}
\overline{s}_0(\vecx)  =   \frac{1}{x} \int s_{0} (\xi) \: \xi \: \fe(\gamma,\etae) \: \frac{\rmd ^3 p}{\gamma},
\ee
where 
\be \label{eq:sfun}
s_{0}(\xi) =  \frac{3}{16\pi} \frac{1}{\xi} \int \frac{\rmd ^3 p_1}{\gamma_{1}} \frac{\rmd ^3 x_1}{x_{1}}
\  F\  \delta^{4} (\fourp_1 + \fourx_1 - \fourp - \fourx)  .
\ee
Using the identity
\be \label{eq:delta_ident}
\delta(\gamma_{1}+x_1-\gamma-x)=\gamma_1\
\delta\left( \fourx_1 \cdot (\fourp+ \fourx) - \fourx \cdot \fourp\right) 
\ee
and taking the integral over $\vecp_1$ in equation (\ref{eq:sfun}), we get
\beq \label{eq:sfun2}
s_{0}(\xi) &= &  \frac{3}{16\pi} \frac{1}{\xi} \int  \frac{\rmd ^3 x_1}{x_{1}} \  F\ 
 \delta\left( \fourx_1 \cdot (\fourp+ \fourx) - \fourx \cdot \fourp\right)   \nonumber \\
&= &  \frac{3}{16\pi} \frac{1}{\xi} \int \xi_1 \rmd \xi_1\ \rmd \mu_0\ \rmd \phi_0 \ 
\  F\ \delta\left[\xi_1 + \xi\xi_1(1-\mu_0) - \xi \right]\    \nonumber \\
&=&   \frac{3}{8\xi^2} \int_{\xi/(1+2\xi)}^{\xi} \ F \ \rmd \xi_1 ,
\eeq
where we used invariant $x_1\rmd x_1 \rmd^2 \omega_1 = \xi_1 \rmd \xi_1\  \rmd\mu_0\ \rmd \phi_0$ and the fact that $F$ does not depend on azimuthal angle $\phi_0$.  
Substituting $F$ from  Equation~(\ref{eq:kn}) we get (\citealt*{LLVol4}; NP94)
\be \label{eq:s_0}
s_{0}(\xi) =  \frac{3}{8\xi^2} \left[ 4 + \left( \xi - 2 - \frac{2}{\xi} \right) \ln(1+2\xi)
+ 2\xi^2 \frac{1+\xi}{(1+2\xi)^2} \right] .										
\ee
Putting $\xi\rightarrow x$, we, of course, get the total Klein-Nishina cross-section 
for a  photon of energy $x$ on electrons at rest.   

To obtain the total scattering cross-section on an anisotropic electron distribution,  
we have to calculate the angular integrals over incoming electron directions in Equation~(\ref{eq:ovlsigma}). 
We introduce the cosines between electron momentum and photons: 
\be 
\zeta=\Omega\cdot \vomega, \quad \zeta_1= \Omega\cdot \vomega_1  
\ee
so that 
\be \label{eq:zetaxi}
\xi= x(\gamma-p\zeta), \quad \xi_1= x_1 (\gamma-p\zeta_1) .
\ee
We choose the spherical coordinate system and measure the polar angle from the direction of the  {\it initial} photon $\vomega$, 
we get
\be \label{eq:sigmaxeta}
\ovl{s_{0}}(\vecx)\!  =\!  \ovl{s_{0}}(x,\eta) \! =\!  \frac{1}{x} \int\limits_{0}^{\infty}\!\! \frac{p^2 \rmd p}{\gamma} \!\!
\int \limits_{-1}^{1} \!\! \rmd \zeta \ s_{0}(\xi) \, \xi \!\!  \int \limits_{0}^{2\pi} \!\! \rmd \Phi \: \fe(\gamma,\etae).  
\ee
Azimuth $\Phi$ is now defined as the difference between the azimuths of the electron momentum direction
and the vector $\vl_3$ in a frame with $z$-axis along $\vomega$.
The angular variable $\etae$  in the expansion (\ref{eq:fe}) is expressed in this frame as
\be \label{eq:etaephi}
\etae = \eta  \zeta + \sqrt{1-\eta^2} \sqrt{1-\zeta^2} \cos\Phi	.	
\ee 
The physical meaning of Equation~(\ref{eq:sigmaxeta}) can be also understood if we consider a monoenergetic beam of electrons along $\vl_3$ axis: $\fe(\gamma,\etae)=\delta(\etae-1) \delta(\gamma-\gamma_0) /(2\pi p \gamma)$. Then 
\be \label{eq:sigma_beam}
\ovl{s_{0}}(x,\eta) =   \left(1- \beta \eta\right) s_{0}(x'),   
\ee
where $x'=x \gamma(1- \beta\eta)$ is the photon energy in the electron rest frame (we omitted subscript 0 in $\gamma$ and $\beta$). 
The factor $1- \beta \eta$ in Equation~(\ref{eq:sigma_beam}) accounts for the reduced number of collision per unit length. 

\begin{figure}
\centerline{\epsfig{file= 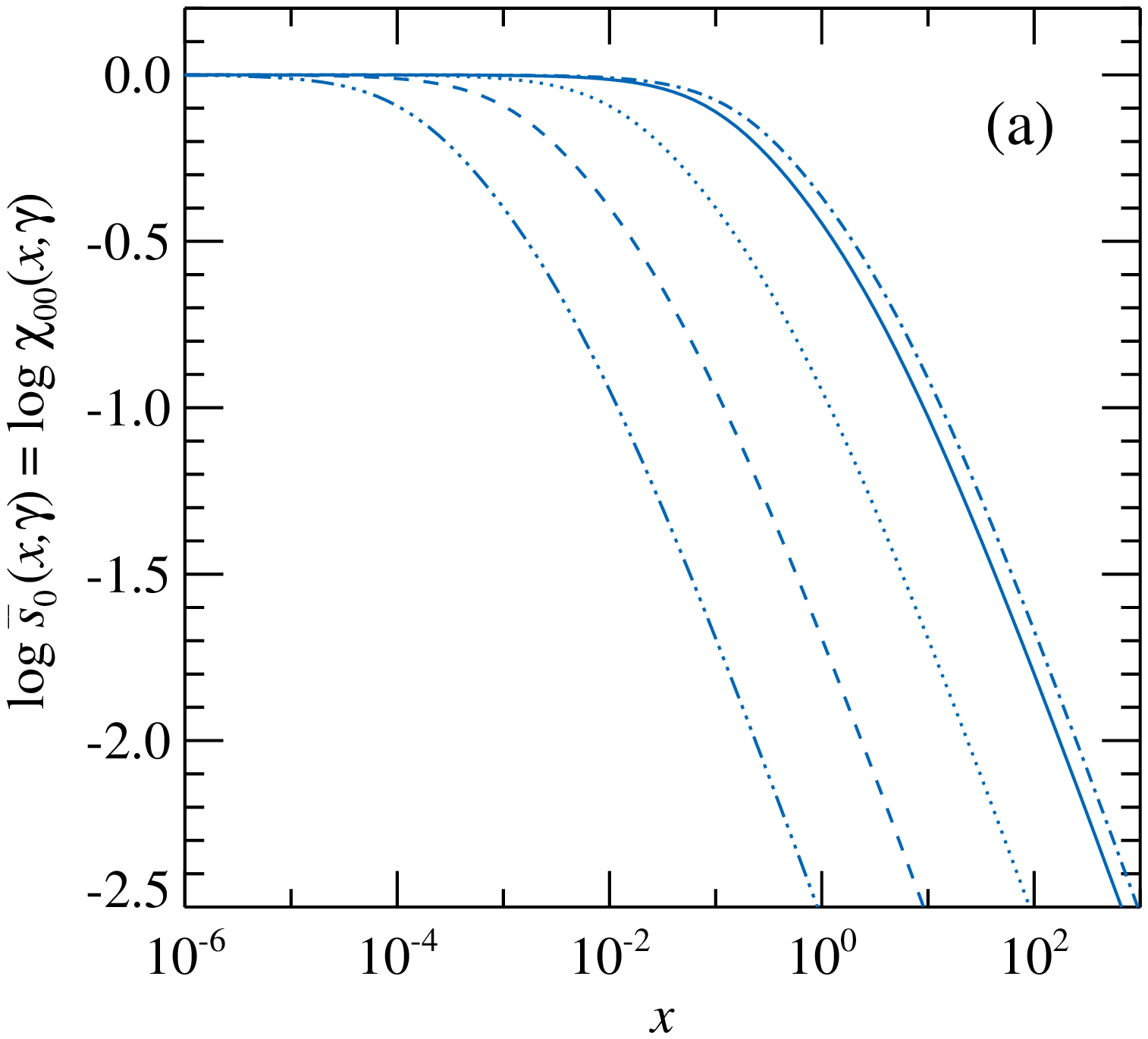,width=7.5cm}} 
\centerline{\epsfig{file= 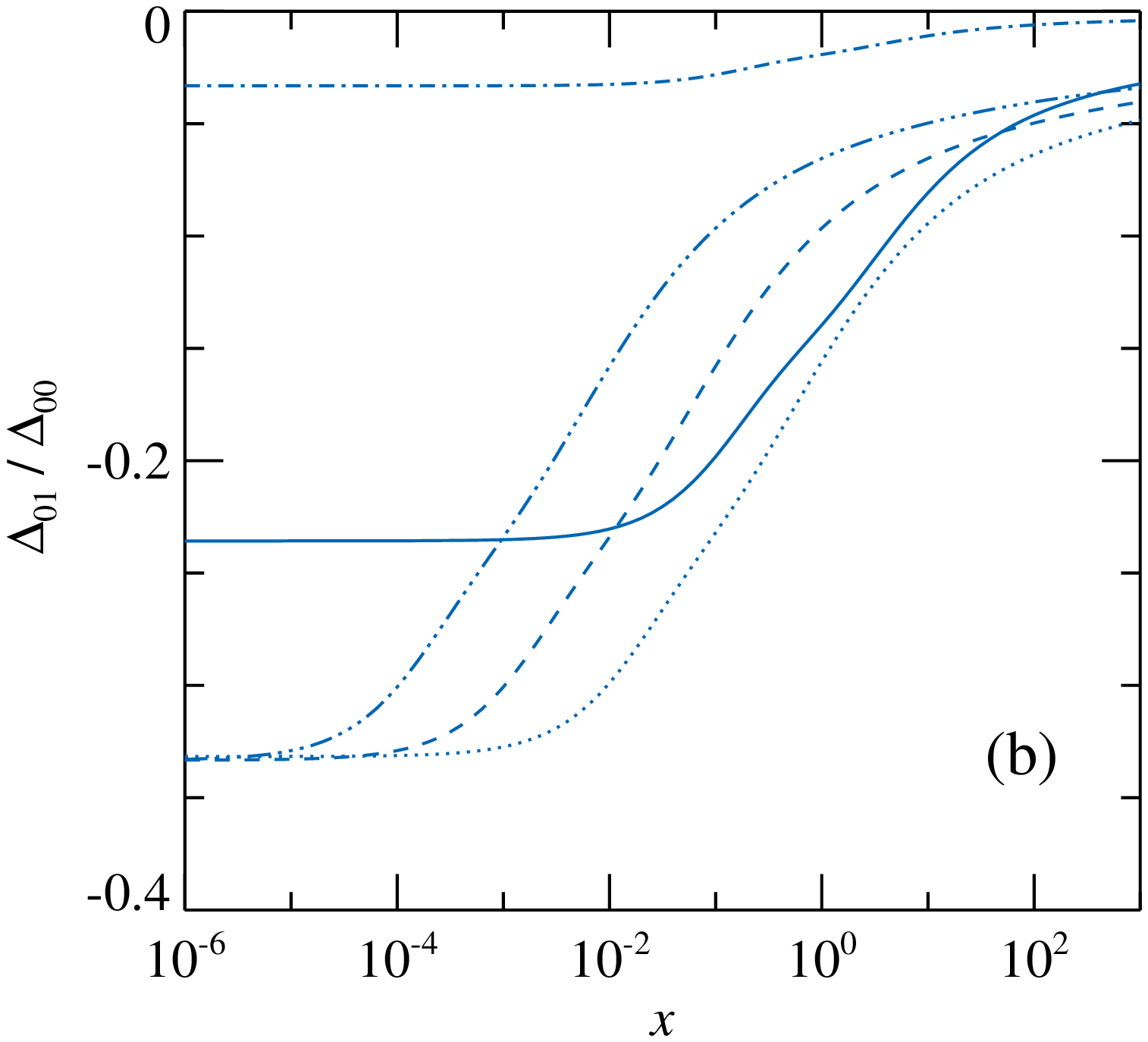,width=7.5cm}} 
\centerline{\epsfig{file= 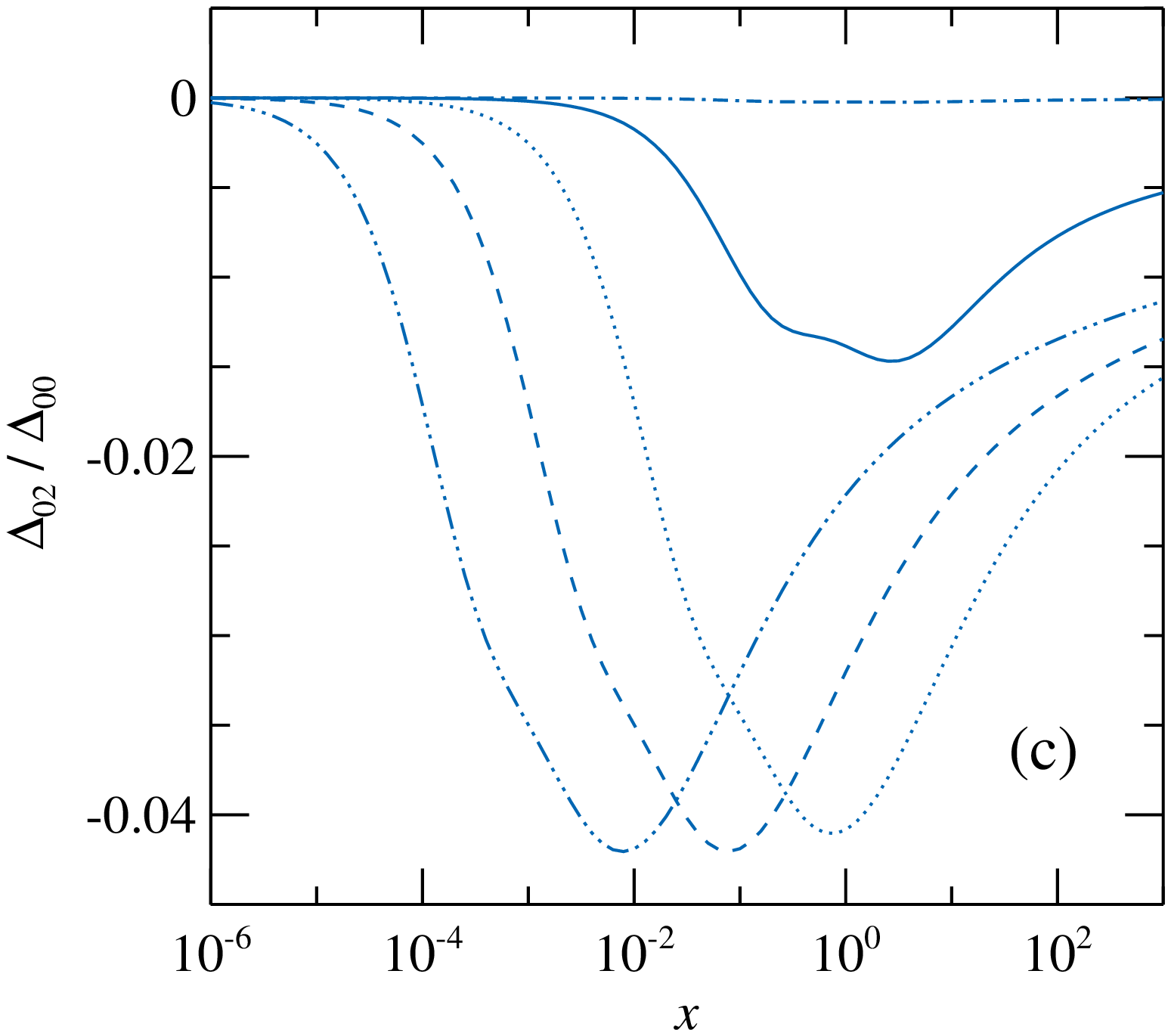,width=7.5cm}} 
\caption{(a) Total cross-section $\chi_{00}=\Delta_{00}$ for isotropic mono-energetic electrons of various momenta 
$p=0.1, 1, 10, 10^2, 10^3$ (from top to bottom, dot-dashed, solid, dotted, dashed,  and triple-dot-dashed curves) 
as a function of photon energy $x$. 
(b) Relative correction to the cross-section arising from the dipole term in the electron distribution (\ref{eq:fe}) 
for the same $p$ as in panel (a).   
At small $x$ the curves approach the asymptotic value in Thomson limit $-\beta /3$. 
(c) Relative correction to the cross-section arising from the quadrupole term $\Delta_{02}/\Delta_{00}$.  
}
\label{fig:chi_j0}
\end{figure}

When calculating the azimuthal integral in equation (\ref{eq:sigmaxeta}) we just have to integrate $P_k(\etae)$ with $\etae$ given by 
Equation (\ref{eq:etaephi}).  
The properties of the Legendre polynomials give us the average 
\be\label{eq:etaave}
\ovl{P_k(\etae)} = 	P_k(\eta)\ P_k(\zeta),			
\ee
so that the averaged distribution function becomes
\be  \label{eq:fe_ave_pk} 
\ovl{\fe} (\gamma,\eta,\zeta) =\frac{1}{2\pi} \int_{0}^{2\pi}\!\!\!\! \rmd \Phi \: \fe(\gamma,\etae) = 
\sum_{k=0}^2 f_k(\gamma) P_k(\eta)\ P_k(\zeta) . 
\ee
The cross-section now can be written as 
\beq \label{eq:s0_d_chi}  
\ovl{s_{0}}(x,\eta)  = 4\pi \sum_{k=0}^{2} P_k(\eta)\, \int_{1}^{\infty}  p\gamma\ \rmd \gamma \    f_k \,  \Delta_{0k} (x, \gamma) , 
\eeq
where 
\be \label{eq:delta0k}
\Delta_{0k} (x,\gamma)  =    \frac{1}{2\gamma x } \int_{-1}^{1}  P_k(\zeta)\ \xi \ s_{0}(\xi) \  \rmd\zeta . 
\ee
Changing the integration variable and expressing $P_k$ through $\xi$ using Equation~(\ref{eq:zetaxi}),
we get
\beq \label{eq:delta0k_chi}
\Delta_{0k} (x,\gamma) & = & \sum_{n=0}^{k} \ b_{nk}\ \chizeron , 
\eeq
where  
\be \label{eq:chi0n}
\chi_{0n} (x,\gamma)= \frac{1}{2\gamma p\, x^{2+n}} \int_{x(\gamma-p)}^{x(\gamma+p)} 
\xi^{n+1} \ s_{0}(\xi) \  \rmd\xi . 
\ee
and 
\beq \label{eq:coef_bn}
b_{00} &=& 1, \quad  b_{01}  =  \frac{\gamma}{p}, \quad b_{11}  =  - \frac{1}{p} , \\
 b_{02} & =& \frac{1}{2 p^2} \, (2\gamma^2 + 1) , 
\quad b_{12} = - \frac{3 \gamma}{p^2}  , \quad
b_{22} = \frac{3}{2 p^2}  . \nonumber
\eeq
The zeroth function $\Delta_{00}=\chi_{00}$ coincides with the function $\Psi_0(x,\gamma)$ from NP94.  When electron distribution is isotropic (i.e. $f_1=f_2=0$),   expression (\ref{eq:s0_d_chi}) for the total cross-section is reduced to equation (3.4.1) from NP94 and  the dependence on $\eta$ obviously disappears. Functions  $\chi_{0n}$ of two arguments can be presented through the functions of one argument:  
\be \label{eq:chi0n_exp}
\chi_{0n}(x,\gamma) =   \frac{1}{2\gamma p } \left. \frac{u^{2+n}}{2+n} \psi_{n+1,0}(xu) \right|_{u=\gamma-p}^{u=\gamma+p}, 
\ee
where 
\be \label{eq:psi_ij}
\psi_{i0}(\xi)= \frac{i+1}{\xi^{i+1}}  \int_0^{\xi} t^i s_{0} (t) \rmd t. 
\ee
The explicit expressions for the functions $\psi_{i0}(\xi)$ ($i=1,2,3$) can be found in Appendix \ref{app:psi_ij} (see also NP94). Thus the total cross-section is given by a single integral over the electron energy  (\ref{eq:s0_d_chi}) with all functions under the integral given by analytical expressions. Numerical calculations of functions $\Delta_{0k}$ can be separated into three regimes: (1) in Thomson regime, $x\gamma\ll 1$, the series expansion (see Appendix \ref{app:chi_ij}) can be used; (2) for $p\ll 1$, but $x$ not sufficiently small for regime (1), we numerically take  the integral in Equation~(\ref{eq:delta0k}) using 5-point Gaussian quadrature to reach accuracy better than 1\%; (3) in other cases, we use the sum in Equation~(\ref{eq:delta0k_chi})  and analytical expressions (\ref{eq:chi0n_exp}) for $\chi_{0n}$. 
 
For mono-energetic electron distribution of Lorentz factor $\gamma$ given by Equation (\ref{eq:fe_mono}), we can introduce the cross-section 
analogously to Equation~(\ref{eq:s0_d_chi}): 
\be \label{eq:s0_d_chi_gam}
\ovl{s_{0}}(x,\gamma,\eta) = \sum_{k=0}^2 \ \frac{f_k}{f_0} P_k(\eta) \Delta_{0k} (x,\gamma) .
\ee
For isotropic mono-energetic electrons, the total cross-section is shown in Figure~\ref{fig:chi_j0}a. The relative corrections arising due to the dipole and quadrupole term in the electron distribution are shown  in Figures~\ref{fig:chi_j0}b and \ref{fig:chi_j0}c, respectively. 
These have to be multiplied by the angle- and, possibly, the energy-dependent factor $P_k(\eta)f_k/f_0$ to obtain the final correction. 
In the Thomson limit, at small $x \gamma \ll 1$, the cross-section takes the form (see Appendix \ref{app:chi_ij})
\be \label{eq:d01d00}
\ovl{s_{0}}(x,\gamma,\eta) \approx 1 - \frac{1}{3} \frac{f_1}{f_0}\, \eta\, \beta , 
\ee
where the correction to unity term can be easily obtained by averaging the transport cross-section over electron directions  (i.e. integrating $(1-\beta\zeta)\etae$ over the angles). This corresponds to the flattening in Figure~\ref{fig:chi_j0}b at $\Delta_{01}/\Delta_{00}= -\beta /3$.
The correction from the quadrupole term in this regime as well as for non-relativistic electrons becomes negligible: 
\be \label{eq:d02d00}
\Delta_{02}/\Delta_{00} \approx  - \frac{4}{15} \ \beta^2 \ (x\gamma) . 
\ee

\subsection{Mean powers of scattered photon energy}

In some situations, the full relativistic kinetic equations can be substituted by the approximate one obtained in Fokker-Planck approximation. This requires knowledge of various moments of the  redistribution function, such as total cross-section, the mean energy and dispersion of the scattered photons (see NP94, \citealt{VP09}). It is often time-consuming to compute 
numerically the integrals of the redistribution function and instead direct calculations of the moments are preferable. 
Below we obtain analytical expressions for the mean energy and dispersion of the energy of scattered photons in frame $E$ as a function of the initial photon energy $x$ and the direction of its propagation relative to a symmetry axis of the electron distribution $\vl_3$. 

Following NP94, we define the mean of powers of energy of scattered photons: 
\be \label{eq:ovl_moments}
\overline{x_1^j} \overline{s}_0(\vecx)  =   \frac{1}{x} \int 
\langle x_1^j \rangle s_{0} (\xi) \: \xi \: \fe(\gamma,\etae) \: \frac{\rmd ^3 p}{\gamma},
\ee
where now 
\beq \label{eq:sfun_xj}
\lefteqn{ \langle x_1^j \rangle s_{0}(\xi) =  \frac{3}{16\pi} \frac{1}{\xi}\! \int \!\frac{\rmd ^3 p_1}{\gamma_{1}} \frac{\rmd ^3 x_1}{x_{1}}
 F \,  x_1^j \, \delta^{4} (\fourp_1 + \fourx_1 - \fourp - \fourx) } \\
&\!=\!&  \frac{3}{16\pi} \frac{1}{\xi}\! \int\!\!  x_1^{j+1} \rmd x_1 \rmd^2\omega_1 
 \,  F  \  \delta\!\left\{ x_1 [\gamma+ x - \vomega_1\!\cdot\! ( p\vOmega\!+\!x\vomega) ] - \xi \right\} . \nonumber
\eeq

\subsubsection{Averaging over photon directions}

Quantities (\ref{eq:sfun_xj}) are not invariants (except for $j=0$), and we have to compute the scattered photon energy is a certain frame, which we choose to be frame $E$. Because of the additional term $x_1^j$ under the integral, a simple change of variables to the electron rest frame as in Equation~(\ref{eq:sfun2}) is not possible.  Instead, we use the $\delta$-function to take the integral over $x_1$: 
\be \label{eq:sfun_xj2}
\langle x_1^j \rangle s_{0}(\xi) = \frac{3}{16\pi} \frac{1}{\xi^2} \int  x_1^{j+2}\ \rmd^2\omega_1 \  F \ .
\ee
Now we change the variables to those in the electron rest frame (with subscript 0). We choose the coordinate system with the polar axis along the direction of the incoming photon $\vomega_0$. In this frame, the cosine of the angle between the electron momentum and the incoming photon is $\zeta_0$. The cosine of the  angle between the outgoing photon momentum and the electron is then  $\zeta_{10}=\zeta_0\mu_0+ \sqrt{1-\zeta_0^2} \sqrt{1-\mu_0^2}\cos\phi_0$.

We use invariants $x_1^2 \rmd^2\omega_1=\xi_1^2 \rmd \mu_0\rmd \phi_0$ and the energy conservation law in the electron rest frame $\xi_1=\xi/(1+\xi[1-\mu_0])$, to get (see NP94) 
\be\label{eq:invar_x1omega} 
x_1^2 \rmd^2\omega_1=\rmd \xi_1 \rmd \phi_0.
\ee 
Finally, we have 
\be \label{eq:sfun_xj3}
\langle x_1^j \rangle s_{0}(\xi) = \frac{3}{16\pi} \frac{1}{\xi^2} \int_{\xi/(1+2\xi)}^{\xi}  
  F\ \rmd \xi_1  \int_{0}^{2\pi} x_1^j  \rmd \phi_0 .
\ee
Because $\xi_1$ is the energy of scattered photon in the electron rest frame, the Doppler effect gives us 
\be\label{eq:x1j}
x_1 \!\! = \!  \xi_1 (\gamma+p\zeta_{10} )  \!= \! \xi_1 \! \left( \gamma  \!+ \!  p \zeta_0\,\mu_0 \!+  \! p\! \sqrt{1-\zeta_0^2} \!\sqrt{1-\mu_0^2}\cos\phi_0 \right) , 
\ee
where we now can substitute  
\be \label{eq:mu0_pzeta0}
\mu_0=1 + \frac{1}{\xi} - \frac{1}{\xi_1} , \quad 
p \zeta_0 = \frac{x}{\xi} - \gamma , 
\ee
which are consequences of the conservation law and of the Lorentz transformation $x=\xi (\gamma+p\zeta_0)$, respectively.  The terms containing a linear combination of square roots and $\cos\phi_0$ will disappear after averaging over $\phi_0$.  

We now introduce moments of the invariant cross-section
\be \label{eq:sjxi} 
s_{j}(\xi)=\frac{3}{8\,\xi^{j+2}}\,\int _{\xi/(1+2\,\xi)}^\xi
 \,\xi_1^j  F\,\rmd \xi_1.
\ee
For $j=0$, we get of course the total cross-section $s_{0}$ given by Equation~(\ref{eq:s_0}). 
NP94 derived  the corresponding expressions for $j=1, 2$: 
\beq \label{eq:s1xi} 
 s_{1}(\xi)&=&\frac{3}{8\,\xi^3}\left( l_\xi+\frac{4}{3}\,\xi^2-\frac{3}{2}\xi -\frac{\xi}{2}\,R_\xi-\frac{\xi^2}{3}\,R_\xi^3 \right) \ ,  \\ \label{eq:s2xi} 
s_{2}(\xi)&=& \frac{R_\xi}{16} \left(9 + R_\xi+3\, R_\xi^2 +3 \,R_\xi^3 \right) , 
\eeq
 where $l_\xi=\ln{(1+2\,\xi)}$, and $R_\xi=1/(1+2\,\xi)$.

For the mean energy of the scattered photon we have then 
\be \label{eq:x1s} 
\langle x_1 \rangle \,s_{0}(\xi)=\xi\, \left[\gamma\,S_{1} (\xi)+x\,S_{2} (\xi) \right],
\ee
and for the mean square of energy
\be \label{eq:x12s} 
\langle x_1^2 \rangle \,s_{0}(\xi)=\gamma^2\,\xi^2\,S_{4} (\xi)-\gamma\,x\,\xi\,S_{5}(\xi)+
x^2\,S_{6}(\xi)-\xi^2\,S_{7} (\xi) ,
\ee
where 
\beq \label{eq:Ss12} 
S_{1}(\xi) & =& \left[s_{0} (\xi)-s_{1} (\xi) \right]/\xi,\quad  S_{2} (\xi)=\left[s_{1} (\xi)-S_{1} (\xi)\right]/\xi,  \nonumber\\ 
S_{3}(\xi)&=& \left[s_{1} (\xi)-s_{2} (\xi) \right]/\xi,\quad  S_{4} (\xi)= \left[S_{1} (\xi)-S_{3} (\xi) \right]/\xi,  \nonumber\\
S_{5}(\xi)& =& 3\,S_{4} (\xi)-4\,S_{3} (\xi),\quad S_{7} (\xi)=S_{3} (\xi)-S_{4} (\xi)/2, \nonumber\\
S_{6}(\xi)&=&s_{2}(\xi)-3\,S_{7} (\xi). 
\eeq
All  functions $S_{j}(\xi)$ are elementary. In addition, they are defined in such a way so that not to become zero at $\xi=0$. The series expansion  of functions $s$ and $S$ for small arguments are presented in Appendix \ref{app:funsS}.

\begin{figure}
\centerline{\epsfig{file=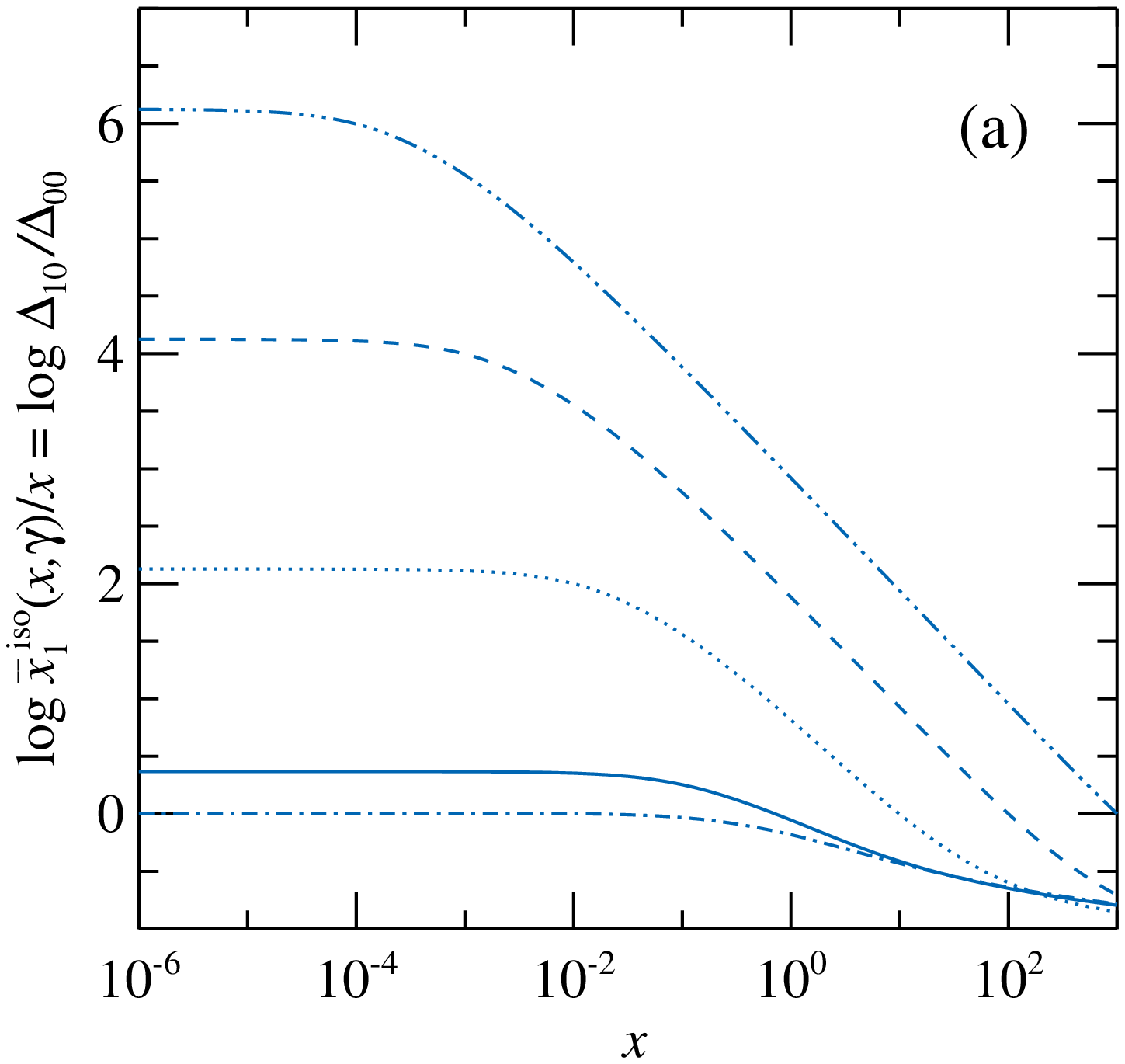,width=7.5cm} }
\centerline{\epsfig{file= 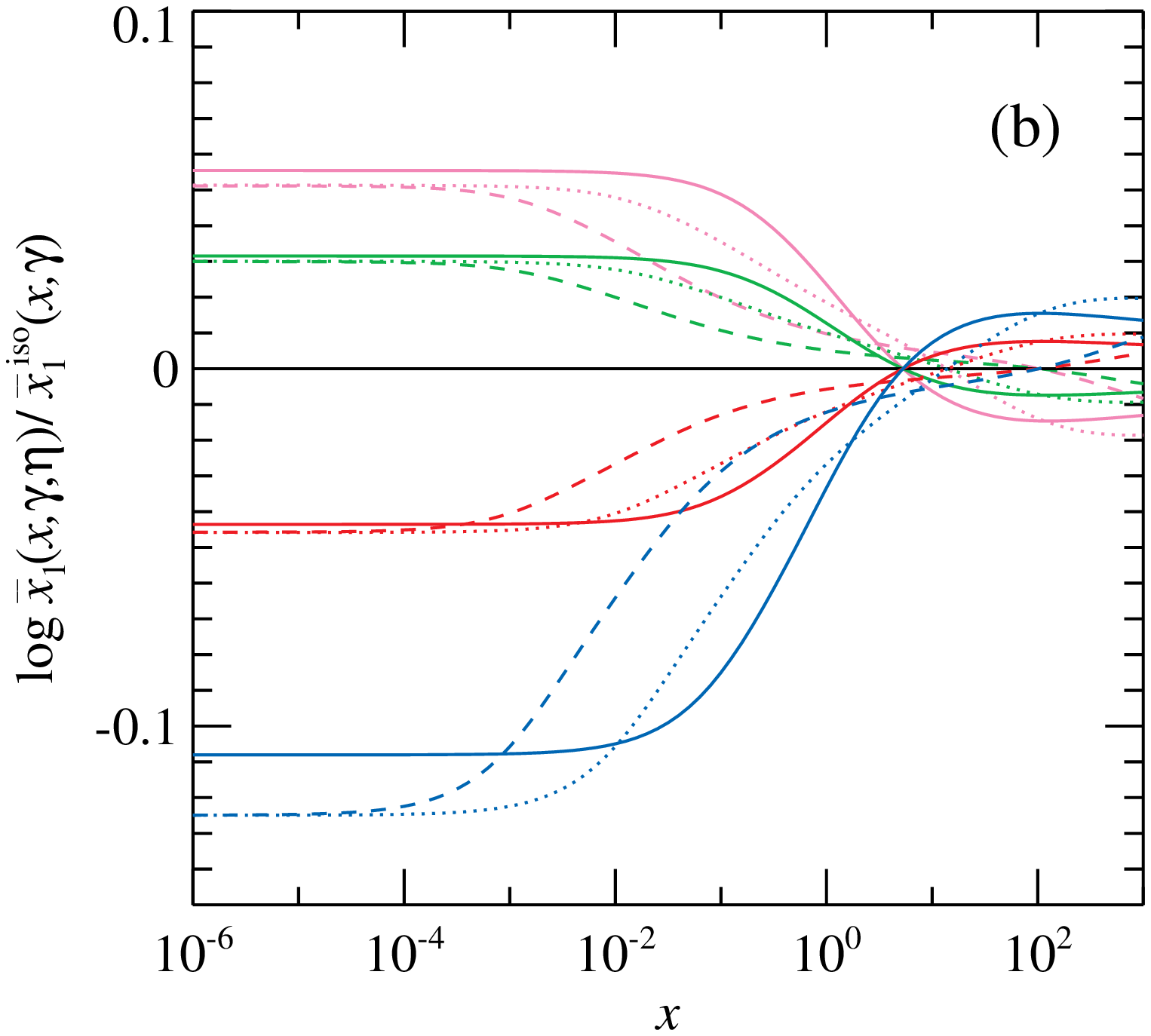,width=7.5cm} }
\centerline{\epsfig{file= 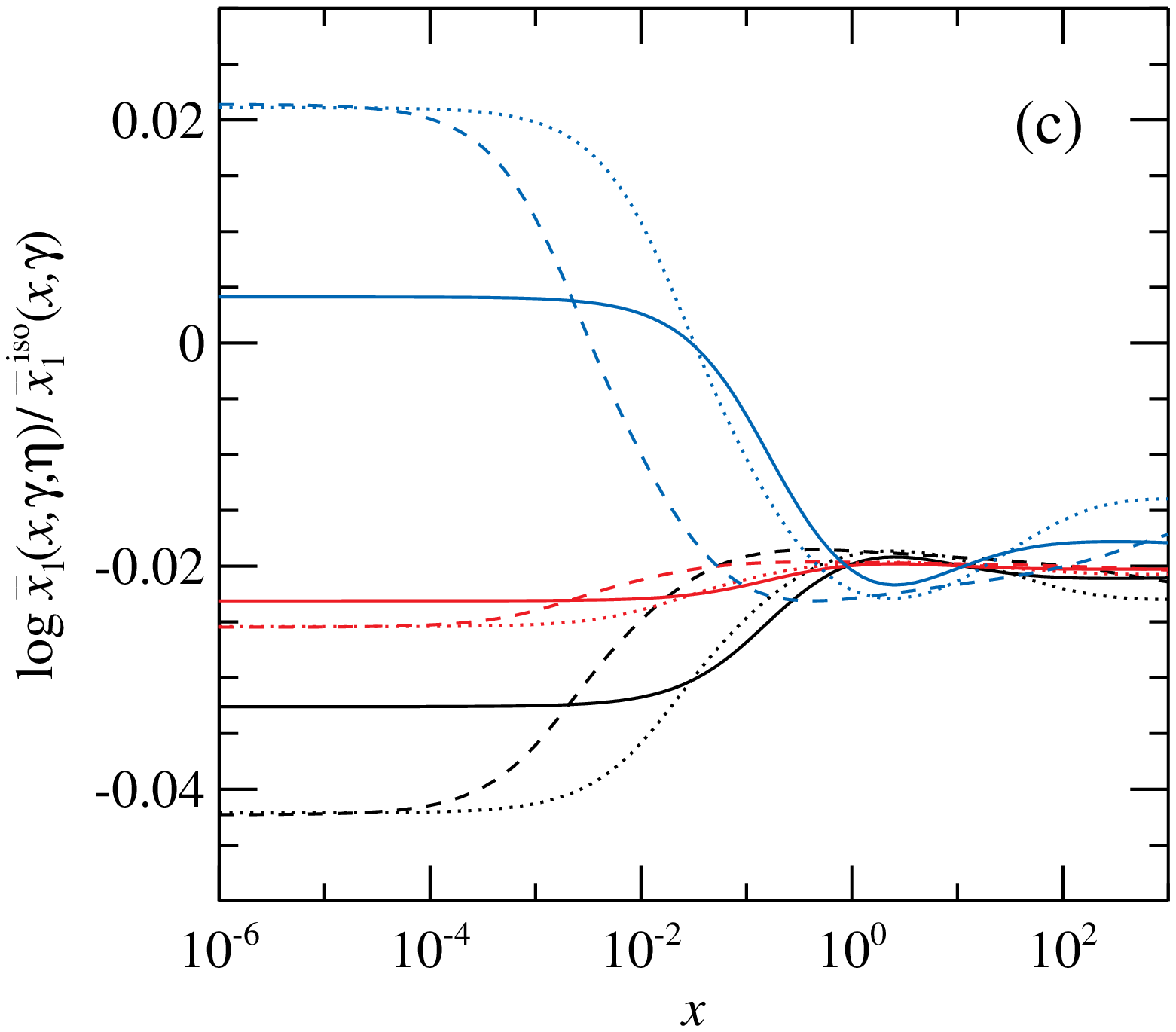,width=7.5cm} }
\caption{(a) Mean energy of scattered photons in units of the incident photon energy 
$\overline{x_1}^{\rm\, iso} (x,\gamma)/x= \Delta_{10}/\Delta_{00}$ as a function of $x$ 
for isotropic mono-energetic electrons of various momenta 
$p=0.1, 1, 10, 10^2, 10^3$ (from bottom  to top, dot-dashed, solid, dotted, dashed,  and triple-dot-dashed curves). 
The asymptotic value at small $x$ in Thomson approximation is $1+\frac{4}{3} p^2$. 
(b) A correction to the mean energy (in units of $\overline{x_1}^{\rm\, iso}$) arising from the linear term in the electron distribution
(\ref{eq:fe}) with $f_1/f_0=1$.  
Solid, dotted and dashed curves correspond for $p=1, 10, 100$, respectively. 
The curves from top to bottom correspond to $\eta=-1, -0.5, 0, 0.5, 1$. 
At small $x$, the curves approach the limiting value given by Equation~(\ref{eq:thoms_freq1}).  
(c) Same as (b), but for the quadrupole term in the electron distribution (\ref{eq:fe}) with $f_2/f_0=1$.   
These are even functions of $\eta$, the curves from the bottom to  the top correspond to $\eta=0, 0.5, 1$.  
At small $x$, the curves approach the limiting value given by Equation~(\ref{eq:thoms_freq2}). 
}
\label{fig:energy}
\end{figure}

\subsubsection{Averaging over electron directions}

We need to integrate in Equation~(\ref{eq:ovl_moments}) over   anisotropic electron distribution. We follow the derivation of the total cross-section that lead from Equation~(\ref{eq:sigmaxeta}) to Equation~(\ref{eq:s0_d_chi}). Representing integral over electron momentum $\rmd^3 p=p^2 \rmd p\, \rmd \zeta\,\rmd\Phi$ we get: 
\be \label{eq:s0_j_d_chi}
\ovl{x_1^j}\ovl{s_{0}}(x,\eta) = 
 4\pi\ x^j \sum_{k=0}^{2} P_k(\eta)\ \int_{1}^{\infty}  p\gamma\ \rmd \gamma \    f_k \  \Delta_{jk} (x,\gamma) ,
\ee
where 
\be\label{eq:Delta_j_xetagam}
\Delta_{jk} (x,\gamma) = 
 \frac{1}{2\gamma x^{1+j} } \int_{-1}^{1}  P_k(\zeta)\ \xi \ \langle x_1^j \rangle \, s_{0}(\xi) \  \rmd\zeta 
 = \sum_{n=0}^{k} \ b_{nk}\ \chijn  
\ee
and 
\be \label{eq:chi_jn}
\chijn (x,\gamma)= \frac{1}{2\gamma p x^{2+j+n}} \int_{x(\gamma-p)}^{x(\gamma+p)} 
\langle x_1^j \rangle s_{0}(\xi) \  \ \xi^{n+1}\  \rmd\xi  .
\ee

Functions $\chijzero=\Delta_{j0}$ coincide with functions $\Psi_j(x,\gamma)$ introduced by NP94, while functions $\chi_{0n}$ are given  by Equation~(\ref{eq:chi0n_exp}).  The explicit expressions for the function  $\chijn$ for $j=1, 2$ (which are analogous to functions $\Psi_{1}$ and $\Psi_{2}$ from NP94)  can be obtained using expression for mean powers of energies (\ref{eq:x1s}) or (\ref{eq:x12s}): 
\beq
\chi_{\!1n}(x,\gamma) & =&   \frac{1}{2\gamma p } \left. \frac{u^{3+n}}{3+n} 
\left[ \gamma \,   \Psi_{2+n,1}(xu) + x\,  \Psi_{2+n,2}(xu) \right] \right|_{u=\gamma-p}^{u=\gamma+p},  \\
\chi_{\!2n}(x,\gamma) & =&    \frac{1}{2\gamma p } \left.  \left[ 
\gamma^2 \frac{u^{4+n}}{4+n}  \Psi_{3+n,4}(xu) 
- \gamma \frac{u^{3+n}}{3+n}  \Psi_{2+n,5}(xu)  \right. \right. \nonumber \\
&+&  \left.  \left. \frac{u^{2+n}}{2+n}  \Psi_{1+n,6}(xu) 
- \frac{u^{4+n}}{4+n} \Psi_{3+n,7}(xu) \right] \right|_{u=\gamma-p}^{u=\gamma+p}, 
\eeq
where 
\be \label{eq:ppsi_ij}
\Psi_{ij}(\xi)  = \frac{i+1}{\xi^{i+1}}  \int_0^{\xi} t^i S_{j} (t) \rmd t .
\ee
These are related to functions 
\be \label{eq:Psi_ij}
\psi_{ij}(\xi) =  \frac{i+1}{\xi^{i+1}}  \int_0^{\xi} t^i s_{j} (t) \rmd t ,
\ee
because  functions $S_{j}$ are expressed through $s_{j}$. The explicit expressions for both type of these functions as well as their series expansions for small arguments are given in Appendices~\ref{app:psi_ij}.

As in the case of functions $\Delta_{0k}$, for calculating $\Delta_{jk}$, we consider  three regimes: (1) $x\gamma\ll 1$, when we use the series expansion (see Appendix \ref{app:chi_ij}); (2) for $p\ll 1$ we numerically take  the integral in Equation~(\ref{eq:Delta_j_xetagam}) using Gaussian quadrature; (3) in other cases, we use the sum in Equation~(\ref{eq:Delta_j_xetagam})  and analytical expressions for $\chijn$.

For   mono-energetic electron distribution 
(\ref{eq:fe_mono}) of Lorentz factor $\gamma$, we can introduce the mean powers of photon energy 
analogously to Equation~(\ref{eq:s0_j_d_chi}): 
\be \label{eq:s0_j_d_chi_gam}
\ovl{x_1^j}\, \ovl{s_{0}}(x, \gamma,\eta) = \ x^j  
\sum_{k=0}^2 \frac{f_k}{f_0}\, P_k(\eta)\, \Delta_{jk} . 
\ee
The mean energy of scattered photons  for such electrons for a scattering act is given by the ratio of Eqs. (\ref{eq:s0_j_d_chi_gam}) and 
(\ref{eq:s0_d_chi_gam}). 
It is shown in Figure~\ref{fig:energy}a. In the low-energy (Thomson) limit the energy gain factor is given by  a well known expression $\ovl{x_1}^{\rm iso}/x = 1+4p^2/3$, which translated to $\frac{4}{3} \gamma^2$ at large $\gamma$.   
The relative corrections arising due to the dipole and quadrupole term in the electron distribution are shown  in Figures~\ref{fig:energy}b and \ref{fig:energy}c, respectively.  
Using asymptotic expansions of $\Delta_{jk}$ in the Thomson limit (see Appendix \ref{app:chi_ij}),  we get the 
asymptotic value 
\be 
\frac{\ovl{x_1}(x,\gamma,\eta)}{\ovl{x_1}^{\rm\, iso}(x,\gamma) } 
= \frac{\gamma^2\left(1+\frac{1}{3}\beta^2\right)  -\frac{2}{3}\beta \gamma^2 \frac{f_1}{f_0} \eta + \frac{2}{15} \beta^2\gamma^2  \frac{f_2}{f_0}P_2(\eta)}
{ \gamma^2\left(1+\frac{1}{3}\beta^2\right)  \left(1- \frac{1}{3}\beta \frac{f_1}{f_0}\eta \right)} . 
\ee
Thus in non-relativistic limit $\beta\ll 1$, the correction is negligible. In the relativistic limit $\gamma\gg 1$, 
the relative corrections arising from the two terms  are 
\beq \label{eq:thoms_freq1}
\frac{\ovl{x_1}(x,\gamma,\eta)}{\ovl{x_1}^{\rm\, iso}(x,\gamma) }  &=& \frac{1-\frac{1}{2} \frac{f_1}{f_0} \eta }{1-\frac{1}{3} \frac{f_1}{f_0}\eta },  \\ 
\label{eq:thoms_freq2}
\frac{\ovl{x_1}(x,\gamma,\eta)}{\ovl{x_1}^{\rm\, iso}(x,\gamma) } &=&  1 + \frac{1}{10}\frac{f_2}{f_0}P_2(\eta). 
\eeq

\begin{figure}
\centerline{\epsfig{file=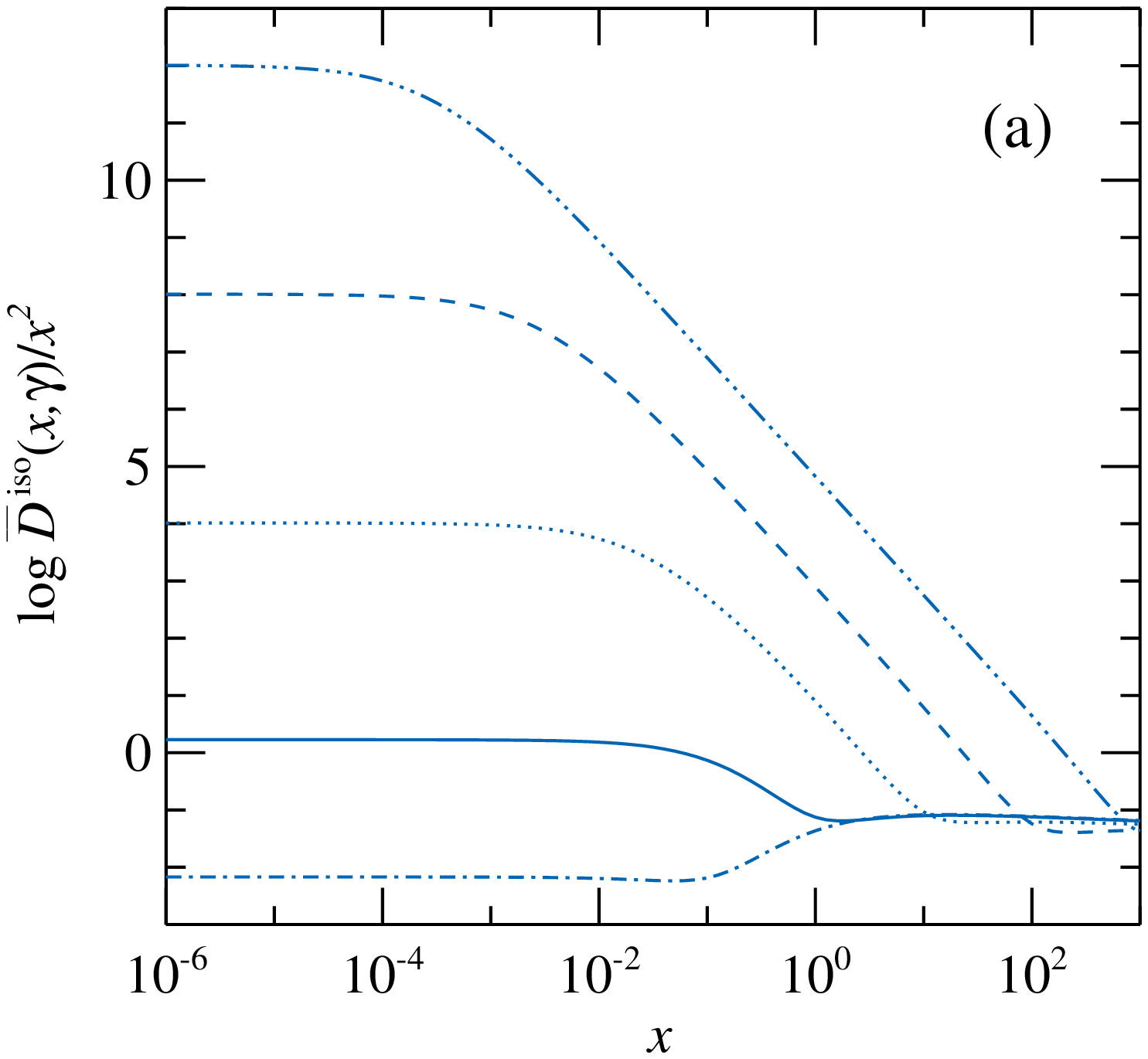,width=7.5cm} }
\centerline{\epsfig{file= 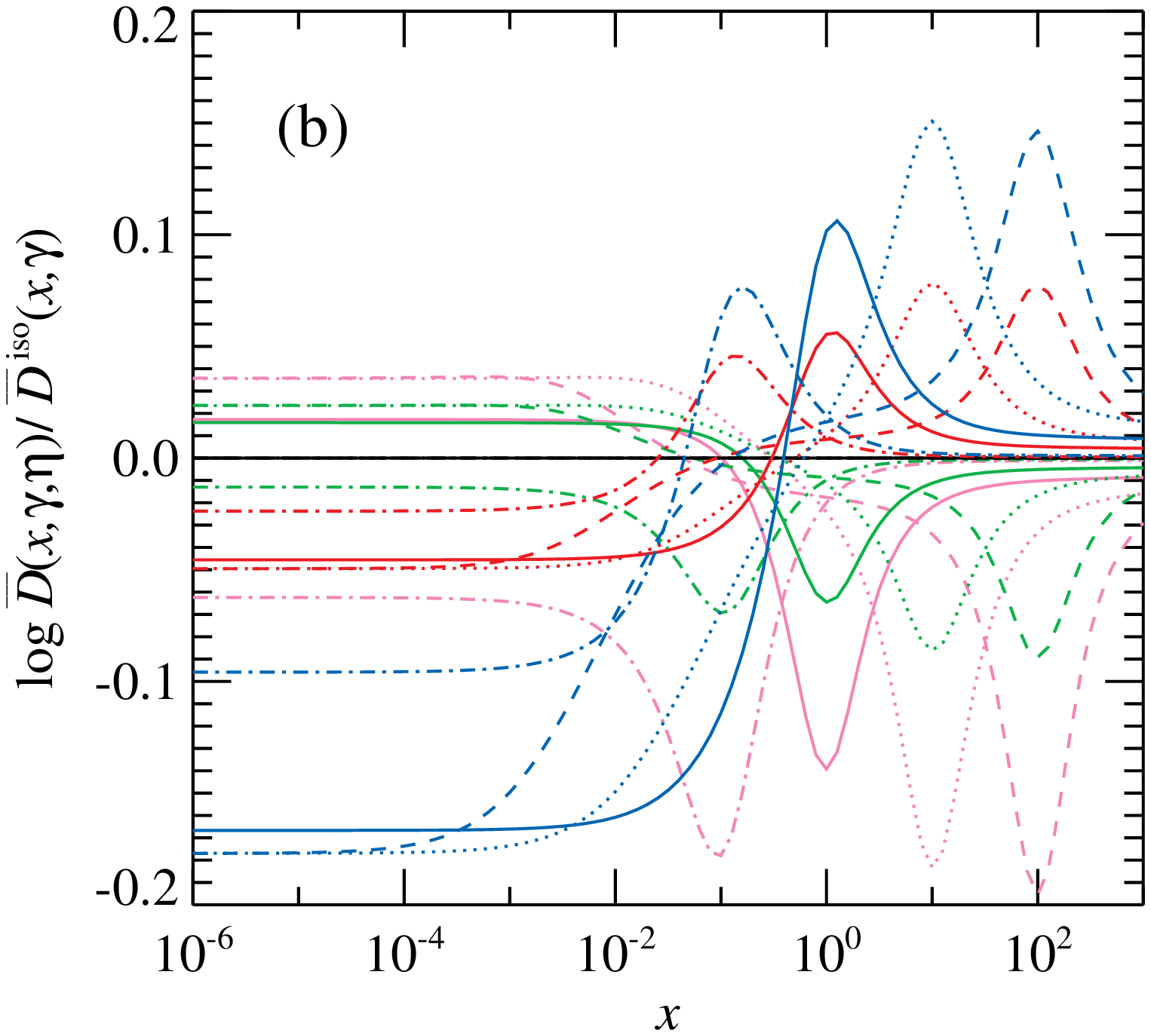,width=7.5cm} }
\centerline{\epsfig{file= 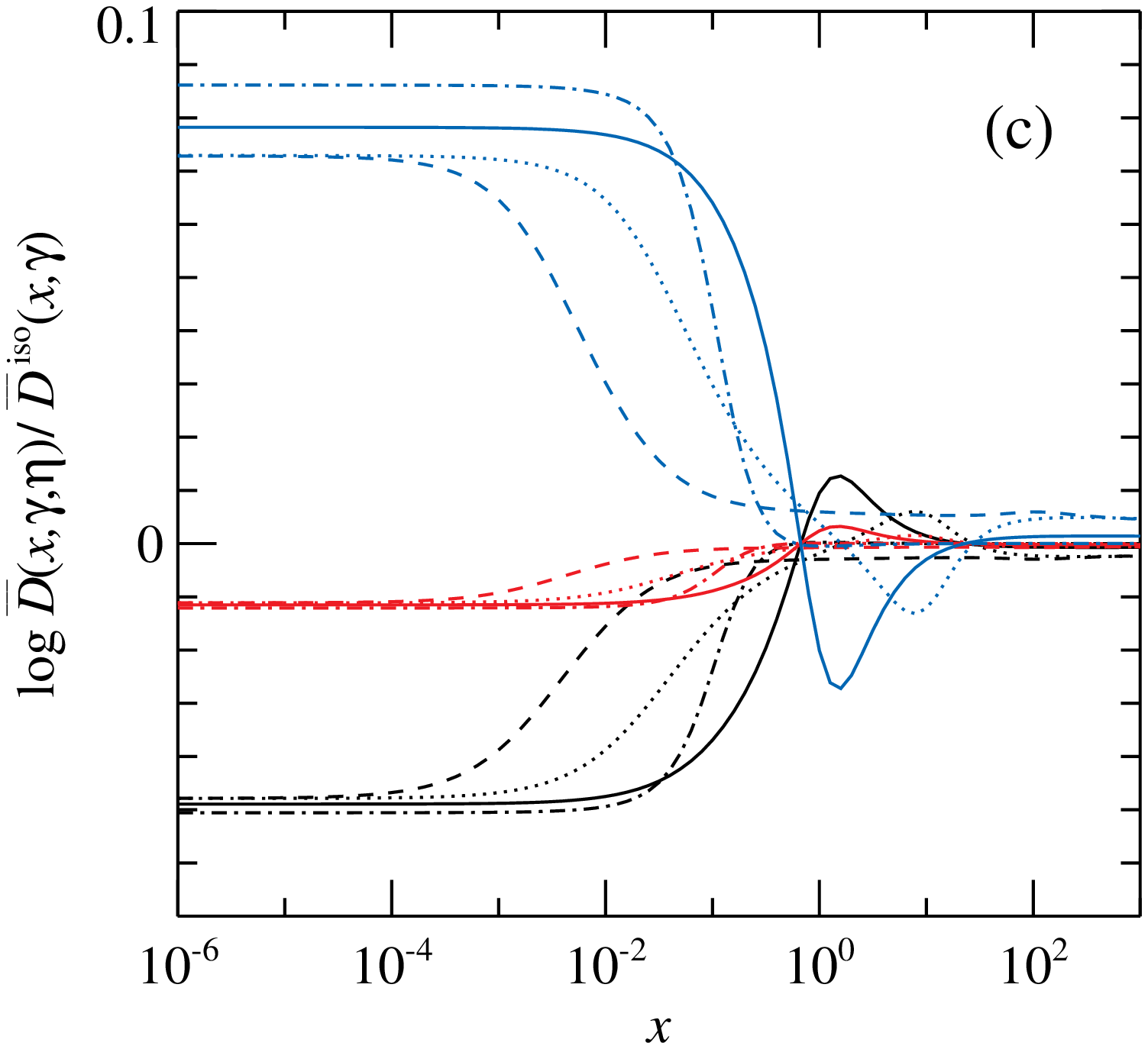,width=7.5cm} }
\caption{(a) Dispersion of the energy of scattered photons (in units of  $x^2$) 
for isotropic mono-energetic electrons of $p=0.1, 1, 10, 10^2, 10^3$ (from bottom  to top). 
The asymptotic value of $\ovl{D}/x^2$ at small $x$ in Thomson approximation is $\frac{2}{45}(23\gamma^2-8)p^2$.   
(b) A correction to the dispersion (in terms of the isotropic quantity) arising from the linear term in the electron distribution
(\ref{eq:fe}) with $f_1/f_0=1$. 
Solid, dotted and dashed curves correspond for $p=1, 10, 100$, respectively. 
The curves from top to bottom correspond to $\eta=-1, -0.5, 0, 0.5, 1$. 
(c) A correction to the dispersion  arising from the quadrupole term in the electron distribution
(\ref{eq:fe}) with $f_2/f_0=1$ for the same $p$ and $\eta$  as in panel (b).  
These are even functions of $\eta$, the curves from bottom to  top correspond to $\eta=0, 0.5, 1$.  
}
\label{fig:dispersion}
\end{figure}

\subsection{Energy exchange and dispersion}

The difference of the photon energies before and after scattering $x-x_1$ is of course just the energy transfer to the electron gas. For the fixed angle between electrons and incident photons $\zeta$ (and fixed electron energy $\gamma$),  the  energy loss averaged over the directions of scattered photons is $x - \langle x_1 \rangle$. The product $(x - \langle x_1 \rangle)\Ne\sigmat \,  s_{0}(\xi)$ is then 
the energy loss on a unit length. From Equation~(\ref{eq:x1s}) we can easily get  (see also NP94): 
\be \label{eq:x-x1_angle}
\left( x \! - \! \langle x_1 \rangle \right) s_{0}(\xi)\! =\!   x s_{0}(\xi) - \gamma \xi S_{1}(\xi) - x \xi S_{2}(\xi) = 
(x+ x \xi - \gamma \xi) S_{1}(\xi). 
\ee
The corresponding energy loss (per unit length and in units $\Ne\sigmat$) averaged over the electron directions (and integrated over electron energies)  becomes [see Eqs. (\ref{eq:s0_d_chi})  and  (\ref{eq:s0_j_d_chi})]: 
\be \label{eq:energy_exchange}
\left( x - \ovl{x_1} \right) \ovl{s_{0}}(x,\eta)  = 
4\,\pi x \sum_{k=0}^{2} \! P_k(\eta) \int_{1}^{\infty} \!\!\! p\gamma\ \rmd \gamma \
 f_k  \, \left( \Delta_{0k}  - \Delta_{1k}  \right) .   
\ee
The heating rate per unit volume is then 
\be  \label{eq:heating_rate}
\dot{E} = \Ne\sigmat \int \rmd x \int \rmd^2\omega\ I (x,\vomega) \left( 1 - \frac{\ovl{x_1}}{x} \right) \ovl{s_{0}}(x,\eta) , 
\ee
where $I (x, \vomega)$ is the specific intensity of radiation in a given direction. This expression can be positive (so called Compton heating) when the photons typically have larger energies than the electron gas, or negative (Compton cooling) when one considers cooling of the relativistic electron gas by soft radiation.  
 
The dispersion  of the scattered photon  energy is given by the usual expression $\ovl{D}(x)=\ovl{x_1^2}-\ovl{x_1}^2$, which of course depends on the electron momentum distribution.  For mono-energetic electrons we can define the dispersion as 
\be 
\ovl{D}(x,\gamma,\eta) = \ovl{x_1^2}(x,\gamma, \eta)-\ovl{x_1}^{\,2}(x,\gamma, \eta) , 
\ee
where $\ovl{x_1^j}(x,\gamma, \eta)$ are given by Equation~(\ref{eq:s0_j_d_chi_gam}). 
The dispersion  for isotropic electrons is shown in Figure~\ref{fig:energy}a. 
The low-energy (Thomson) limit for $x\ll 1/\gamma$ is  (see NP94) 
\be
\ovl{D}(x,\gamma) = x^2  \frac{2}{45}\,\left(23\,\gamma^2-8 \right)\,p^2 .
\ee  
The relative corrections arising due to the dipole and quadrupole term in the electron distribution reach about 50 per cent and 
are shown  in Figures~\ref{fig:dispersion}b and \ref{fig:dispersion}c, respectively.

\subsection{Radiation force}

Now we would like to derive analytic expression for the radiation force acting on the electron gas. NP94 have developed a formalism appropriate for    isotropic electron distribution, when the averaged transferred momentum is along the momentum of the incoming photons, because of the symmetry. For the electron distribution  described by Equation~(\ref{eq:fe}), the momentum is transferred in the plane containing the initial photon momentum  and the symmetry axis $\vl_3$. If the incident photons are axially symmetric around $\vl_3$, then obviously, the total momentum transferred to the electrons has be parallel to $\vl_3$ by symmetry. We derive here more general formulae for the total momentum transferred by a beam of photons propagating along direction $\vomega$ (such as $\vomega\cdot\vl_3=\eta$), as well as its projections to $\vl_3$ and perpendicular direction. 

Let us introduce the vector basis: 
\be \label{eq:basis_l3_omega}  
\ve_1(\vomega,\vl_3)=\frac{\vl_3-\eta\,\vomega}{\sqrt{1-\eta^2}},\,\,\,
\ve_2(\vomega,\vl_3)=\frac{\vomega \times \vl_3}{\sqrt{1-\eta^2}},\,\,\,
\ve_3=\vomega . 
\ee
In a single scattering act  the momentum transferred is 
\be 
\vecQ = x\vomega-x_1 \vomega_1. 
\ee
The components of the momentum transferred to the electron gas along and 
perpendicular to $\vomega$ are: 
\beq \label{eq:Q3_def}
Q_3 & =&  x -  x_1\,  \vomega_1 \cdot \ve_3 = x- x_1 \mu, \\
\label{eq:Q1_def}
Q_1 & = & - x_1\, \vomega_1 \cdot \ve_1  = - x_1 \frac{\eta_1-\eta\mu}{\sqrt{1-\eta^2 }}. 
\eeq
Analogously to Equation~(\ref{eq:ovl_moments}), we define the mean transferred momentum as
\be \label{eq:ovl_pressure}
\overline{\vecQ} \ \overline{s}_0(\vecx)  =   \frac{1}{x} \int 
\langle \vecQ  \rangle s_{0} (\xi) \: \xi \: \fe(\gamma,\etae) \: \frac{\rmd ^3 p}{\gamma},
\ee
where 
\beq \label{eq:sfun_Q}
\lefteqn{ \langle \vecQ  \rangle s_{0}(\xi) =  \frac{3}{16\pi} \frac{1}{\xi} \int \frac{\rmd ^3 p_1}{\gamma_{1}} \frac{\rmd ^3 x_1}{x_{1}}
\  F \  \vecQ \ \delta^{4} (\fourp_1 + \fourx_1 - \fourp - \fourx) } \\
&\!=\!&  \frac{3}{16\pi} \frac{1}{\xi}\! \int\!\!  x_1 \rmd x_1 \rmd^2\omega_1 
 \,  F \, \vecQ  \  \delta\!\left\{ x_1 [\gamma+ x - \vomega_1\!\cdot\! ( p\vOmega\!+\!x\vomega) ] - \xi \right\} . \nonumber
\eeq

\subsubsection{Averaging over photon directions}

Let us introduce a vector basis defined by the photon and electron momenta: 
\be \label{eq:omeOme} 
\ve_1(\vomega,\vOmega)=\frac{\vOmega-\zeta\,\vomega}{\sqrt{1-\zeta^2}},\,\,\,
\ve_2(\vomega,\vOmega)=\frac{\vomega \times \vOmega}{\sqrt{1-\zeta^2}},\,\,\,
\ve_3=\vomega,
\ee
where $\zeta=\vomega\cdot\vOmega$, therefore $\vOmega=\sqrt{1-\zeta^2}\,\ve_1(\vomega,\vOmega)+
\zeta\,\ve_3$. Fixing the angle $\arccos\zeta$ between the  electrons of momentum $p\,\vOmega$ and the incident photon momentum and averaging over directions of scattered photons, we can get the mean momentum transmitted in the direction  $\vomega$: 
\be \label{eq:xx1mu}
\langle Q_3 \rangle \,s_{0}(\xi) = \langle x-x_1\rangle \,s_{0}(\xi)+\langle x_1\,(1-\mu)\rangle \,s_{0}(\xi). 
\ee

The first term is given by Equation~(\ref{eq:x-x1_angle}), the second term is 
 \beq \label{eq:xxmuS}
\lefteqn{  
\langle x_1(1-\mu)\rangle \,s_{0}(\xi)=
\frac{3}{16\pi\xi^2} \int x_1^3(1-\mu)F\rmd^2 \omega_1 } \nonumber \\
&= &\frac{3}{16\,\pi\,x\,\xi} \int \xi_1\,(1-\mu_0)\,F\,\rmd \xi_1\,\rmd \phi_0 \\
&=&   \frac{3}{8\,x\,\xi^2} \int_{\xi/(1+2\,\xi)}^\xi \!\!\!\!\! (\xi-\xi_1)\,F\,\rmd \xi_1 
=\frac{\xi}{x}\,[s_{0}(\xi)-s_{1}(\xi)]=\frac{\xi^2}{x}\,S_1(\xi), \nonumber 
\eeq
where we have used the invariant given by Equation (\ref{eq:q_kk1}) and changed the variables according to Equation (\ref{eq:invar_x1omega}). Thus, for the fixed electron and photon energies and the angle between their momenta, the mean momentum transmitted to the electron gas in the direction of the initial photon propagation $\vomega$, in accordance with (\ref{eq:Ss12}), (\ref{eq:x1s}) and (\ref{eq:xxmuS}), is (NP94)
\be \label{eq:xx1m}
\langle Q_3  \rangle s_{0}(\xi) \!=\! 
\langle x- x_1\,\mu\rangle s_{0}(\xi) \!=\! \left(x+x\,\xi-\gamma\,\xi+\frac{\xi^2}{x} \right)S_{1}(\xi) .
\ee

In contrast to NP94, we are now interested to know the total momentum transfer. Obviously, by symmetry, it has to lie in the $(\vOmega, \vomega)$ plane. Averaging expression (\ref{eq:Q1_def}) for $Q_1$ over angles is not easy, but we can compute the momentum transferred along the electron momentum: 
$ Q_{\Omega}\equiv x\zeta-x_1 \zeta_1$.  Using identity $x_1 \zeta_1= (\gamma x_1-\xi_1)/p$ and substituting  Eqs. (\ref{eq:x1j}) and  (\ref{eq:mu0_pzeta0}), similarly to Equation~(\ref{eq:sfun_xj3}), we get 
\beq \label{eq:x1zeta1}
\lefteqn{  \langle x_1 \zeta_1 \rangle s_{0}(\xi) = \frac{3}{16\pi} \frac{1}{\xi^2} \int_{\xi/(1+2\xi)}^{\xi}  
  F\ \rmd \xi_1  \int_{0}^{2\pi} x_1\zeta_1  \rmd \phi_0 } \nonumber \\
&=& \frac{3}{8\xi^2} \int_{\xi/(1+2\xi)}^{\xi}  
  F\ \rmd \xi_1 \frac{1}{p} \left[ \left( \frac{\gamma x}{\xi} +   \frac{\gamma x}{\xi^2} -  \frac{\gamma^2}{\xi} -1 \right)\xi_1 
  + \gamma \left( \gamma  -\frac{x}{\xi}  \right)\right]
\nonumber \\
&=&   \frac{1}{p} \left( \gamma x +   \frac{\gamma x}{\xi} -  {\gamma^2} -\xi \right) s_{1}(\xi)  
  + \frac{\gamma}{p} \left( \gamma  -\frac{x}{\xi}  \right) s_{0}(\xi) 
\nonumber \\
&=&  \frac{1}{p} \left[ \gamma^2 \xi\, S_{1}(\xi) +  \gamma x\, \xi\,  S_{2}(\xi) -  \xi \, s_{1}(\xi) \right], 
\eeq
and using identity $x\zeta=(\gamma x - \xi)/p$, we finally obtain
\be \label{eq:xzeta-x1zeta1}
\langle Q_{\Omega}  \rangle s_{0}(\xi) = \langle x\zeta- x_1 \zeta_1 \rangle s_{0}(\xi) =
  \frac{1}{p} \left( \gamma x +  \gamma x\, \xi - \gamma^2 \xi - \xi^2 \right) S_{1}(\xi). 
\ee
The momentum along $\ve_1 (\vomega,\vOmega)$ is then simply
\beq \label{eq:Q1_angle}
\lefteqn{ \langle Q_{1} \rangle s_{0}(\xi) =
\frac{  \langle Q_{\Omega} \rangle  -  \zeta\ \langle Q_{3} \rangle  }{\sqrt{1-\zeta^2}}   s_{0}(\xi) } \nonumber \\
&=&    \frac{S_{1}(\xi)}{p\!\sqrt{1-\zeta^2}} \left(  \xi \!-\! 2 \gamma\frac{\xi^2}{x} \!+ \!\frac{\xi^3}{x^2}  \right) 
 = - p \sqrt{1-\zeta^2} \, \xi \, S_{1}(\xi) .
\eeq
Thus the total transferred momentum $\vecQ$ averaged over directions of scattered photons can be decomposed into two components along basis vectors: 
\be
\langle \vecQ  \rangle 
=   \langle Q_{1} \rangle \ \ve_1 (\vomega,\vOmega) + \langle Q_{3} \rangle \ \ve_3 .
\ee

\subsubsection{Averaging over electron directions}

As in previous sections, we choose to measure azimuth of the electron momentum  $\Phi$ in the frame defined by Eqs. (\ref{eq:basis_l3_omega}) from the projection of vector $\vl_3$ onto the plane perpendicular to $\vomega$. Therefore, 
\be
\langle \vecQ  \rangle 
= \langle Q_{1} \rangle \left[ \cos\Phi \ \ve_1 (\vomega,\vl_3)  + \sin\Phi\ \ve_2 (\vomega,\vl_3) \right] + 
 \langle Q_{3} \rangle \  \ve_3 .
\ee
The total momentum transfer averaged over the electron distribution is obtained from Equation~(\ref{eq:ovl_pressure}): 
\be \label{eq:ovl_vecQ}
\ovl{\vecQ}\, \ovl{s_{0}}(x,\eta)  = \frac{1}{x} \int_{1}^{\infty} \! \! p\,\rmd \gamma \! \! 
\int_{-1}^{1} \! \! \rmd\zeta \, \xi \!  \int_0^{2\pi} \! \!  \rmd\Phi \ \langle \vecQ\rangle s_{0}(\xi)  \, {\fe}(\gamma,\etae). 
\ee
Obviously, the component along $\ve_2 (\vomega,\vl_3)$ becomes zero, as $\fe$ is an even function of $\Phi$. 
The term along $\ve_1 (\vomega,\vl_3)$ involves integration of $\fe$ over azimuth with the weight $\cos\Phi$, and its averaged value is
\be 
\ovl{ \fe \cos\Phi} 
= \sum_{k=1}^2 \frac{f_k}{k(k+1)} P_k^1(\eta) P_k^1(\zeta) , 
\ee
where $P_k^1$ are the associated Legendre functions: 
\beq
P_1^1(u)= \sqrt{1-u^2}, \quad P_2^1(u)= 3\, u\sqrt{1-u^2} .
\eeq
 
It is worth mentioning that the isotropic component of the electron distribution $f_0$ does not contribute to the momentum transferred perpendicular to $\vomega$ by symmetry. 
Substituting expression (\ref{eq:Q1_angle}) to Equation~(\ref{eq:ovl_vecQ}), we get the first component of vector
\be\label{eq:ovl_Q1}
\ovl{Q}_1 \ovl{s_{0}}(x,\eta)  
=  4\pi x   \sum_{k=1}^{2} P_k^1(\eta) \int_{1}^{\infty} \! \! p\gamma\ \rmd \gamma \,
 f_k \  \Delta_{1k}^{\bot} (x,\gamma), 
\ee
where 
\be \label{eq:Psi_perp}
\Delta_{1k}^{\bot}(x,\gamma)  =  \frac{1}{k(k+1)}  
\frac{1}{2\gamma x^{2} } \int_{-1}^{1}  P_k^1(\zeta)\ \xi \ \langle Q_1 \rangle \, s_{0}(\xi) \  \rmd\zeta .
\ee
Changing the integration variable to $\xi$, and introducing a set of functions
\beq \label{eq:chi_bot}
\chionen^{\bot}  (x,\gamma) &=  & \frac{1}{2\gamma p\ x^{3+n}}   
\int_{x(\gamma-p)}^{x(\gamma+p)} \! \! \sqrt{1-\zeta^2} \ 
 \langle Q_{1} \rangle s_{0}(\xi) \  \ \xi^{n+1}\  \rmd\xi   \nonumber \\
 &=&  \frac{1}{2\gamma p}  \frac{1}{p} 
 \left.  \left[  \frac{u^{3+n}}{3+n}  \Psi_{2+n,1}(xu) 
- 2 \gamma \frac{u^{4+n}}{4+n}  \Psi_{3+n,1}(xu)  \right. \right. \nonumber \\
&+& \left. \left.  \frac{u^{5+n}}{5+n}  \Psi_{4+n,1}(xu) 
 \right] \right|_{u=\gamma-p}^{u=\gamma+p} , \ n=0,1 ,
\eeq
we get
\be
\Delta_{11}^{\bot} = \frac{1}{2}\, \chi_{10}^{\bot} , \quad 
\Delta_{12}^{\bot} =  \frac{1}{2p} \left( \gamma\chi_{10}^{\bot}-\chi_{11}^{\bot}\right) .
\ee

\begin{figure}
\centerline{\epsfig{file=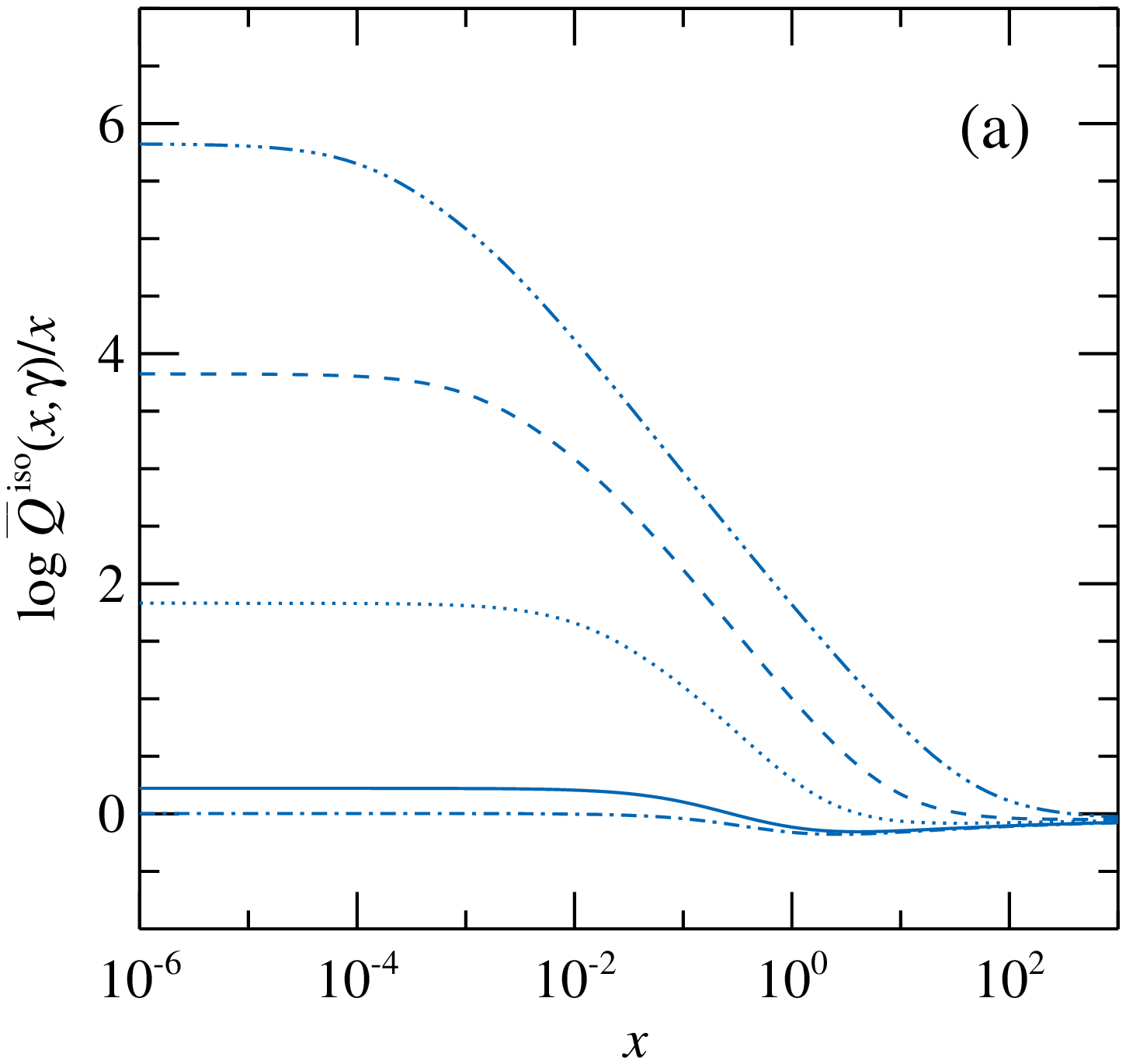,width=7.5cm} }
\centerline{\epsfig{file= 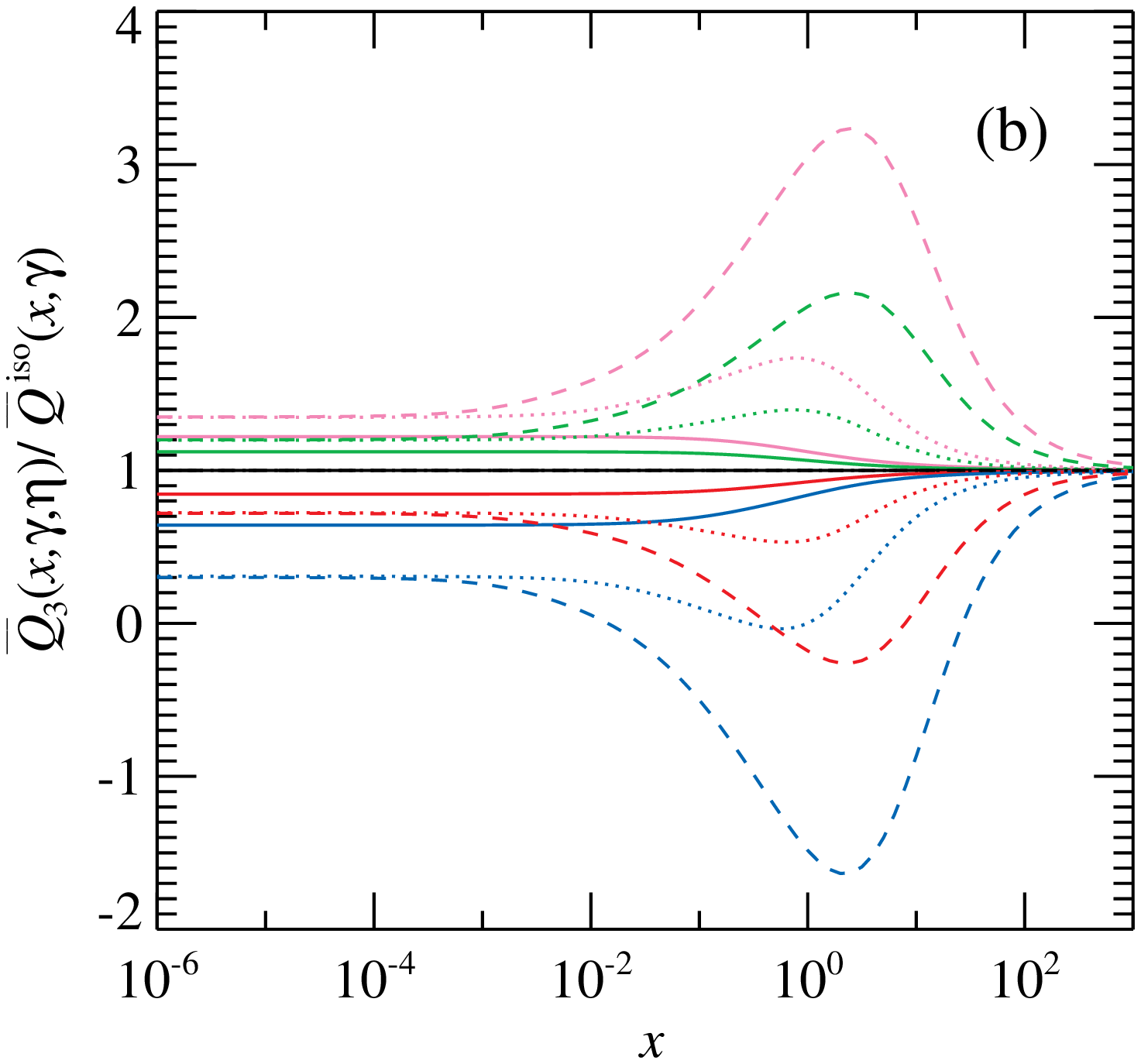,width=7.5cm} }
\centerline{\epsfig{file= 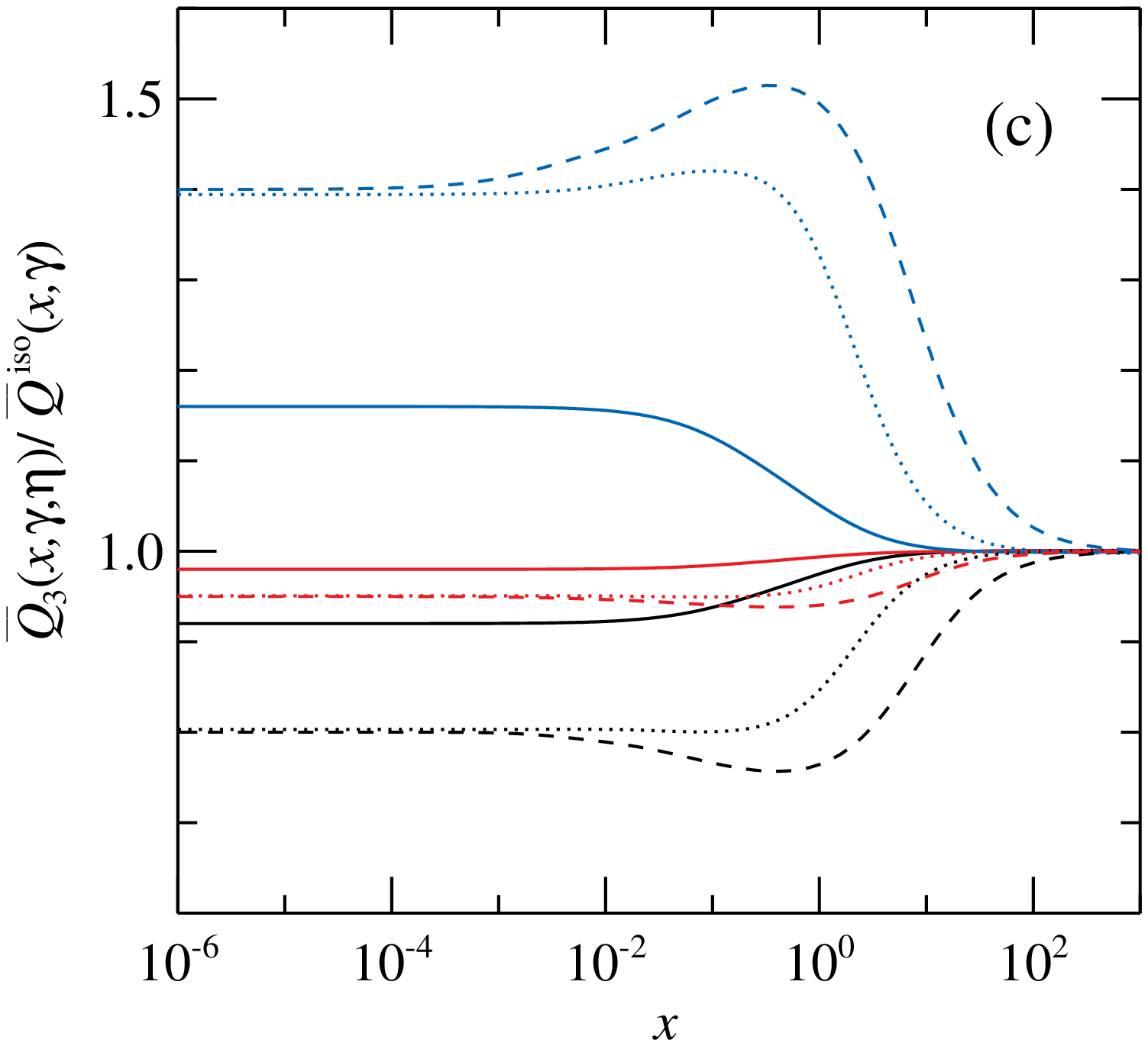,width=7.5cm} }
\caption{(a) Average momentum transferred to the electron gas per scattering $\ovl{Q}^{\rm\, iso}$ (in units of  $x$)  for isotropic mono-energetic electrons (with $f_1=f_2=0$). Curves from bottom  to top correspond to the electron momenta $p=0.1, 1, 10, 10^2, 10^3$. The asymptotic value at small $x$ in Thomson approximation is $1+2 p^2/3$ [NP94; see Equation~(\ref{eq:ovl_Qomega_gam_Thom})].    
(b) The momentum transferred along $\vomega$ for   anisotropic electrons with $f_1/f_0=1$ in units of the isotropic quantity $\ovl{Q}^{\rm\, iso}$.  Solid, dotted and dashed curves correspond to $p=1, 10, 100$, respectively. 
The curves from top to bottom correspond to $\eta=-1, -0.5, 0, 0.5, 1$ (pink, green, black, red, and blue curves, respectively). 
(c) Same as (b), but for the electron distribution (\ref{eq:fe})  with the quadrupole term with $f_2/f_0=1$ for the same $p$ and $\eta$  as in panel (b).  These are even functions of $\eta$, the curves from the bottom to  the top correspond to $\eta=0, 0.5, 1$.  
The flat parts of the curves correspond to the Thomson limit given by Equation~(\ref{eq:ovl_Qomega_gam_Thom}).
}
\label{fig:radpress}
\end{figure}

Now let us evaluate the component of vector (\ref{eq:ovl_vecQ}) along $\ve_3$. Because $ \langle Q_{3} \rangle$ does not depend on azimuth $\Phi$, the azimuthal integration just gives the averaged electron distribution given by Equation~(\ref{eq:fe_ave_pk}).  
Thus we get 
\be\label{eq:ovl_Qomega}
\ovl{Q}_{3}\, \ovl{s_{0}}(x,\eta)  \! =  \! 
4\pi x \sum_{k=0}^{2} P_k(\eta) \!\!\int\limits_{1}^{\infty} \!  p\gamma\ \rmd \gamma 
 f_k\,  \left(\Delta_{0k}-\Delta_{1k}+\Delta_{1k}^*\right), 
\ee
where 
\be \label{eq:delta*}
\Delta_{1k}^*(x,\gamma) \!=\! 
 \frac{1}{2\gamma x^{2} } \!\!\int\limits_{-1}^{1}\!\!  P_k(\zeta)\, \xi \, \langle x_1(1-\mu) \rangle s_{0}(\xi) \,  \rmd\zeta  
 \!=\! \sum_{n=0}^{k} \ b_{nk} \, \chionen^{*}  ,
\ee
and 
\beq \label{eq:chi_vert}
\chionen^{*}  (x,\gamma) &=  & \frac{1}{2\gamma p\ x^{3+n}}   
\int_{x(\gamma-p)}^{x(\gamma+p)} \! \!  \langle x_1(1-\mu)\rangle \,s_{0}(\xi) 
   \ \xi^{n+1}\  \rmd\xi   \nonumber \\
 &=&  \frac{1}{2\gamma p} \left.  \frac{u^{4+n}}{4+n}  \Psi_{3+n,1}(xu)  \right|_{u=\gamma-p}^{u=\gamma+p}, \ n=0,1,2. 
\eeq

For isotropic electron distribution, the only function of interest is $\Delta_{10}^*$ which coincides 
with function $\Psi_1^*(x,\gamma)$ introduced by NP94. 
Now combining Eqs. (\ref{eq:ovl_Q1}) and  (\ref{eq:ovl_Qomega}), we get the momentum transfer along the symmetry axis of the electron distribution $\vl_3$ and perpendicular to it: 
\beq \label{eq:ovlQ_l3_perp}
\ovl{Q}_{\Vert}  &=& \sqrt{1-\eta^2}\, \ovl{Q}_{1}  + \eta \,\ovl{Q}_{3} , \\ 
\ovl{Q}_{\bot} & = &-  \eta\, \ovl{Q}_{1}  +\sqrt{1-\eta^2} \,\ovl{Q}_{3} . 
\eeq
Expressions (\ref{eq:ovl_Qomega}), (\ref{eq:ovl_Q1}) and (\ref{eq:ovlQ_l3_perp}) give the momentum transferred to the electron gas (in terms of one integral over the electron energy) along $\vomega$, perpendicular to that direction as well as along vector $\vl_3$ and  perpendicular to it. 

Similarly to Equation~(\ref{eq:heating_rate}), we can also get the two components of the momentum transfer rate per unit volume:  
\be  \label{eq:momentum_transfer_rate}
\dot{P}_{1,\,3} = \frac{\Ne\sigmat}{c} \int \frac{\rmd\, x}{x} \int \rmd^2\omega\ 
I (x,\eta)   \ \ovl{Q}_{1,\,3} \,  \ovl{s_{0}}(x,\eta) . 
\ee

\begin{figure}
\centerline{\epsfig{file=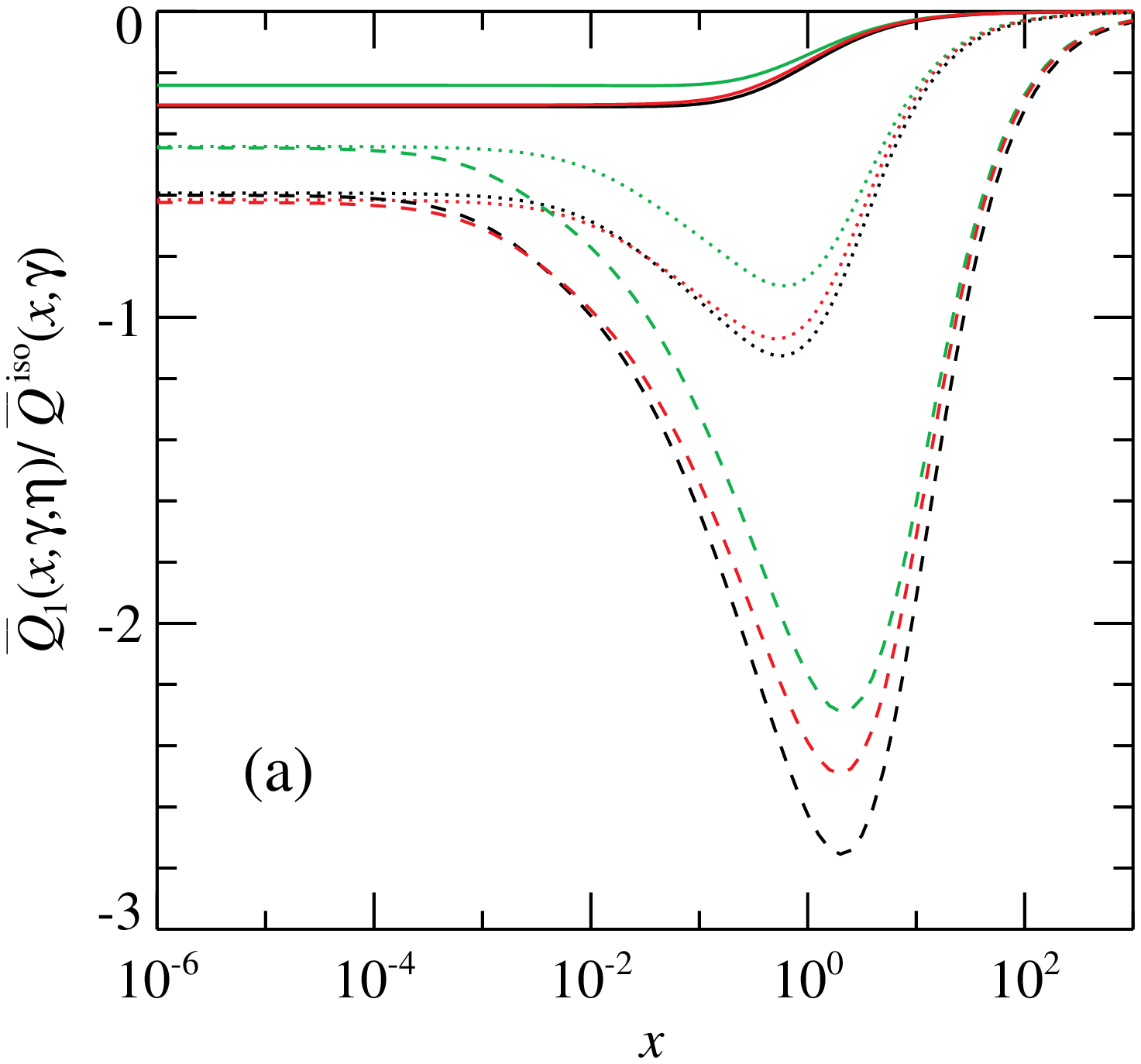,width=7.5cm} }
\centerline{\epsfig{file= 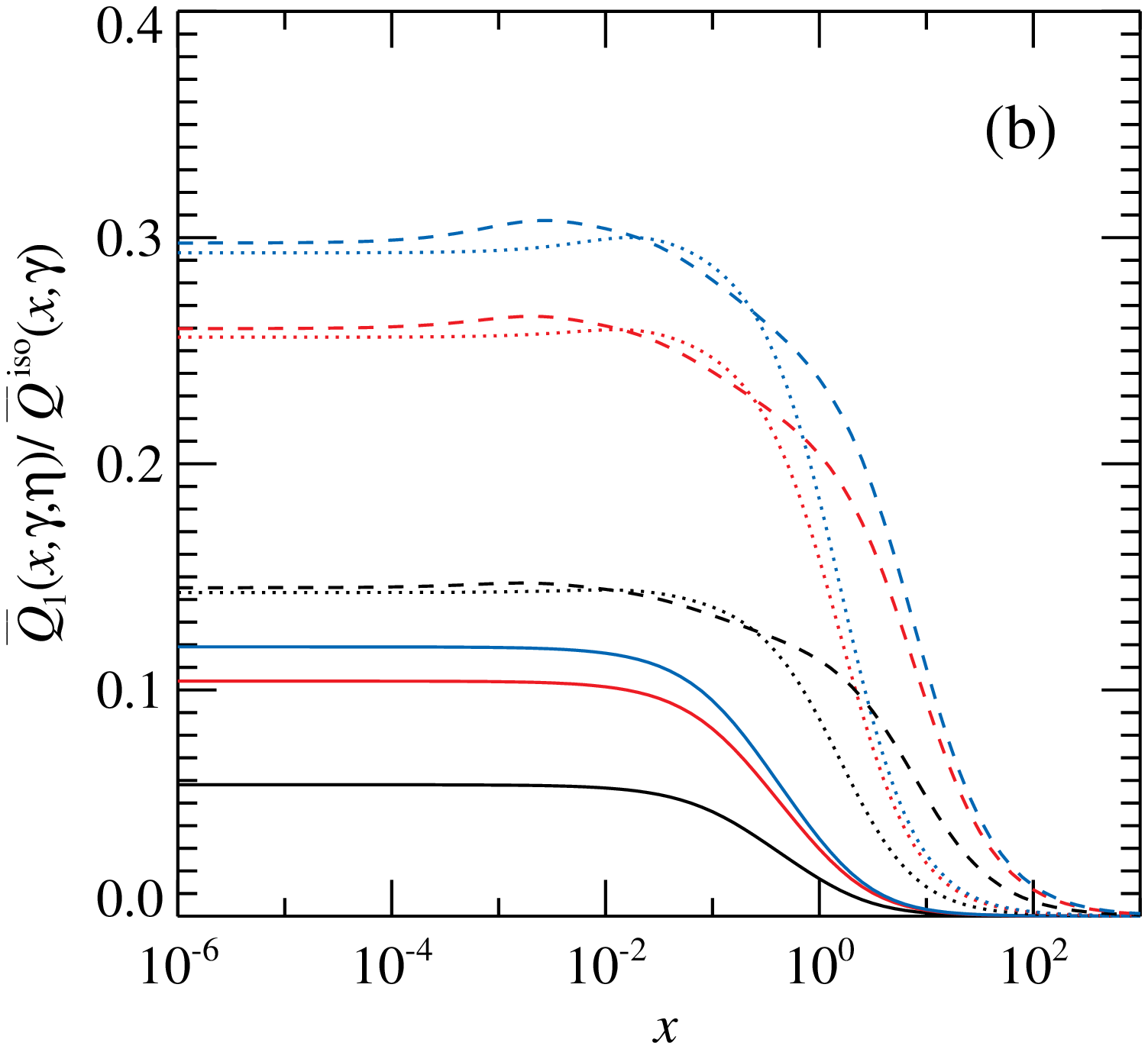,width=7.5cm} }
\caption{(a)  Momentum transferred in the direction  perpendicular to $\vomega$ (in units of the momentum 
along $\vomega$ for isotropic distribution) arising from the linear term in the electron distribution (\ref{eq:fe}) with $f_1/f_0=1$.  
For $\eta=\pm1$, the momentum is zero by symmetry. 
From the top to the bottom curves correspond to $\eta=-0.5$ (green), 0.5 (red), 0 (black).  
Solid, dotted and dashed curves correspond to $p=1, 10, 100$, respectively. 
(b)  Momentum transferred in the direction  perpendicular to $\vomega$ 
arising from the quadrupole term in the electron distribution (\ref{eq:fe}) with $f_2/f_0=1$ for the same $p$ as in panel (a).
The curves from bottom to top correspond to $\eta=\pm0.25, 0.5, 0.75$ (black, red, blue curves). 
The momentum is zero for $\eta=-1, 0, 1$  because of the symmetry.
The flat parts of the curves correspond to the Thomson limit given by Equation~(\ref{eq:ovl_Q1_gam_Thom}).
}
\label{fig:radperp}
\end{figure}

For   mono-energetic electron distribution of Lorentz factor $\gamma$ given by Equation (\ref{eq:fe_mono}), the momenta transferred along $\omega$ and 
perpendicular to it are 
\beq \label{eq:ovl_Qomega_gam}
\ovl{Q}_{3}\, \ovl{s_{0}}(x,\gamma,\eta)  &=& x 
\sum_{k=0}^2  \frac{f_k}{f_0} \ P_k(\eta) \ \left( \Delta_{0k}-\Delta_{1k}+\Delta_{1k}^*\right), \\
\label{eq:ovl_Q1_gam}
\ovl{Q}_{1}\, \ovl{s_{0}}(x,\gamma,\eta)  &=& x 
\sum_{k=1}^2  \frac{f_k}{f_0} \ P_k^1(\eta) \ \Delta_{1k}^{\bot},
\eeq
where we kept the notations for the functions $\ovl{Q}_{3}$ and $\ovl{Q}_{1}$, but added the argument $\gamma$.  
To get the average momentum transferred in a single scattering act, one needs to divide these expression by the total cross-section $\ovl{s_{0}}(x,\gamma,\eta)$. The $\omega$ component of the transferred momentum for isotropic electrons is shown in Figure~\ref{fig:radpress}a. As shown by NP94, the low-energy (Thomson) limit is given by  $x\, (1+2 p^2/3)$. 
The angular dependent corrections arising due to the dipole and quadrupole term in the electron distribution are shown  in Figures~\ref{fig:radpress}b and \ref{fig:radpress}c, respectively. While the component perpendicular to $\omega$ is zero for   isotropic electrons,
a substantial momentum component arises in the anisotropic case.  For a large linear term of the electron distribution with $f_1/f_0=1$, 
the  momentum transferred  in that direction is shown in Figure~\ref{fig:radperp}a. Similar results in case of the quadrupole term with  $f_2/f_0=1$, are shown in  Figure~\ref{fig:radperp}b. In the Thomson limit, we get  (see Appendix \ref{app:chi_ij})
 \beq \label{eq:ovl_Qomega_gam_Thom}
\ovl{Q}_{3}\, \ovl{s_{0}}(x,\gamma,\eta)  &=& x \, 
\left[ 1+\frac{2}{3} p^2 - \frac{f_1}{f_0} \eta\,  \beta\, \frac{2}{15} ( 4\gamma^2+1) \right. \nonumber \\ 
& +& \left.  \frac{f_2}{f_0} \, P_2(\eta) \, \frac{4}{15}\, p^2  \right]  ,  \\
\label{eq:ovl_Q1_gam_Thom}
\ovl{Q}_{1}\, \ovl{s_{0}}(x,\gamma,\eta)  &\!=\!\!& x \, \gamma^2 \beta \sqrt{1-\eta^2} \nonumber \\ 
&\times& \left[ -  \frac{f_1}{f_0} \frac{1}{3}  \left( 1+\frac{\beta^2}{5}\right) +
\frac{2}{5}  \frac{f_2}{f_0}  \eta\,   \beta   \right]  . 
\eeq



 
\section{Redistribution functions for anisotropic electrons}
\label{sec:redistr}

We would like to reduce the expression for the  redistribution function (\ref{eq:rk1k})  to a 
form suitable for calculations. For the electron distribution of the form (\ref{eq:fe}), this function should depend on the energies of incoming and scattered photons $x_1$ and $x$, the corresponding (cosines of) polar angles $\eta_1$ and  $\eta$ as well as the difference in azimuth $\phi-\phi_1$ (or cosine of the scattering angle $\mu$). 

\subsection{Integration over electron directions}

The three-dimensional integral over $\vecp$ in Equation  (\ref{eq:rk1k}) disappears
due to the $\delta$-function. For further 
simplifications we can also use the identity
\begin{equation}  
\delta(\gamma_{1}+x_1-\gamma-x)=\gamma 
\delta\left( \fourx \cdot (\fourp_1+ \fourx_1) -
\fourx_1 \cdot \fourp_1\right) . 
\ee
At this stage, we drop subscript 1 with the electron quantities 
and get
\be \label{eq:rk1k_red}
R({\vecx}_1 \rightarrow {\vecx})  =  \frac{3}{16\ \pi}\,
\int \frac{\rmd^3 p}{\gamma}\  \delta( \Gamma) \ \fe(\gamma,\etae) \ F .
\ee 
where 
\be
\Gamma= \gamma(x_1-x) - p (x_1 \vomega_1 -x \vomega ) \cdot \vOmega - q . 
\ee

The angular integrals in Equation~(\ref{eq:rk1k_red}) need the introduction of a suitable coordinate system. Often  the polar axis is taken along the direction of the scattered photon $\vomega$ \citep[see e.g.][]{NP93,NP94}. However, the easiest and the most transparent way, is to choose the polar axis along the direction of the transferred momentum as was proposed by \citet{AA81} (see also \citealt{PKB86})
\be 
\vn\equiv \left( x_1 \vomega_1 -x \vomega\right) / Q ,
\ee
where 
\be \label{eq:Qcap}
Q^2=(x_1 \vomega_1 -x \vomega)^2= x^2+x_1^2-2xx_1\mu = (x-x_1)^2+2q. 
\ee 
With this definition we get: 
\be
\cos \kappa \equiv \vn\cdot\vomega=  \left( x_1\mu-x \right) / Q  ,  
\quad \sin \kappa=   x_1\sqrt{1-\mu^2} / Q  
\ee
and 
\be \label{eq:cosalpha}
\cos\alpha\equiv \vn\cdot \vl_3= \left( x_1\eta_1-x\eta \right) / Q .
\ee

Thus one of the integration variables becomes $\cos\theta = \vOmega\cdot\vn$ and another is azimuth $\Phi$. The redistribution function (\ref{eq:rk1k_red}) then can be written as
\be\label{eq:red_sim}
 R(\vecx_1 \rightarrow  \vecx)  \!= \! \frac{3}{16\pi} \!\int\limits_{1}^{\infty} \!\! p \rmd \gamma \!\!
 \int \limits_{-1}^{1} \!\! \rmd\cos\theta  
\!\!\int \limits_{0}^{2\pi} \!\! \rmd\Phi \, \fe (\gamma,\etae) \, F \, \delta(\Gamma),	
\ee
where now
\be 
\Gamma= \gamma(x_1-x) - q -  p Q \cos\theta .
\ee 
Integrating first  over $\cos\theta$ using  the $\delta$-function we get 
\be \label{eq:red_phi}
R(\vecx_1 \rightarrow  \vecx)   =  \frac{3}{16\ \pi}  \int_{ \gamma_{*}}^{\infty}  \rmd \gamma 
 \frac{1}{Q} \int_{0}^{2\pi} \rmd\Phi \: \fe (\gamma,\etae) \: F ,	
\ee
where we need to substitute 
\be \label{eq:costheta}
\cos\theta= \frac{\gamma(x_1-x) - q}{pQ} 
\ee
to the expressions for $\etae$ and $F$ (see below). 
This yields
\be \label{eq:sintheta}
\sin\theta=  \frac{b}{\sqrt{r}\  pQ}, 
\ee
where
 \be \label{eq:def_b}
b = \sqrt{r}\sqrt{p^2 Q^2 - [\gamma(x_1-x)-q]^2} , \quad r=\frac{1+\mu}{1-\mu}.
\ee
The lower limit for the integral over $\gamma$ comes from the requirement that 
$|\cos\theta|\le1$:
\be \label{eq:gammamin}
\gamma \ge \gamma_{*} (x,x_1,\mu)= \frac{1}{2}\left( x-x_1+Q\sqrt{1+2/q} \: \right) .
\ee


\subsection{Integration over the azimuth}

In order to calculate the azimuthal integral in Equation~(\ref{eq:red_phi}) we have to express  $\xi$ and $\xi_1$ (that enter the expression for $F$) and $\etae$ in terms of the integration variable $\Phi$. We measure the azimuth $\Phi$ from the projection of $\vomega$ onto the plane normal to $\vn$, so that in this system
\be
\vomega=(\sin\kappa , 0 , \cos\kappa ) 
\ee
and the unit vector along the electron momentum is
\be 
\vOmega=(\sin\theta \cos \Phi, \sin\theta \sin \Phi, \cos\theta ) .
\ee 
Thus we can express the angle between the electron momentum and $\vl_3$ (see Fig.~\ref{fig:geom1})
through $\Phi$: 
\be \label{eq:etae}
\etae \equiv \vOmega \cdot \vl_3 = 
\cos\theta\ \cos\alpha +  \sin\theta\ \sin\alpha  \cos(\chi-\Phi),
\ee
where $\chi$   is the  azimuth of the vector $\vl_3$ in the $\vn$ frame.
We can also write 
\be 
\eta \equiv \vomega\cdot\vl_3 = \cos\kappa\ \cos\alpha +  \sin\kappa\ \sin\alpha  \cos\chi,
\ee 
and use this expression to obtain $\cos\chi$. 
Substituting it to Equation~(\ref{eq:etae})
we thus express the electron polar angle $\etae$ in Equation~(\ref{eq:fe}) 
through the integration variable $\Phi$.

The kernel $F$ depends on the four-products $\xi$ and $\xi_1$, which can be rewritten as  
\be \label{eq:xixi1_kap}
\xi_1= x(\gamma-p\zeta), \quad \xi = q +\xi_1,			
\ee
where 
\be
\zeta\equiv \vOmega\cdot \vomega = \cos\theta\ \cos \kappa +  \sin\theta\sin \kappa\cos\Phi .
\ee
Equation~(\ref{eq:xixi1_kap}) then can be transformed to 
\be \label{eq:xixi1_phi}
\xi_1 = \frac{q}{Q^2} (d_--b\cos\Phi), \quad   \xi = \frac{q}{Q^2} (d_+-b\cos \Phi) ,		
\ee 
where we defined
\beq \label{eq:dd1}
d_- & = & \gamma(x+x_1) - x(x-x_1 \mu)  \,  ,  \nonumber\\
  d_+ &=&   \gamma(x+x_1) + x_1(x_1-x \mu) = d_- + Q^2, 
\eeq
which have the following property: 
\beq \label{eq:dd1_aa}
\left(d_-^2-b^2\right)/Q^2 & = &   (\gamma-x)^2 + r \equiv a_-^2 , \nonumber \\
\left(d_+^2-b^2\right)/Q^2  & = &   (\gamma+x_1)^2 + r \equiv a_+^2 .
\eeq

Function $F$ in the azimuthal integral in Equation~(\ref{eq:red_phi}) is an even function of $\Phi$. Therefore the terms in $\fe$ containing  $\sin\Phi$ give zero contribution. 
Neglecting these terms we can express the azimuthal integral as 
\beq
\int_0^{2\pi} \etae F\rmd\Phi & = & \int_0^{2\pi} \ovl{\etae}  F\rmd\Phi , \\
\int_0^{2\pi}  \etae^2 F\rmd\Phi & = & \int_0^{2\pi} \ovl{\etae^2}  F\rmd\Phi  ,
\eeq 
where 
\beq  \label{eq:etae_ave}
\ovl{\etae} & =&  \cos\theta\ \cos\alpha +  \sin\theta\ \sin\alpha\ \cos\chi\ \cos\Phi , \\
\label{eq:etae2_ave}
\ovl{\etae^2} &=& 
\cos^2\theta\ \cos^2\alpha +  \sin^2 \theta   \sin^2 \alpha\sin^2\chi \nonumber \\
&+& 2 \sin\theta \sin\alpha\ \cos\theta\ \cos\alpha \ \cos\chi\ \cos\Phi \nonumber \\
&+&  \sin^2 \theta \  \sin^2 \alpha\ \cos 2\chi \ \cos^2\Phi .				
\eeq
Thus the expansion (\ref{eq:fe}) (with $\etae$ and $\etae^2$ substituted by $\ovl{\etae}$ and $\ovl{\etae^2}$, respectively)  is a quadratic function of   $\cos\Phi$. Expressing 
\beq
\cos\Phi & =&  -\frac{Q^2}{2b}\frac{\xi+\xi_1}{q} +\frac{d_-+d_+}{2b}  , \\
\cos^2\Phi & =&  \frac{Q^4 }{b^2}\frac{\xi\xi_1}{q^2} - \frac{Q^2(d_++d_-)}{2b^2} \frac{\xi+\xi_1}{q} \nonumber
+\frac{d_+^2+d_-^2}{2b^2}  
\eeq
and using the identity $\xi=\xi_1+q$, we obtain an expansion of $\fe$ that is symmetric in $\xi$ and $\xi_1$: 
\be \label{eq:fe_ave}
\ovl{\fe}(\gamma) = c_{0} + c_{\Sigma} \frac{\xi+\xi_1}{q} +
c_{\Pi} \frac{\xi\xi_1}{q^2}. 						
\ee
The coefficients $c_{0}$, $c_{\Sigma}$ and $c_{\Pi}$
can be represented in the form: 
\beq \label{eq:c012}
c_{0} &=& f_0 + c_{01} f_1 + c_{02} f_2, \nonumber \\
c_{\Sigma} &=& c_{11} f_1 + c_{12} f_2,   \\
c_{\Pi} &=&  c_{22} f_2, \nonumber 
\eeq
where the coefficients in front of $f_{0,1,2}$ can be easily derived 
after lengthy but straightforward calculation: 
\beq \label{eq:c01}
c_{01}  &=& \frac{2\rho - \epsilon + \epsilon_1}{2p(1+\mu)} , \nonumber \\
c_{11} &=& - \frac{\epsilon + \epsilon_1}{2p (1+\mu)} , \nonumber \\
c_{02}  &=& \frac{3}{4p^2 (1+\mu)^2} 
\left[ (\epsilon -\rho)^2 + (\epsilon_1 +\rho)^2 
+  \lambda (a_-^2+a_+^2) \right] - \frac{1}{2} ,  \nonumber \\
\label{eq:xixi11}
c_{12} &=& - \frac{3}{4p^2  (1+\mu)^2}
 \left[ (\epsilon + \epsilon_1)(2 \rho-\epsilon + \epsilon_1) 
+ \lambda (d_-+d_+) \right] ,	\nonumber \\	
\label{eq:c22}
c_{22} &= & \frac{3}{2p^2  (1+\mu)^2} 
\left[ (\epsilon + \epsilon_1)^2 +  \lambda Q^2 \right].
\eeq
Here we defined
\beq \label{eq:defin}
\epsilon &\equiv & x(\eta_1 -\eta\mu)\,, \quad 
\epsilon_1 \equiv x_1(\eta -\eta_1 \mu)\,, \quad 
\rho = \gamma(\eta + \eta_1)\, , \nonumber \\
 \lambda &=& \mu^2 + \eta^2 + \eta_1^2 - 2 \mu\eta \eta_1 -1 . 
\eeq

The redistribution function (\ref{eq:red_phi}) is then expressed as
\be \label{eq:rx1x_cr}
R(x_1,\mbox{\boldmath $\omega$}_1 \rightarrow x,\mbox{\boldmath $\omega$}) = 
\frac{3}{8} \!\!
\int\limits_{\gamma_*(x,\,x_1,\,\mu) }^{\infty} \!\!\!\!\!\! \rmd \gamma 
 \left[  c_{0} R_0 + c_{\Sigma} R_{\Sigma} + c_{\Pi} R_{\Pi} 
 \right] ,
\ee
where we have introduced three functions 
\beq \label{eq:r0}
R_0 (x,x_1,\mu,\gamma) &=& \frac{1}{\pi Q  } \int_{0}^{\pi}  \: F \ \rmd \Phi, \\
 \label{eq:rsigma}
R_{\Sigma} (x,x_1,\mu,\gamma) &=& \frac{1}{\pi Q\, q} \int_{0}^{\pi}  (\xi+\xi_1) F\ \rmd \Phi,  \\
 \label{eq:rpi}
R_{\Pi} (x,x_1,\mu,\gamma) &=& \frac{1}{\pi Q\, q^2} \int_{0}^{\pi}   \xi \xi_1 F\ \rmd \Phi .
\eeq 
Alternatively, we can represent the redistribution function as a sum of three terms 
arising from the corresponding three terms in the electron distribution: 
\be \label{eq:rxx_alt}
R(x_1,\mbox{\boldmath $\omega$}_1 \rightarrow x,\mbox{\boldmath $\omega$}) =
\frac{3}{8}\!\! \int \limits_{\gamma_*(x,\,x_1,\,\mu)}^{\infty} \!\!\!\!\!\!\rmd \gamma 
 \left[  f_0 R_0 + f_1 R_{1} + f_2 R_{2} 
 \right] , 
\ee
where 
\beq \label{eq:R12_alt}
R_1 (x, \eta; x_1, \eta_1; \mu; \gamma) & = & c_{01} R_0 + c_{11} R_{\Sigma} \ ,  \\
R_2  (x, \eta; x_1, \eta_1; \mu; \gamma)  & = &  c_{02} R_0 + c_{12} R_{\Sigma} + c_{22} R_{\Pi} \ . 
\nonumber 
\eeq

Using the Klein-Nishina cross-section (\ref{eq:kn}) in the   form
\be \label{eq:F1_simple}
F=2+\frac{q^2-2q-2}{q}\left( \frac{1}{\xi_1} -\frac{1}{\xi} \right) +
\frac{1}{\xi^2}+\frac{1}{\xi_1^2} ,
\ee
(and remembering that $\xi=\xi_1+q$), we see that the integrals (\ref{eq:r0})--(\ref{eq:rpi}) involve integrals of types
\be\label{eq:integr_type}
\int_0^{\pi} \xi^s \: \rmd \Phi, \quad \int_0^{\pi} \xi_1^{s} \: \rmd \Phi, 
\ee 
where $s=-2,..,2$. The integrals over non-negative powers of $\xi$ and $\xi_1$ are trivial. For the negative powers, using Equations~(\ref{eq:xixi1_phi}) and (\ref{eq:dd1_aa}) we get \citep[see][ for details]{NP93}:
\be
\int_0^{\pi} \frac{\rmd \Phi}{\xi_1} = \frac{\pi Q}{q}  \frac{1}{a_-} , \quad 
\int_0^{\pi} \frac{\rmd \Phi}{\xi_1^2} = \frac{\pi Q}{q^2} \frac{d_-}{a_-^3 }  \label{eq35}
\ee
and similar equations for $\xi$ which we get by substituting $\xi$, $a_+$ and $d_+$
for $\xi_1$, $a_-$ and $d_-$, respectively.
 
After some straightforward algebra we get the expressions for $R_0$, $R_{\Sigma}$ and $R_{\Pi}$:
\be \label{eq:r0f}
R_0 = \frac{2}{Q} + \frac{q^2-2q-2}{q^2} \left( \frac{1}{a_-} - \frac{1}{a_+} \right)
+ \frac{1}{q^2} \left( \frac{d_-}{a_-^3} + \frac{d_+}{a_+^3} \right), 	
\ee
which was obtained by \citet{AA81} \citep[see also][]{NP93},
\be 	\label{eq:rsf}
R_{\Sigma} \! = \!\frac{2}{Q^3} \left( d_- + d_+ \right)\! +\! \left( 1- \frac{2}{q}\right) \left( \frac{1}{a_-} \!+\! \frac{1}{a_+} \right) + \frac{1}{q^2} \left( \frac{d_-}{a_-^3} \!- \!\frac{d_+}{a_+^3} \right)
\ee
and
\be\label{eq:rpf}
R_{\Pi} = \frac{2}{Q^5} \left(d_- d_+ + \frac{b^2}{2} \right) + 
 \left( 1- \frac{2}{q}\right) \frac{1}{Q}
+ \frac{1}{q^2} \left( \frac{1}{a_-} - \frac{1}{a_+} \right).	
\ee
Equation (\ref{eq:rx1x_cr}) (or alternatively equations [\ref{eq:rxx_alt}] and [\ref{eq:R12_alt}]) together with our computed redistribution functions (\ref{eq:r0f})--(\ref{eq:rpf}) and the coefficients (\ref{eq:c012})--(\ref{eq:defin}) give the full analytical solution for the redistribution function describing scattering of arbitrary photons from the electron gas which anisotropy can be described by Equation~(\ref{eq:fe}).

\subsection{Alternative redistribution functions}

An alternative expression for the redistribution function $R(\vecx_1\rightarrow \vecx)$ can be obtained if we 
compute the moments 
\beq \label{eq:Rphi_phi}
\lefteqn{ R_{\phi} (x,x_1,\mu,\gamma) =  \frac{1}{\pi Q}  \int_{0}^{\pi} \cos\Phi\ F\ \rmd \Phi  } \nonumber \\
&=&  \frac{q^2-2q-2}{q^2} \frac{1}{b} \left( \frac{d_-}{a_-} - \frac{d_+}{a_+} \right) 
+ \frac{b}{q^2} \left( \frac{1}{a_-^3} + \frac{1}{a_+^3} \right) , \\
\label{eq:Rphi_phiphi}
\lefteqn{ R_{\phi\phi} (x,x_1,\mu,\gamma) =  \frac{1}{\pi Q}  \int_{0}^{\pi} \cos^2\Phi\ F\ \rmd \Phi } \nonumber\\
&=&  \frac{1}{Q} + \frac{Q^3}{b^2} \left( 1- \frac{2}{q}\right) + 
 \frac{q^2-2q-2}{q^2}  \frac{1}{b^2} \left( \frac{d_-^2}{a_-} - \frac{d_+^2}{a_+} \right) \nonumber\\
 & - & \frac{Q^2}{q^2b^2} \left( \frac{d_-}{a_-} + \frac{d_+}{a_+} \right) 
 + \frac{1}{q^2} \left( \frac{d_-}{a_-^3} + \frac{d_+}{a_+^3} \right) . 
\eeq
The expressions for $R_1$ and $R_2$ then take the form: 
\beq \label{eq:R12_phiphi}
R_1   & = & d_{01} R_0 + d_{11} R_{\phi} \ ,  \\
R_2   & = &  d_{02} R_0 + d_{12} R_{\phi} + d_{22} R_{\phi\phi} \ ,
\nonumber 
\eeq
where 
\beq \label{eq:coef_d}
d_{01} &=& \cos\theta\ \cos\alpha , \nonumber \\ 
d_{11} &=&  \sin\theta\ \sin\alpha\ \cos\chi   
= \frac{b}{p(1+\mu)Q^2} \left( \epsilon + \epsilon_1 \right) , \nonumber \\ 
d_{02} &=&   \frac{3}{2} \left( \cos^2\theta \ \cos^2\alpha + \sin^2\theta\ \sin^2\alpha\ \sin^2\chi \right) - \frac{1}{2} \nonumber \\ 
&=& \frac{3}{2} \left( 2 d_{01}^2-d_{11}^2+\sin^2\theta-\cos^2\alpha \right) -   \frac{1}{2} ,  \\
d_{12} &=&  3  \cos\theta\ \cos\alpha   \sin\theta\ \sin\alpha\ \cos\chi = 3\ d_{01} \ d_{11}  , \nonumber \\ 
d_{22} &=& \frac{3}{2} \sin^2\theta\ \sin^2\alpha\ \cos2\chi  \nonumber \\
 &=& \frac{3}{2} \left(2 d_{11}^2-d_{01}^2-\sin^2\theta+\cos^2\alpha\right) .  \nonumber
\eeq

\subsection{Approximate redistribution functions}
\label{sec:appr_rf}

Approximate forms of Equations~(\ref{eq:r0f})--(\ref{eq:rpf}) can be obtained by making certain simplifying assumptions about the scattering. For example, in the Thomson regime in the electron rest frame the  Klein-Nishina kernel $F$ is just $1+\mu_0^2$. 
Assuming further isotropic scattering in that frame and substituting $F$ by  $4/3$, we now get for
the integrals (\ref{eq:r0})--(\ref{eq:rpi}): 
\beq \label{eq:r0_app}
R_0 & \approx& \frac{4}{3Q},    \\
\label{eq:rsigma_app} 
R_{\Sigma} & \approx& \frac{4}{3Q^3} \left( d_- + d_+\right),   \\
\label{eq:rpi_app}
R_{\Pi} & \approx& \frac{4}{3Q^5} \left(d_- d_+ + \frac{b^2}{2} \right).  
\eeq
The expression for $R_0$ was derived by \citet{an80}.
For the alternative functions (\ref{eq:Rphi_phi}), (\ref{eq:Rphi_phiphi}), we then have
\beq \label{eq:rphi_app}
R_{\phi} &=& 0,   \qquad
R_{\phi\phi} = \frac{2}{3Q}  .
\eeq  
These then give 
\beq \label{eq:R1app}
 {R}_1  & \approx &   {d_{01}} R_0 = \cos\theta\  {\cos\alpha}\ R_0, \\ 
\label{eq:R2app}
 {R}_2  & \approx  &  \left(   {d_{02}}+ \frac{1}{2}  {d_{22}}\right) \ R_0 
=  P_2(\cos\theta)\  {P_2(\cos\alpha)} \ R_0 ,
\eeq
with $\cos\theta$ and $\cos\alpha$  given by Equations (\ref{eq:costheta}) and (\ref{eq:cosalpha}), respectively. 
The approximate expressions are better than 50 per cent accurate  in the Thomson regime for $x_1\gamma<0.1$ 
at all scattered photon energies.  

\subsection{Relation to  the mean powers of photon energies}

The relation between the redistribution function averaged over any electron distribution and the mean powers of photon energies follows directly from their definitions (\ref{eq:rk1k}), (\ref{eq:totcross}) and (\ref{eq:ovl_moments}): 
\be \label{eq:ovlxRx}
\ovl{x_1^j}\,\ovl{s_{0}}(x,\eta)= \frac{1}{x} \int x_1^{j+1}\rmd x_1\, \int \rmd^2\omega_1\  R(\vecx \rightarrow \vecx_1).
\ee
This relation is valid for any electron distribution. Comparing Equations~(\ref{eq:rxx_alt})  and (\ref{eq:s0_j_d_chi}),  we get a relation between the functions depending on the electron energy: 
\beq \label{eq:RF012_MPE}
\lefteqn{ \Delta_{jk}(x,\gamma) P_k(\eta)  = \frac{3}{32\pi \gamma p x^{j+1}}  } \nonumber \\
&\times&   
\! \int_{x^- (x,\gamma)}^{x_{\rm m} (x,\gamma)} \!\! x_1^{j+1}\rmd x_1\!\! \int \!\! 
R_k (x_1,\eta_1;x,\eta; \mu;\gamma) \, \rmd^2\omega_1 , 
\eeq
where $\eta_1=\eta\mu+\sqrt{1-\eta^2}\sqrt{1-\mu^2}\cos\Phi$ and 
$R_0$ depends only on the scattering angle $\mu$, but not $\eta, \eta_1$. 
The integrals over the solid angle can be represented as the integrals over $\rmd\mu$ and $\rmd\Phi$, where 
$\Phi\in [0,2\pi]$ and the limits on $\mu$, $\mu_{\rm m}(x_1,x,\gamma)$ and $\mu^+(x_1,x,\gamma)$, are given by Equations~(\ref{eq:mulim})--(\ref{eq:Dm}) with the arguments $x$ and $x_1$ reversed. 
Using Equations~(\ref{eq:ovl_Qomega}) and (\ref{eq:Psi_perp}), we also get
\be \label{eq:Rstar_mom}
\Delta_{1k}^* (x,\gamma)  P_k(\eta) = 
\frac{3}{32\pi \gamma p x^2}   \! \int \!\!   x_1^{2}\rmd x_1 \! \int  (1-\mu)\
R_k \  \rmd^2\omega_1 , 
\ee
\be
\label{eq:Rbot_mom}
\Delta_{1k}^{\bot} (x,\gamma)  P_k^1(\eta) = 
\frac{3}{32\pi \gamma p x^2}  \! \! \int \!\!  x_1^{2}\rmd x_1 \!\!  \int \!\!\! \!  \sqrt{1-\mu^2}\cos\Phi \,
R_k   \rmd^2\omega_1 .
\ee
In order to check the accuracy of our derivations we compared the left hand sides of Equations~(\ref{eq:RF012_MPE})--(\ref{eq:Rbot_mom})  to the right hand sides, where  the integrals were performed numerically and obtained consistent results.

\begin{figure}
\centerline{\epsfig{file= 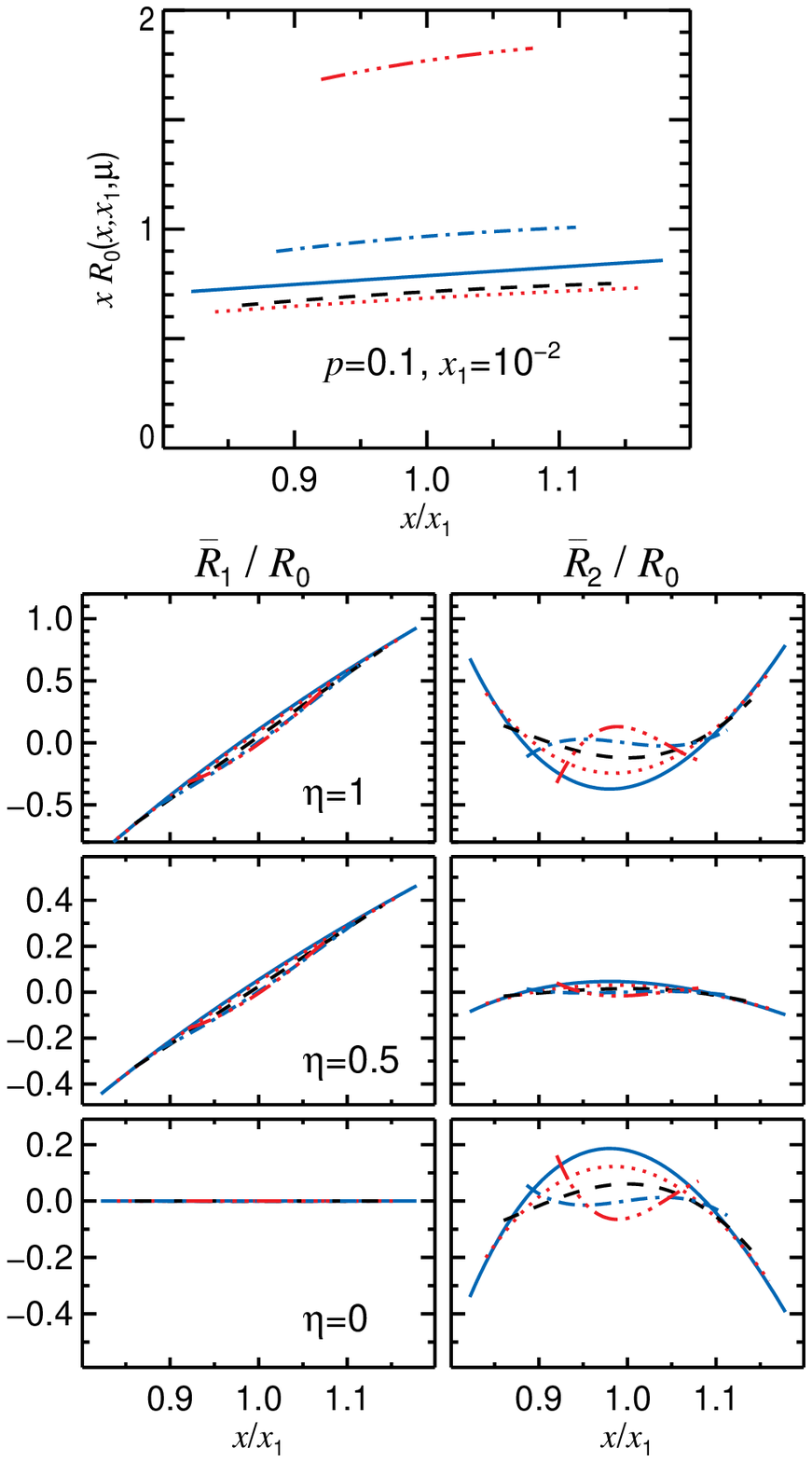,width=8.2cm}} 
\caption{Redistribution functions for anisotropic electrons at a given scattering angle. 
The incident photon energy is $x_1 = 10^{-2}$ and the electron momentum $p=0.1$. 
The upper panel shows the photon (number) emissivity for   isotropic electrons. 
The solid, dotted, dashed, dot-dashed  and dot-dot-dashed curves correspond to the cosine of scattering angle 
$\mu=-2/3, -1/3, 0 , 1/3, 2/3$, respectively. 
The lower left panels show the ratio $\overline{R}_1/R_0$, while the right panels show $\overline{R}_2/R_0$ 
as a function of the ratio of the scattered to the incident photon energies. 
The three row of panels corresponds to the different observer directions $\eta=1, 0.5, 0$.  
} 
\label{fig:rf_p0.1}
\end{figure}

 \section{Applications}
\label{sec:numer}

\subsection{Examples of redistribution functions}

Now we demonstrate the properties of the derived redistribution functions.  We consider a volume filled by electrons with the angular distribution given by Equation~(\ref{eq:fe}).   The emissivity in a direction $\vomega$ at energy $x$ can be obtained from the radiative transfer equation (\ref{eq:rte}) and is given by the integral over the redistribution function 
\be
\epsilon(\vecx) = \sigmat \Ne\, x^2\!\! \int_0^{\infty} \!\! \frac{\rmd x_1}{x_1^2} \!\! \int\!\! \rmd^2\omega_1 \,
I (\vecx_1)\ R (x_1, \vomega_1 \rightarrow  x, \vomega) ,
\ee
where $I(\vecx_1) = 2\me (\me c^2/h)^3 x_1^3 n(\vecx_1)$ is the specific intensity of the incident radiation normalized to the photon density as 
\be 
\Nph =  \frac{1}{\me c^3} \int \rmd^2\omega \ \int I(\vecx)  \ \frac{\rmd  x }{x}.
\ee

Let us consider mono-energetic (with energy $\gamma$) electron distribution (\ref{eq:fe_mono}). Consider also a monochromatic source of isotropic seed photons at energy $x_1$  with total photon number density $\Nph$. According to Equation~(\ref{eq:rxx_alt}) we can write the  emissivity at an observer direction $\eta$ for a given scattering angle as 
\be \label{eq:emis_x_eta}
\overline{\epsilon}(x,\eta,\mu) = \frac{3}{32\pi}\me c^3 \sigmat \Ne \Nph  \frac{x^2}{x_1} \frac{1}{p\gamma}
 \left[ R_0 + \frac{f_1}{f_0} \overline{R}_1 +  \frac{f_2}{f_0}  \overline{R}_2 \right] , 
\ee 
which is related to the scattering angle-averaged emissivity as $\epsilon(\vecx)  = \frac{1}{2}\int \overline{\epsilon}(x,\eta,\mu) \rmd \mu$, 
and where (for $k=1,2$)
\be\label{eq:Rkave}
\overline{R}_k  (x,\eta;x_1;\mu;\gamma)= \frac{1}{2\pi}\!
\int_0^{2\pi} \!\!\!\! \rmd \Phi \ R_k (x,\eta;x_1,\eta_1;\mu; \gamma) 
\ee
and $\eta_1=\eta\mu+\sqrt{1-\eta^2}\sqrt{1-\mu^2}\cos\Phi$. 
These functions obviously possess symmetry properties: 
\beq
 \overline{R}_1   (x,-\eta;x_1;\mu;\gamma) & =&   - \overline{R}_1   (x,\eta;x_1;\mu;\gamma) , \\
  \overline{R}_2  (x,-\eta;x_1;\mu;\gamma) & =&   \overline{R}_2  (x,\eta;x_1;\mu;\gamma)  . 
\eeq

\begin{figure}
\centerline{\epsfig{file= 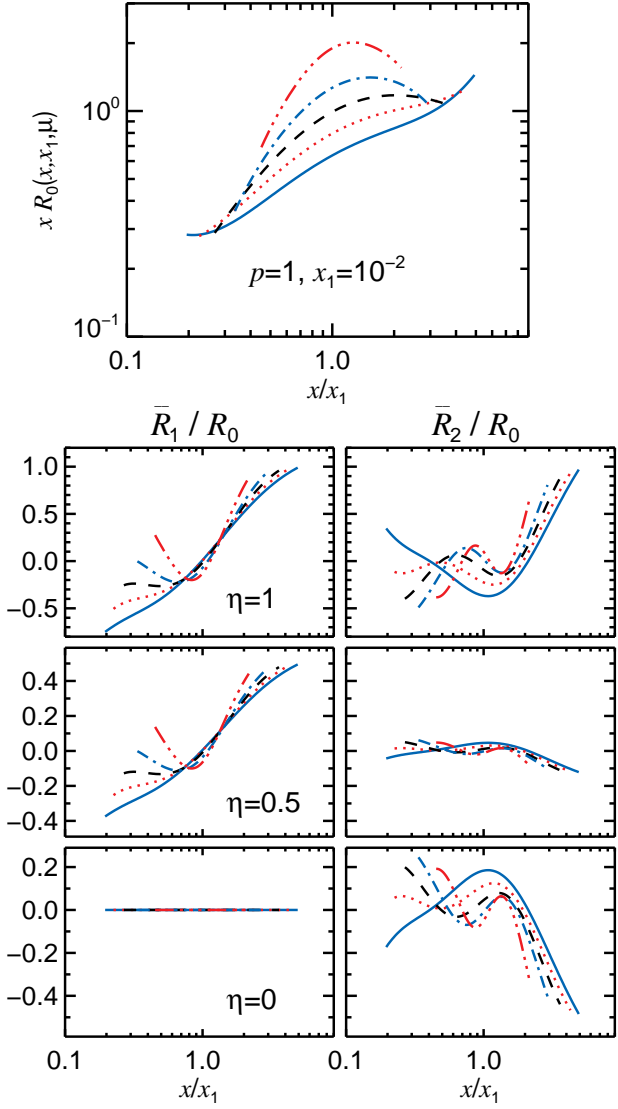,width=8.2cm}} 
\caption{Same as Figure~\ref{fig:rf_p0.1}, but for $p=1$. Note, that here the axes are in logarithmic units.
 } 
\label{fig:rf_p1}
\end{figure}

\begin{figure}
\centerline{\epsfig{file= 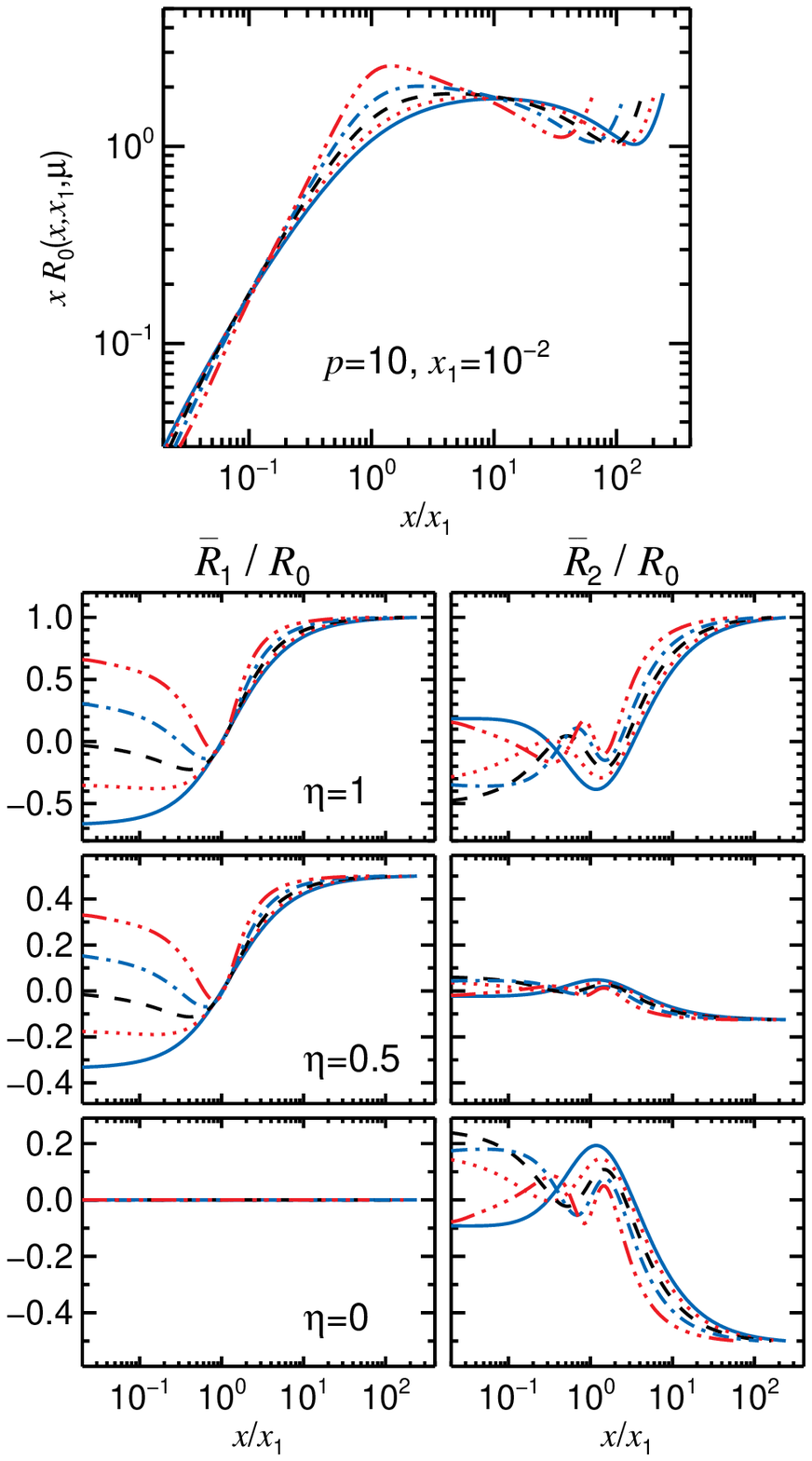,width=8.2cm}} 
\caption{Same as Figure~\ref{fig:rf_p1}, but for $p=10$. } 
\label{fig:rf_p10}
\end{figure}

\begin{figure}
\centerline{\epsfig{file= 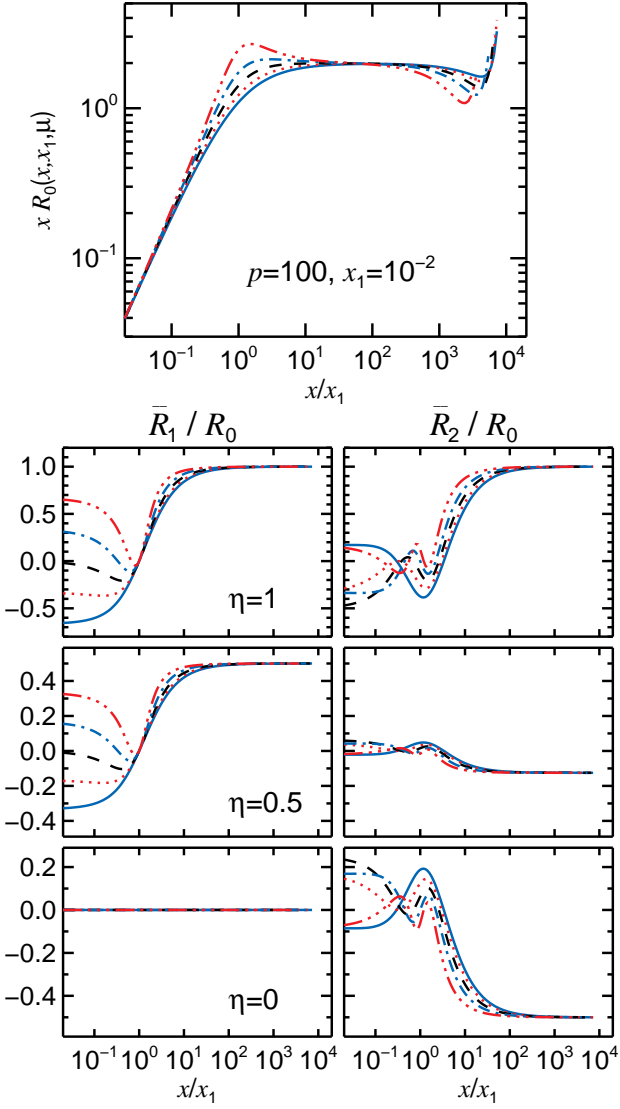,width=8.2cm}} 
\caption{Same as Figure~\ref{fig:rf_p1}, but for $p=100$. 
 } 
\label{fig:rf_p100}
\end{figure}

We compute separately the emissivities resulting from three terms in the electron distribution, i.e. functions $R_0, \overline{R}_1, \overline{R}_2$ (see Equation~[\ref{eq:emis_x_eta}]), and show in  Figures~\ref{fig:rf_p0.1}--\ref{fig:rf_p100} the function $R_0$ multiplied by $x$ (i.e. quantity proportional to the photon number emissivity) for better visibility as well as the ratios $\overline{R}_1/R_0$ and $\overline{R}_2/R_0$.  The main behavior of the functions can be easily understood using formulae  (\ref{eq:R1app})--(\ref{eq:R2app})  derived in Thomson limit and isotropic scattering approximation.  Averaging them over the azimuth and using relation $\overline{P_k(\eta_1)}=P_k(\eta)P_k(\mu)$, we get
\beq \label{eq:aveR1}
\frac{\overline{R}_1 }{{R}_0} & \approx &  \frac{ x_1 \mu-x}{Q} \  \eta\ \cos\theta\  ,  \\
\label{eq:aveR2}
\frac{\overline{R}_2 }{{R}_0} & \approx &  \frac{x_1^2P_2(\mu)- 2x x_1 \mu + x^2 }{Q^2} \ 
P_2(\eta)\ P_2(\cos\theta)\  . 
\eeq
These approximate expressions become extremely accurate for high $p$ (i.e. accuracy is  about $10^{-3}$ at $p=100$).

For a small electron momentum $p=0.1$ and low photon energies $x_1=10^{-2}$, the exact redistribution functions are shown in  Figure~\ref{fig:rf_p0.1}. In this regime, scattering is nearly coherent with the scattered photon energies bounded by (see Equation (\ref{eq:x1_limits})) $x^{\pm}/x_1\approx 1 \pm p \sqrt{2(1-\mu)}$.  In this regime, $|x-x_1|^2 \ll q\ll x,x_1$ and $\cos\theta\approx (x_1-x)/pQ$ is a nearly linear function of $x/x_1$, because $Q/x_1\approx \sqrt{2(1-\mu)}$. For $\mu$ not too close to 1, the azimuth averaging of $\cos\alpha$ gives $-\eta\sqrt{(1-\mu)/2}$ and thus $\overline{R}_1/R_0\approx (1-x/x_1)\eta/(2p)$. 
For $\eta=0$, the function is always zero, because of the symmetry. Similarly, the nearly quadratic dependence of $\overline{R}_2/R_0$ on energy results from the $\cos^2\theta$ term, while at $\mu\approx1/3$ the function becomes more complicated because of the cancellation in the $\overline{P_2(\cos\alpha)}$ term.

In the opposite limit of the relativistic electrons (see Figures~\ref{fig:rf_p10} and \ref{fig:rf_p100}),  the approximation (\ref{eq:r0_app}) for the function $R_0$ works fine up to $x_1\gamma\lesssim 0.1$, while as said above the ratios  $\overline{R}_1/R_0$ and  $\overline{R}_2/R_0$ are very close to those given by Equations (\ref{eq:aveR1}) and (\ref{eq:aveR2}) for any photon and electron energies. At small scattered photon energies $x\ll x_1$, $xR_0\propto x/x_1$,  and 
\be 
\overline{R}_1/R_0\approx \eta \mu \cos\theta, \quad \overline{R}_2/R_0\approx P_2(\eta)P_2(\mu) P_2(\cos\theta),
\ee
with $\cos\theta\approx 1-x/x_1$. 
At high scattered photon energies $x\gg x_1$, the photons are scattered at large angles in the electron rest frame  and therefore they are beamed in the direction of the incoming electrons. In that case, the angular distribution of the scattered photons resemble that of the electrons. In this regime $xR_0\propto \mbox{const}$, and ${R}_1/R_0\approx \eta$ and ${R}_2/R_0\approx P_2(\eta)$, which gives the flat dependences 
clearly seen in Figures~\ref{fig:rf_p10} and \ref{fig:rf_p100}, and $\epsilon(\vecx) \propto \fe(\gamma,\eta)\, x/x_1 $.

\newpage

\subsection{Sunyaev--Zeldovich effect}
 \label{sec:sz}
 
Let consider a cloud of isotropic Maxwellian electrons of temperature $\Theta\equiv k\Te/\me c^2$, which moves  with velocity $c\betab$ (corresponding Lorentz factor $\Gammab$)  through the isotropic cosmic microwave background of temperature $\Thetacmb\equiv k\Tcmb/\me c^2$.  We compute the thermal and kinematic  Sunyaev--Zeldovich effects \citep{ZS69,SZ72CoASP}, i.e. the spectrum of the scattered radiation (and resulting deviations from the black body) as a function of $\Theta$ and the angle between the line of sight and the direction of motion. 

One approach would be to make a Lorentz transformation of the incident radiation to the comoving frame, compute the Compton scattered radiation using the kernel corresponding to   isotropic electron distribution, and  then to Lorentz transform it back to the observer frame. Another way is to compute the electron distribution in the observer frame, approximate it by the expansion (\ref{eq:fe}) and compute directly the Compton scattered radiation in the observer frame. The second approach might be favorable from numerical point of view if the object velocity is variable in space and/or time, as allows to pre-compute redistribution functions at a fixed grid of angles and photon energies. 

\begin{figure*}
\centerline{\epsfig{file= 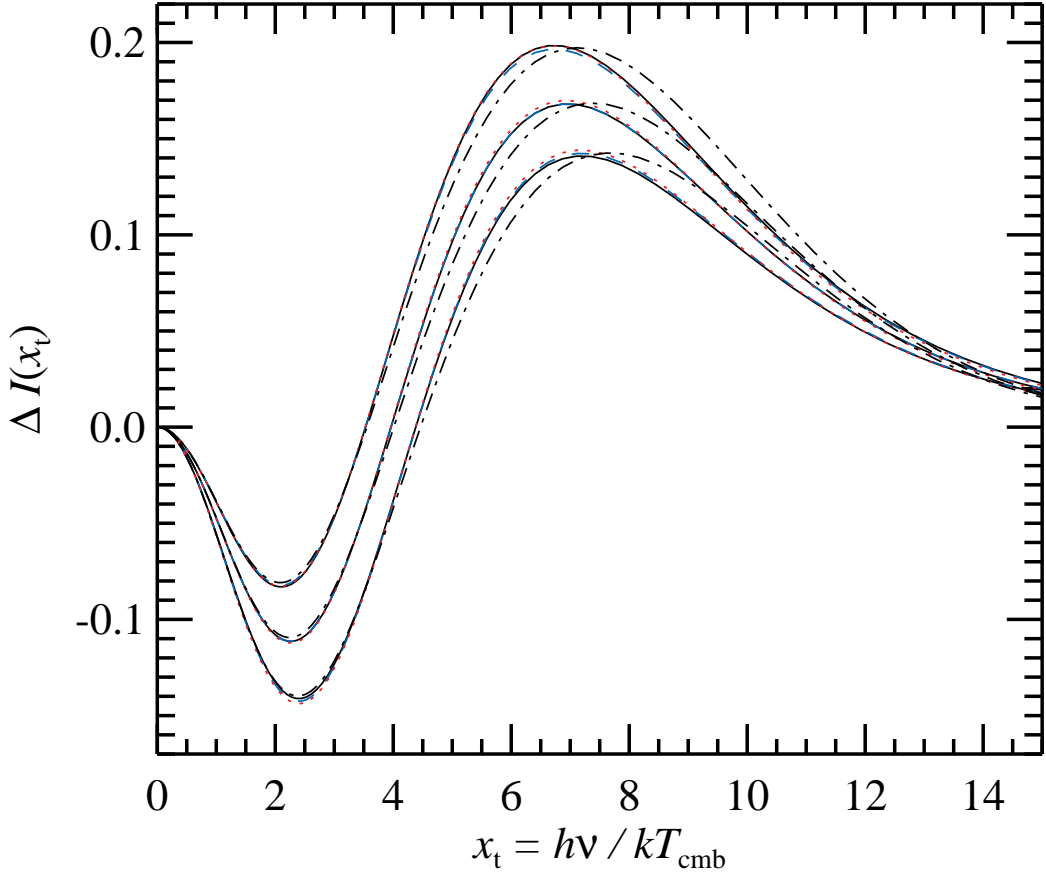,width=14cm}} 
\caption{Deviation from the black body spectrum of cosmic microwave background radiation with $\Tcmb=2.7$ K
resulting from the Compton scattering in a moving cloud of isotropic hot electrons (thermal and kinematic Sunyaev-Zeldovich effects). 
The electron temperature is $\Theta=0.03$ and the cloud velocity $\betab=0.01$. 
The solid curves are computed using Equations (\ref{eq:sz_spec_corr1})--(\ref{eq:sz_source1}), considering scattering in the cloud frame, where electrons are  isotropic. The  dashed curves show the results using the formalism developed in this paper for   anisotropic electrons, given by Equations  (\ref{eq:sz_spec_corr2})--(\ref{eq:sz_source2}). The dotted curves correspond to the semi-analytical approximation of the angle-averaged redistribution function given by Equations  (\ref{eq:intRomega})--(\ref{eq:intRgamma}). The three different methods give nearly identical results. 
The three curves from bottom to the top correspond to the three viewing angles with $\eta=-0.98, 0, +0.98$.  
The dash-dotted curves show analytical approximation of \citet{SaSu98}, which includes terms up to second order in $\betab$ and $\Theta$, as well as a cross-term $\betab\Theta$. It works reasonably well up to the temperatures $\Theta<0.02$, but fails at higher temperatures in Wien tail. 
} 
\label{fig:sz}
\end{figure*}

\subsubsection{Scattering in the comoving frame}

Let us first compute the scattered radiation by a standard method considering scattering in the comoving frame. 
The relativistic Maxwellian distribution of electrons in the comoving frame (quantities with primes) is given  by 
\be\label{eq:sz_fe_pr}
\fe'(\vecp') = \Ne'
\frac{\exp(-\gamma'/\Theta)}{4\pi \, \Theta\, K_2(1/\Theta)} ,   
\ee
where $K_2$ is the modified Bessel function and $\Ne'$ is the electron density in that frame. 
The  incident   black body radiation occupation number is 
\be
\noccx_{\rm bb}(\vecx)  = \frac{1}{\exp(x_t) - 1} ,
\ee
where $x_t= x/\Thetacmb= h\nu/k\Tcmb$. 
From the radiative transfer equation (\ref{eq:rte}), in the limit of small optical depth, we get the correction to the 
black body spectrum:  
\be\label{eq:sz_spec_corr1}
\Delta \noccx (x,\eta) =  \noccx (x,\eta) - \noccx_{\rm bb}(x) = 
- \taut  \overline{s}_0(x')\noccx_{\rm bb}(x)  + S(x,\eta) ,
\ee
where $\taut$ is the Lorentz invariant optical depth for Thomson scattering, 
\be \label{eq:sz_source1}
S(x,\eta) = \taut \frac{1}{x'}  
\int_{0}^{\infty} x'_1 \rmd x'_1 \int  \rmd^2  \omega'_1 \: R^{\rm iso}(\vecx'_1 \rightarrow \vecx') \noccx_{\rm bb}(x_1) 
\ee
is the source function, and we used here the fact that the photon occupation number is Lorentz invariant. 
The energy transformation is given by Doppler shift $x=x' {\cal D}$ and $x_1=x'_1  {\cal D}_1$ with the Doppler factors  
\be 
{\cal D} = \frac{1}{\Gammab(1-\betab \eta)} , \quad 
{\cal D}_1 =  \Gammab(1+\betab \eta'_1) . 
\ee 
The relation between the angles is given by the aberration formula: 
\be
\eta'= \frac{\eta-\betab}{1-\betab \eta} .
\ee
We note here that $\overline{s}_0$ is equal to unity with high accuracy, because scattering is in deep Thomson regime. 
The calculation of the redistribution function $R^{\rm iso}$ involves numerical integration over the Maxwellian distribution (see Equation [\ref{eq:rxx_alt}]; note that $f_1=f_2=0$)  $f_0=\fe'(\gamma)=\fe'(\vecp)/\Ne'$ given by Equation (\ref{eq:sz_fe_pr}). Thus the source function 
(\ref{eq:sz_source1})  involves 4-dimensional integral to be taken numerically, which is rather time-consuming.

\subsubsection{Scattering in the external frame}

We can also compute the same effect directly in the external frame.  The electron Lorentz factor in the comoving frame is related to the electron four-momentum in the external frame as 
\be
\gamma' = \Gammab (\gamma - p \betab \etae) ,
\ee
where $\etae$ is the cosine of the angle between the electron momentum and the direction of cloud motion. Because the  distribution function is Lorentz invariant,  we easily get the electron distribution in the external frame: 
\beq \label{eq:el_external}
\lefteqn{\fe (\vecp) =\fe'(\vecp') = \Ne'
\frac{\exp(-\gamma\Gammab/\Theta) }{4\pi\, \Theta \, K_2(1/\Theta)} 
\exp(p\, \Gammab\, \betab\, \etae/\Theta) }\\
&\approx&  \Ne' \frac{\exp(-\gamma/\Theta)}{4\pi\, \Theta \, K_2(1/\Theta)} 
\left[  1 + \frac{\beta_t ^2}{6} (p^2-3\gamma\Theta) + \beta_t p \etae + 
 \frac{\beta_t^2p^2}{3}P_2(\etae) \right] ,  \nonumber
\eeq
where $\beta_t = \betab/\Theta$ and we expanded the expression up to  the second order in $\betab$.
The electron density in that frame is: 
\be
\Ne = \int \fe (\vecp) \rmd^3 \vecp = \Gammab\Ne' . 
\ee
The corresponding terms $f_k$ of the electron distribution 
can be obtained from Equation (\ref{eq:el_external}) noting  that $\fe(\gamma,\etae)=\fe(\vecp)/\Gammab\Ne'$.
The change to the occupation number is: 
\be\label{eq:sz_spec_corr2}
\Delta \noccx (x,\eta) =   - \taut  \overline{s}_0(x,\eta)\noccx_{\rm bb}(x)  + S(x,\eta) . 
\ee
The scattering cross-section is given by Equation (\ref{eq:s0_d_chi}) and in Thomson limit is just $\overline{s}_0(x,\eta) \approx 1-\betab\eta$. 
The source function is now 
\be 
\label{eq:sz_source2}
S(x,\eta) =  \taut \frac{1}{x}  
\int_{0}^{\infty} x_1 \, \noccx_{\rm bb}(x_1)\, \rmd x_1 \int \rmd^2  \omega_1 \: R(\vecx_1 \rightarrow \vecx) ,
\ee
where the redistribution function $R$ given by Equation (\ref{eq:rxx_alt}) is averaged over directions of incident photons, but still depends on the scattered photon direction $\eta$. This form of the source function is more favorable compared to Equation (\ref{eq:sz_source1}) from numerical point of view, as it can be tabulated in advance at a given grid of photon energies and angles.  Computed directly it still involves  numerical calculations of 4-dimensional integrals.

\subsubsection{Isotropic scattering in Thomson regime in the electron rest frame}

In Thomson limit (as in the case  of Sunyaev-Zeldovich effect), the calculations in the external frame can be dramatically simplified. 
We can use  the azimuthally averaged approximate expression  (\ref{eq:r0_app}), (\ref{eq:aveR1}), and (\ref{eq:aveR2}) for the redistribution functions: 
\be \label{eq:intRomega}
\int  \!  \!\rmd^2   \omega_1 \: R(\vecx_1 \rightarrow \vecx) \!= \!2\pi\!\!
\int_{-1}^1 \! \!  \!\rmd \mu  \frac{3}{8}  \!\!\!\!\!\!\! 
\int\limits_{\gamma_*(x,x_1,\mu)}^\infty \!\!\!\!\!\!\rmd \gamma \left[ f_0 R_0 + f_1 \overline{R}_1 +f_2 \overline{R}_2 \right] , 
\ee
Interestingly, $R_0$ does not depend on $\gamma$ and in expressions for $\overline{R}_1$ and $\overline{R}_2$ it comes only through $\cos\theta \approx (x_1-x) \gamma/Q p$ (because $q\ll x,x_1$, see eq. [\ref{eq:costheta}]). For the electron distribution given by Equation (\ref{eq:el_external}), the integrals over $\gamma$ thus can be taken analytically: 
\beq \label{eq:intRgamma}
\int \rmd \gamma f_0 R_0 & = &C  \frac{1}{\Gammab}
\left[ 1- \frac{\betab^2}{6} \left( \frac{1}{\Theta^2} + 1 + \frac{\gamma_*}{\Theta} - 
\left(\frac{\gamma_*}{\Theta}\right)^2 \right) 
\right] , \nonumber
\\
\int \rmd \gamma f_1 R_1 & =& C  
\ \betab\ \eta\ \frac{x_1\mu-x}{Q} \ \frac{x_1-x}{Q} \left(1 + \frac{\gamma_*}{\Theta} \right) ,
\\ 
\int \rmd \gamma f_2 R_2 & = & C  
\ \frac{\betab^2}{3}\  P_2(\eta)\ \frac{x_1^2P_2(\mu)- 2x x_1 \mu + x^2 }{Q^2} 
\nonumber \\
&\times& \left[ P_2\left(\frac{x_1-x}{Q}\right) \left( 2 + 2\frac{\gamma_*}{\Theta} + \left(\frac{\gamma_*}{\Theta}\right)^2  
\right) +  \frac{1}{2\Theta^2} \right] , \nonumber
\eeq
where the proportionality coefficient $C = R_0$ $\exp(-\gamma_*/\Theta)/$ $[4\pi K_2(1/\Theta)]$.
The zeroth order term in $\betab$ was derived by \citet{Pou94PhD}, see also \citet{PS96}.

Evaluation of the source function (\ref{eq:sz_source2}) now involves only two numerical integrations over the photon energy $x_1$ 
and cosine of the scattering angle $\mu$, reducing  the computational time by 2-3 orders of magnitude. 

For all three methods we numerically compute the correction function for the black body intensity 
\be
\Delta I(x) = \frac{ 1}{ \taut} x_t^3 \Delta \noccx (x,\eta) , 
\ee
and compare the results of calculations in Figure~\ref{fig:sz}. The three different methods give nearly identical results.

\section{Conclusions}
\label{sec:concl}

We have developed the exact analytical theory of  Compton scattering by anisotropic distribution of electrons that  can be represented by a second order polynomial over cosine of some angle (dipole and quadrupole anisotropy). 
For the total cross-section, we reduce the 9-dimensional integral to a single integral over the electron energy. 
Analogous expressions have been derived for the mean energy of the scattered photons and its dispersion. We 
also obtained analytical expressions for the radiation pressure force acting on the electron gas. These moments 
can be used for analytical estimations as well as for the numerical solutions of the kinetic equations in the Fokker-Planck 
approximation \citep[see e.g.][]{VP09}.
 
Furthermore, the expression for the redistribution function describing angle-dependent Compton scattering by anisotropic electrons is reduced to a single integral over the electron energy. Exact analytical formulae valid for any photon and electron energy are derived in the case of monoenergetic electrons. We have also derived approximate expressions for the redistribution function, assuming isotropic scattering in the electron rest frame, which are very accurate in the case of relativistic electrons interacting with soft photons in Thomson regime.  
 
We applied the developed formalism to the accurate calculations of the thermal and kinematic Sunyaev-Zeldovich effects for arbitrary electron distributions. A very similar problem arises  in outflowing coronae around accreting black holes and neutron stars, where the bulk motion causes electron anisotropy.  Another application could be  a computation of the radiative transport in the synchrotron self-Compton sources with ordered magnetic field, where the electron distribution can have strong deviations from the isotropy because of pitch angle-dependent cooling. 
These problems will be considered in future publications.

\begin{acknowledgements}
This work was  supported by the CIMO grant TM-06-4630 and the Academy of Finland grants 122055 and 127512. 
\end{acknowledgements}

\begin{table}[h]
\caption{Coefficients $a_{jn}$ and $A_{jn}$.}
\centering
\begin{tabular}{lccc}
\hline\hline
 $n$        &  0     & 1         & 2 \\         
 \hline
$a_{0n}$  & 1     & 1         & 13/10 \\  
$a_{1n}$  & 1     & 3/2     & 47/20 \\   
$a_{2n}$  & 1     & 2         & 15/4  \\ 
$A_{1n}$  & 1     & 21/10 & 147/40 \\ 
$A_{2n}$  & 6/5 & 53/20 & 159/35   \\
$A_{3n}$  & 1     & 14/5   & 47/8 \\ 
$A_{4n}$  & 7/5  & 22/5  & 341/35  \\
$A_{5n}$  & 1/5   & 2       & 401/70 \\
$A_{6n}$  & 1/10 & 1/5   &  207/280 \\
$A_{7n}$  & 3/10 & 3/5   &  281/280 \\
\hline
\end{tabular}
\label{tab:app1}
\end{table}

\appendix

\section{Functions $s_J$ and $S_J$}
\label{app:funsS}

All function  $s_{j}$ and $S_{j}$ can be expanded to the series, which
converge in the region $\xi<1/2$. It is easy to show that
\be \label{eq:sjSj} 
s_{j} (\xi)=\sum _{n=0}^\infty a_{jn}\,(-2\,\xi)^n,\qquad 
S_{j} (\xi)=\sum _{n=0}^\infty A_{jn}\,(-2\,\xi)^n,
\ee
where
\beq \label{eq:a01234} 
a_{0n} =  \frac{3}{8}\, \left[n+2+\frac{2}{n+1}  + \frac{8}{n+2}  - \frac{16}{n+3}  \right] , \quad 
a_{1n} = \frac{1}{8} \left[ n\,(n+5)+ \frac{24}{n+3}  \right] , \quad 
a_{2n} =  \frac{1}{32}\left(n^3+9\,n^2+22\,n+32\right) .  
\eeq
Using Equations~(\ref{eq:Ss12}),  one can obtain the expressions for the coefficients $A$ through $a$:
\beq \label{eq:Anan} 
A_{1\,n}&=&2\,(a_{1\,n+1}-a_{0\,n+1}),\ \ 
A_{2\,n}=2\,(A_{1\,n+1}-a_{1\,n+1}), \  
A_{3\,n}=2\,(a_{2\,n+1}-a_{1\,n+1}), \\
A_{4\,n}&=&2\,(A_{3\,n+1}-A_{1\,n+1}),   \
A_{5\,n}=3\,A_{4\,n}-\,4A_{3\,n}, \   
A_{7\,n}=A_{3\,n}- A_{4\,n}/2, \ 
 A_{6\,n}= a_{2\,n}-3\,A_{7\,n}.  \nonumber
\eeq
The coefficients for $n=0,1,2$ are presented in Table \ref{tab:app1}.

\section{Auxiliary functions $\psiij$ and $\PPsiij$}
\label{app:psi_ij}

The total cross-section and mean powers of energy of scattered photons are expressed through the functions of
one variable 
\beq \label{eq:psiij} 
 \psi_{ij}(\xi) =\frac{i+1}{\xi^{i+1}}
\int_0^\xi x^i\,s_{j}(x)\,\rmd x,\, i>-1,  \quad 
 \psi_{-1j}(\xi)= \frac{1}{\xi}\int_0^\xi
[1-s_{j}(x)]\frac{\rmd x}{x}, \quad 
\Psi_{ij}(\xi)= \frac{i+1}{\xi^{i+1}}
\int_0^\xi x^i\,S_{j}(x)\,\rmd x . 
\eeq
Calculations of  functions $\psi_{ij}$ involve integrals of the following types:
\be
\int \rmd x \: x^{m} \ln(1+2x), \quad  \int \rmd x\: \frac{x^{n}}{(1+2x)^l}, \quad  
g(\xi) =\int_0^\xi \ln(1+2x)\ \frac{\rmd x}{x} , 
\ee
where $m = -1,0,1,2,3$ and $n, l = 1,2,3,4$. All integrals are elementary except $g(\xi)$, 
which is described in details in Appendix \ref{app:g_xi}.

The explicit expressions for the functions $\psi_{ij}$ are the following:
\beq \label{eq:psixi}
 \psi_{-10}(\xi)&=&\frac{1}{8\,\xi} \left[ \frac{4}{\xi^2}+
\frac{2}{\xi}+\left( 8+\frac{3}{\xi}-\frac{3}{\xi^2}-\frac{2}{\xi^3}
\right) l_\xi-3\,R_\xi-\frac{11}{3} \right],  \nonumber \\
 \psi_{-11}(\xi)&=&\frac{1}{8\,\xi} \left[ \frac{7}{3}+
\frac{2}{\xi}-\frac{2}{\xi^2}+\left( 8+\frac{1}{\xi^3}
\right) l_\xi-4\,R_\xi-R_\xi^2 \right],  \nonumber \\
 \psi_{-12}(\xi)&=&\frac{1}{16\,\xi} \left[ 11+16\,l_\xi
-7\,R_\xi-3\,R_\xi^2-R_\xi^3 \right],   \nonumber \\
 \psi_{00}(\xi)&=&\frac{3}{8\,\xi} \left[ g(\xi)-
\frac{2}{\xi}+\left(\frac{1}{2}+\frac{2}{\xi}+\frac{1}{\xi^2}
\right) l_\xi-\frac{1}{2}\,R_\xi- \frac{3}{2} \right],  \nonumber \\
 \psi_{01}(\xi)&=&\frac{3}{8\,\xi} \left[ \frac{1}{\xi}-
\frac{1}{2}+\left(\frac{4}{3}-\frac{1}{2\,\xi^2}
\right) l_\xi-\frac{1}{3}\,R_\xi-\frac{1}{6}\,R_\xi^2 \right],  \nonumber \\
 \psi_{02}(\xi)&=&\frac{1}{32\,\xi} \left[ \frac{7}{2}+
9\,l_\xi-R_\xi-\frac{3}{2}\,R_\xi^2-R_\xi^3 \right],  \\
 \psi_{10}(\xi)&=&\frac{3}{4\,\xi^2} \left[
\left(\xi+\frac{9}{2}+\frac{2}{\xi}
\right) l_\xi-4-\xi+\xi^2\,R_\xi-2\,g(\xi) \right],  \nonumber \\
\psi_{11}(\xi)&=&\frac{1}{4\,\xi^2} \left[ 6+4\,\xi-
3\,\left(\frac{3}{2}+\frac{1}{\xi}
\right) l_\xi-\xi\,(1+\xi)\,R_\xi^2 \right],  \nonumber \\
 \psi_{12}(\xi)&=&\frac{1}{16\,\xi^2} \left[ \frac{1}{2}+
9\,\xi-4\,l_\xi-R_\xi+\frac{1}{2}\,R_\xi^3 \right],  \nonumber \\ 
\psi_{20}(\xi)&=&\frac{9}{32\,\xi^3}\left[ 25\,\xi+
\left( 2 \xi^2 -8\,\xi- 5 \right) \,l_\xi+ \xi\, R_\xi - 8\,g(\xi)\right] , \nonumber \\
 \psi_{21}(\xi)&=&\frac{9}{8\,\xi^3} \left[ g(\xi)+
\frac{2}{3}\,\xi^2-\frac{3}{2}\,\xi-
\frac{1}{4}\,l_\xi-\frac{\xi^2}{6}\,R_\xi^2 \right],  \nonumber \\
 \psi_{22}(\xi)&=&\frac{3}{64\,\xi^3} \left[9\,\xi^2-8\,\xi+
5\,l_\xi-\xi\,(2+11\xi+10\,\xi^2)\,R_\xi^3 \right],  \nonumber \\
\psi_{30}(\xi)&=&\frac{3}{12\,\xi^4}\left[ \frac{\xi^3}{3} +\frac{31}{2} \xi^2 + \frac{31}{4}\,\xi 
+ \left( 2\xi^3 - 6\xi^2 -12\,\xi-\frac{7}{2}\right) \,l_\xi -  \frac{3}{4} \xi\, R_\xi 
\right] , \nonumber \\
\psi_{31}(\xi)&=&\frac{1}{24\,\xi^4}\left[   16\xi^3 - 27 \xi^2 - 45 \,\xi 
+ 3( 7 + 12 \xi) \,l_\xi  + 3\xi \,(1+3\xi)\,R_\xi^2 
\right] , \nonumber \\
\psi_{32}(\xi)&=& \frac{1}{16\,\xi^4}\left[   6\xi^3 - 4 \xi^2 +5 \,\xi 
- 3 \,l_\xi  + \xi \,(1+4\xi+2\xi^2) \, R_\xi^3 
\right] , \nonumber 
\eeq
\beq 
\psi_{40}(\xi)&=&\frac{15}{128\,\xi^5}\left[ \xi^4 + \frac{230}{9} \xi^3 +\frac{23}{6} \xi^2 - \frac{17}{6}\,\xi 
+
\left( 4\xi^4 - \frac{32}{3}\xi^3   -16\xi^2 +\frac{11}{12}\right) \,l_\xi +  \xi\, R_\xi 
\right] , \nonumber \\
\psi_{41}(\xi)&=&\frac{5}{64\,\xi^5}\left[ 8 \xi^4 -12 \xi^3 -9 \xi^2 +  8\,\xi 
+ 
3\left( 4\xi^2   - 1\right) \,l_\xi  - \xi (2+5\xi)\,R_\xi^2 
\right] , \nonumber \\
\psi_{42}(\xi)&=& \frac{5}{256\,\xi^5}\left[  18\xi^4 - \frac{32}{3}\xi^3 +10 \xi^2 -12 \,\xi 
+
 \frac{11}{2} \,l_\xi  + \xi \,(1+7\xi+14\xi^2) \, R_\xi^3 
\right] , \nonumber 
\eeq
where $l_\xi=\ln(1+2\xi)$ and $R_\xi=1/(1+2\xi)$. 

The explicit expressions for $\Psi_{ij}$ can be obtained using definitions~(\ref{eq:Ss12}) for $S_{j}$: 
\beq \label{eq:Psipsi}
\Psi_{11}&=& 2\,(\psi_{00}-\psi_{01})/\xi, \qquad \quad
\Psi_{13}=2\,(\psi_{01}-\psi_{02})/\xi, \qquad 
\Psi_{14}=2\,(2\psi_{-11}-\psi_{-10}-\psi_{-12})/\xi,  \nonumber \\ 
\Psi_{16} & = & \psi_{12}-3\,\Psi_{13}+3\,\Psi_{14}/2, \quad 
\Psi_{21}= 3\,(\psi_{10}-\psi_{11})/2\xi,  \quad 
\Psi_{22}=3\,(\psi_{11}-\Psi_{11})/2\xi, \nonumber  \\
\Psi_{23}&=& 3\,(\psi_{11}-\psi_{12})/2\xi, \qquad \quad
\Psi_{24}=3\,(\Psi_{11}-\Psi_{13})/2\xi, \quad 
\Psi_{25}=3\,\Psi_{24}-4\,\Psi_{23},  \nonumber  \\
\Psi_{26}&=&\psi_{22}-3\,\Psi_{23}+ 3\,\Psi_{24}/2, \quad 
\Psi_{31}=4\,(\psi_{20}-\psi_{21})/3\xi, \quad
\Psi_{32}= 4\,(\psi_{21}-\Psi_{21})/3\xi, \nonumber  \\
\Psi_{33}&=& 4\,(\psi_{21}-\psi_{22})/3\xi, \quad \qquad
\Psi_{34}=4\,(\Psi_{21}-\Psi_{23})/3\xi, \quad 
\Psi_{35}= 3\,\Psi_{34}-4\,\Psi_{33},  \nonumber \\ 
\Psi_{37}&=&\Psi_{33}-\Psi_{34}/2, \qquad \qquad
\Psi_{36}= \psi_{32}-3\Psi_{37} , \qquad \qquad
\Psi_{41}= 5\,(\psi_{30}-\psi_{31})/4\xi, \nonumber \\
\Psi_{42}& =& 5\,(\psi_{31}-\Psi_{31})/4\xi,  \quad \qquad 
\Psi_{43}= 5\,(\psi_{31}-\psi_{32})/4\xi,    \quad
\Psi_{44}= 5\,(\Psi_{31}-\Psi_{33})/4\xi, \nonumber \\
\Psi_{45}&=& 3\,\Psi_{44}-4\,\Psi_{43}, \qquad \qquad
\Psi_{47}= \Psi_{43}-\Psi_{44}/2,   \qquad \quad  
\Psi_{51}= 6\,(\psi_{40}-\psi_{41})/5\xi, \nonumber \\
\Psi_{54}&=& 6\,(\Psi_{41}-\Psi_{43})/5\xi,  \quad \qquad 
\Psi_{57}= 6\,(\psi_{41}-\psi_{42})/5\xi - \Psi_{54}/2 .  
\eeq
In these formulae  the argument $\xi$ is omitted.  
For complete evaluation of these functions we need to compute 18 different functions $\psi_{ij}$ given above. 

To prevent the loss of accuracy if $\xi$ is very small, we can use the series expansions (see NP94) that 
directly follow from the definitions (\ref{eq:psiij})  and Taylor expansions (\ref{eq:sjSj}):  
\beq \label{eq:psiser} 
 \psi_{ij}(\xi) & = & \sum_{n=0}^\infty a_{jn}\,(-2\,\xi)^n\,\frac{i+1}{i+n+1},  \, i>-1, \quad 
 \psi_{-1j}(\xi) = \sum_{n=0}^\infty  a_{j\,n+1}\,(-2\,\xi)^n\,\frac{2}{n+1} , \\ 
\label{eq:ppsiser}
\Psi_{ij}(\xi) & =&  \sum_{n=0}^\infty A_{jn}\,(-2\,\xi)^n\,\frac{i+1}{i+n+1}, 
\eeq
with $a_{jn}$ and $A_{jn}$ given by Equations~(\ref{eq:a01234}) and (\ref{eq:Anan}), respectively.


\section{Auxiliary function $\gxi$}
\label{app:g_xi}

Calculations of function $\psi_{ij}$ from Appendix \ref{app:psi_ij} involve integral 
\be \label{eq:gxi}
g(\xi)=\int_0^\xi \ln(1+2x)\ \frac{\rmd x}{x}.
\ee
We repeat here for completeness  the method of calculations of this integral from NP94.
It is possible to write a relation between the values of this function
on $\xi<1/2$ and $\xi>1/2$. Let us define for that the auxiliary
function for  $\xi\leq 1$
\be \label{eq:gy}
g_*(\xi)=g(\xi/2)=\int_0^\xi \ln{(1+x)}\ \frac{\rmd x}{x}.
\ee
It can be presented by series
\be 
 \label{eq:gyser}
g_*(\xi) \!=\!
\left\{ \begin{array}{ll}
\displaystyle\!\! \sum_{n=1}^{\infty} (-1)^{n-1} \frac{\xi^n}{n^2} & 
\mbox{if}\ \xi\!\leq\! \xi_*\!<\!1, \\
\displaystyle
\!\!\frac{\pi^2} {12}\!+\!\ln2\,\ln \xi\!+\!\!\sum_{k=0}^{\infty}\!\! \frac{(1\!-\!\xi)^{k+2}}{k+2}
\!\sum_{m=1}^{k+1}\!\! \frac{1}{2^m m} & \mbox{if}\ \xi_*\!\leq\! \xi\!\leq\! 1.
\end{array} 
\right.
\ee
As  $\xi_*$ we can take  0.8 -- 0.9. Then
\be \label{eq:gxigy}
g(\xi)=\left\{ \begin{array}{ll}
\strut\displaystyle g_*(2\,\xi) & \mbox{if}\ 0\leq\xi\leq1/2, \\
\strut\displaystyle \frac{\pi^2}{12} & \mbox{if}\ \xi=1/2, \\
\strut\displaystyle \frac{\pi^2}{6}+\frac{1}{2}\ln^2 (2\,\xi) -g_*(1/2\,\xi) & \mbox{if}\ \xi\geq1/2.
\end{array} \right.
\ee

\section{Asymptotic expansions of functions $\chijn$ and $\Deltajn$ in Thomson limit}
\label{app:chi_ij}

Using Taylor expansion (\ref{eq:psiser}) of functions $\psi_{0n}$ for small arguments, 
it is easy to get an expansion of functions $\chizeron$ in Thomson limit $x\gamma\ll 1$:  
\be \label{eq:chi0nser} 
\chizeron(x,\gamma)=\gamma^n \, \sum_{l=0}^{\infty}  \,(-2x \gamma)^l \, a_{0l} \, \kappa_{n+l+1} , 
\ee
where 
\beq 
\kappa_l  &= &  \frac{\left( 1+ \beta\right)^{l+1} - \left( 1- \beta\right)^{l+1}}{2\beta (l+1) }  
= \sum_{k=0}^{{\rm int}(l/2)}\, \frac{l!\, \beta^{2\,k}}{(2\,k+1)!\,(l-2\,k)!} 
=  \frac{(1+\beta)^{l}}{l+1 }  \sum_{k=0}^{l}\, \frac{1}{[\gamma(1+\beta)]^{2k}}  
\eeq
and ${\rm int}(x)$ is the integer  part of $x$. A few first functions are 
\beq 
 \kappa_1&=& 1, \quad 
\kappa_2 = 1+ \frac{1}{3}\beta^2, \quad \kappa_3 = 1+ \beta^2, \quad 
  \kappa_4 =  1+ 2\beta^2 + \frac{1}{5} \beta^4, \quad
 \kappa_5 = 1+ \frac{10}{3}\beta^2 + \beta^4 , \\
 \kappa_6 & =&  1+ 5\beta^2 + 3\beta^4 + \frac{1}{7}\beta^6  , \quad
 \kappa_7= 1+ 7\beta^2 + 7\beta^4 + \beta^6. \nonumber
\eeq

The first three terms of the expansion (\ref{eq:chi0nser}) are as follows: 
\beq \label{eq:chi0nser_three} 
\chi_{\!0n}(x,\gamma)  \approx  \gamma^n\left[ \kappa_{1+n} - 2 x\gamma  \kappa_{2+n} +  
\frac{26}{5} \left(x\gamma\right)^2 \kappa_{3+n}   \right]  .
\eeq
Function $\Delta_{00}$ coincides with $\chi_{00}$, and functions $\Delta_{01}$ and $\Delta_{02}$ can be obtained using 
definitions (\ref{eq:delta0k}) and expansion (\ref{eq:chi0nser}). 
For $\Delta_{01}$, we get  
\beq 
\Delta_{01} \! =\!  - \beta \!  \sum_{l=0}^{\infty}  (-2x \gamma)^l  a_{0l}  \zeta_{l+1} \! 
\approx\!  - \frac{\beta}{3} \left[ 1 \! -\!  4 x \gamma \! +\!  \frac{78}{5}  (x\gamma)^2 \! \left(\! 1\! +\! \frac{\beta^2}{5}\! \right) \right] ,
\eeq 
where 
\beq 
\zeta_l &=&  \frac{\kappa_{l+1}  - \kappa_{l} }{\beta^2}   
= \sum_{k=0}^{{\rm int}[(l-1)/2]}\!\! \frac{2k+2}{(2k+3)!} \, \beta^{2k} \, \frac{l!}{(l-1-2k)!},  \\
\zeta_1 &=& \frac{1}{3}  , \quad \zeta_2= \frac{2}{3}  , \quad \zeta_3= 1 + \frac{1}{5}\beta^2, \quad
\zeta_4= \frac{4}{3}   + \frac{4}{5}\beta^2, \quad 
  \zeta_5 =  \frac{5}{3}   + 2\beta^2 + \frac{1}{7}\beta^4, \quad   
    \zeta_6=  2  + 4\beta^2 + \frac{6}{7}\beta^4 . \nonumber
\eeq
Respectively for $\Delta_{02}$, we have 
\beq 
\Delta_{02}  
= \beta^2 \sum_{l=0}^{\infty}  \,(-2x \gamma)^l \, a_{0l} \,\Lambda_{l+1}  
\approx - \frac{4}{15}\, \beta^2 (x\gamma)\, \left( 1 -  \frac{39}{5} x \gamma   \right) ,  
\eeq 
where 
\beq
\Lambda_l & =& \frac{1}{2\beta^2} 
\left[ 3 \frac{ \kappa_{l}  - 2\kappa_{l+1} +\kappa_{l+2}  }{\beta^2}   - \kappa_{l} \right] 
=  
\sum_{k=1}^{{\rm int}(l/2)}\!\! \frac{l!}{(l-2k)!(2k)!} \frac{2k}{(2k+1)(2k+3)} \, \beta^{2(k-1)} \, , \\
\Lambda_1&=&0, \quad \Lambda_2= \frac{2}{15} , \quad \Lambda_3= \frac{2}{5}, \quad
\Lambda_4= \frac{4}{5} + \frac{4}{35}\beta^2 . \nonumber
\eeq

Similarly, for functions $\chi_{1n}$,  using expansions (\ref{eq:ppsiser}) we get: 
\be \label{eq:chi1nser} 
 \chionen(x,\gamma)  = \gamma^{n+1} \, \sum_{l=0}^{\infty}  
\,(-2x \gamma)^l \, \left( \gamma\, A_{1l} +x\, A_{2l}\right)  \, \kappa_{n+l+2}  
= \gamma^{n} \left[ \gamma^2  \kappa_{2+n} +  
\, \sum_{l=1}^{\infty}  
\,(-2x \gamma)^l \, \left( \gamma^2\, A_{1l}\, \kappa_{n+l+2} - \frac{A_{2\,l-1}}{2} \kappa_{n+l+1} \right) \right]. 
\ee
Functions  $\Delta_{1n}$  can then be obtained using definitions (\ref{eq:Delta_j_xetagam}):   
\beq 
\Delta_{10} & =& \chi_{10} \approx 1 + \frac{4}{3}p^2 - \frac{x\gamma}{5}  \left( 42\gamma^2 - 27 -2 \beta^2 \right) 
, \nonumber \\
\Delta_{11} & =&   -  p \sum_{l=0}^{\infty}  (-2x \gamma)^l 
\left( \gamma\, A_{1l} +x\, A_{2l}\right)  \, \zeta_{l+2} 
\approx
-   \beta \left\{ \frac{2}{3} \gamma^2  
- \frac{x\gamma}{5}  \left[ 21\gamma^2 (1+\beta^2/5)- 4 \right] \right\}  ,  \nonumber \\ 
\Delta_{12} & =&   p \beta \sum_{l=0}^{\infty}  (-2x \gamma)^l 
\left( \gamma\, A_{1l} +x\, A_{2l}\right)  \, \Lambda_{l+2} 
\approx 
\frac{2}{15}  \beta^2 \left[
\gamma^2   - \frac{3 x\gamma}{5}  \left( 21\gamma^2  - 2 \right) \right]  . 
\eeq 

For functions $\chi_{2n}$, we can write the expansion 
\beq 
\label{eq:chi2nser}  
\chitwon(x,\gamma) &=&  \gamma^{n} \, \sum_{l=0}^{\infty}  \,(-2x \gamma)^l  
\left[ \left( \gamma^4 A_{4l} - \gamma^2 A_{7l} \right) \, \kappa_{n+l+3} 
- \gamma^2 A_{5l} \, \kappa_{n+l+2} + A_{6l}\,  \kappa_{n+l+1} \right]    \nonumber \\
&\approx & \gamma^n \left[ \frac{7}{5}   \gamma^4 \kappa_{3+n}  - 
\frac{\gamma^2}{10}   (3 \kappa_{3+n} +2 \kappa_{2+n} ) + \frac{1}{10}  \kappa_{1+n} \right] , 
\eeq
where we kept only the zeroth  term in $x\gamma$ of the series. 
Expansions for  $\Delta_{2n}$  can then be obtained using definitions (\ref{eq:Delta_j_xetagam}):   
\beq 
\Delta_{20} & = & \chi_{20} \approx 1+ \frac{2}{15} p^2  \left( 21 \gamma^2 +4 \right) , \nonumber \\
\Delta_{21} & = & -  \beta \sum_{l=0}^{\infty}  (-2x \gamma)^l 
\left[ \left( \gamma^4 A_{4l} - \gamma^2 A_{7l} \right) \, \zeta_{l+3} 
- \gamma^2 A_{5l} \, \zeta_{l+2} + A_{6l}\,  \zeta_{l+1} \right]   
\approx -\beta \left[ 1+ \frac{2}{75} p^2  \left( 63 \gamma^2 + 34 \right)  \right] , \nonumber \\
\Delta_{22} & = & \beta^2 \sum_{l=0}^{\infty}  (-2x \gamma)^l 
\left[ \left( \gamma^4 A_{4l} - \gamma^2 A_{7l} \right) \, \Lambda_{l+3} 
- \gamma^2 A_{5l} \, \Lambda_{l+2} + A_{6l}\,  \Lambda_{l+1} \right]  
\approx p^2 \frac{1}{75}\left( 42  \gamma^ 2  - 11 \right) . 
\eeq 

Let us now discuss the properties of functions $\chionen^{*}$. The series expansion can be easily obtained 
from the definition~(\ref{eq:chi_vert}) and series (\ref{eq:ppsiser}): 
\be \label{eq:chi1ver_ser}  
\chionen^{*}(x,\gamma)= \gamma^{n+2}  \, \sum_{l=0}^{\infty}  
\,(-2x \gamma)^l \,  A_{1l}   \, \kappa_{n+l+3}  .
\ee
For $\Delta^{*}_{1k}$ we get:
 \beq 
\Delta^{*}_{10} & =& \chi^{*}_{10} \approx \gamma^2 \left(1+\beta^2\right) , \nonumber\\
\Delta^{*}_{11} & =& -\gamma p \sum_{l=0}^{\infty}  (-2x \gamma)^l  \, A_{1l}\,  \zeta_{l+3}
 \approx - \gamma p \, (1+\beta^2/5) , \nonumber \\
\Delta^{*}_{12} & =& p^2 \sum_{l=0}^{\infty}  (-2x \gamma)^l  \, A_{1l}\,  \Lambda_{l+3}
\approx   \frac{2}{5}p^2.
\eeq

The series expansion for functions $\chionen^{\bot}$ are: 
\be \label{eq:chi1bot_ser}  
\chionen^{\bot}(x,\gamma)\! =\!  \frac{\gamma^{n+1}}{p} \! \! \sum_{l=0}^{\infty}  \! 
(-2x \gamma)^l A_{1l}  \left(  \kappa_{n+l+2} - 2\gamma^2\, \kappa_{n+l+3} + \gamma^2
\kappa_{n+l+4}  \right)  
= 
 \frac{2}{3} \gamma^{n+1}p \sum_{l=0}^{\infty}  (-2x \gamma)^l A_{1l}  
\left(\beta^2 \Lambda_{n+l+2} - \kappa_{n+l+2} \right) . 
\ee
For $\Delta^{\bot}_{1k}$ we get: 
 \beq 
\Delta^{\bot}_{11}  &= & \frac{1}{2} \chi^{\bot}_{10} \approx -\frac{1}{3} \gamma p\, (1+\beta^2/5) , \nonumber  \\
\Delta^{\bot}_{12} & =& \frac{1}{2p} \left(  \gamma \chi^{\bot}_{10} - \chi^{\bot}_{11} \right) = \frac{1}{3} p^2 \sum_{l=0}^{\infty}  (-2x \gamma)^l  \, A_{1l}\,  
\left[   \Lambda_{l+2} -\Lambda_{l+3} +  \zeta_{l+2} \right]   \approx   \frac{2}{15} p^2  .
\eeq

\section{Eliminating cancellations in redistribution functions}

If formulae (\ref{eq:r0f}), (\ref{eq:rsf}) and (\ref{eq:rpf}) are used as they stand, numerical  cancellations appear at certain regions of parameter space. For example if $x$ and $x_1$ are small, the quantities $a_-$ and $a_+$, $1/a_-$ and $1/a_+$, are close to each other. Also a combination containing a sum of $d_-/a_-^3$ and $d_+/a_+^3$ minus double the difference $1/a_-$ and $1/a_+$ has a cancellation. Therefore it is useful to rewrite the expressions in a form not containing those cancellations. The cancellations appearing in (\ref{eq:r0f}) were dealt with in \citet{NP93}. Defining
\be \label{eq:def_uv}
u = a_+ - a_- = \frac{(x+x_1)(2\gamma+x_1-x)}{a_-+a_+}, \quad v = a_- a_+,  
\ee 
they got
\be
R_0 =  \frac{2}{Q} + \frac{u}{v}\left(1 - \frac{2}{q}    \right)
+  u\frac{(u^2-Q^2)(u^2+5v)}{2q^2 v^3} + u \frac{Q^2}{q^2 v^2}  .
\ee
Using definitions (\ref{eq:def_uv}), we get from (\ref{eq:rsf}) and (\ref{eq:rpf})
\be
R_{\Sigma} = (a_-+a_+)\left[ \frac{2u}{Q^3} +  \frac{1}{v}\left(1 - \frac{2}{q}    \right)+
\frac{(u^2-Q^2)(u^2+3v)}{2q^2 v^3} + \frac{Q^2}{q^2v^2} \right], 
\ee
\be
R_{\Pi} = \frac{1}{2Q^5}\left[u^2(u^2+4v) + 2b^2 \right]
+ \frac{1}{2Q} \left(1 - \frac{4}{q}    \right) + \frac{u}{q^2 v}.
\ee
Another loss of accuracy occurs in $u^2-Q^2$ term, 
when $\gamma$ is close to $\gamma_*(x,x_1,\mu)$. We can use the following 
formulae \citep{NP93}: 
\beq
u^2-Q^2 =  2 r q C D_u , \quad 
D_u = (\gamma + x_1 - x + \gamma_{*})(\gamma - \gamma_{*}), \quad 
C = 2/[\gamma(\gamma + x_1 - x) + r + xx_1\mu + v ] . 
\eeq

\section{Boundaries}

The redistribution functions $R_0, R_{\Sigma}, R_{\Pi} $ and $R_1, R_2$ are defined within the interval of photon and electron energies and scattering angles satisfying the relation $|\cos\theta|\le1$, where $\cos\theta$ is given by Equation~(\ref{eq:costheta}). 
These limits were discussed in NP94, but we repeat them here for completeness. For fixed photon energies and scattering angle, we already got the limits on the electron energies given by Equation~(\ref{eq:gammamin}), $\gamma\geq \gamma_*(x,x_1,\mu)$. If we are interested in the interval of scattered photon energies for the fixed $x_1, \gamma$ and $\mu$, we have then $x^- \leq x \leq x^+ $, where
\be \label{eq:xpm}
x^{\pm}(x_1,\gamma,\mu)=x_1\,
\frac{\mu+\gamma(\gamma+x_1)(1-\mu) \pm p\,(1-\mu)\,a_+}{1+2\,\gamma\,x_1 \,(1-\mu)+x_1^2\ (1-\mu)^2}. 
\ee
If the energy of the scattered photon $x$ is fixed,  the initial photon $x_1$ lies in the interval 
\be \label{eq:x1_limits}
\begin{array}{ll}
x^- _1 \leq x_1 \leq x^+ _1 & \mbox{if}\  0 \leq x\,(1-\mu) \leq \gamma-p, \\
x_1 > x^-_1 & \mbox{if}\    \gamma-p \leq x\,(1-\mu) \leq \gamma+p ,
\end{array}
\ee
where
\be \label{eq:x1pm}
x^{\pm}_1 (x,\gamma,\mu) =x\,
\frac{\mu+\gamma(\gamma-x)(1-\mu) \pm p\,(1-\mu)\,a_-}{1-2\,\gamma\,x \,(1-\mu)+x^2\ (1-\mu)^2}.
\ee
In Equations (\ref{eq:xpm}) and (\ref{eq:x1pm}), the quantities $a_{\pm}$ are defined by Equations (\ref{eq:dd1_aa}).
If $|x-x_1|\leq 2xx_1$, the quantity $\gamma_*(x,x_1,\mu)$ as a function of $\mu$ has a  minimum 
\be \label{eq:ggg}
 \gamma_{\rm min} =  1+(x-x_1+|x-x_1|)/2 
\ee
at  $\mu=\mu_{\rm min}=1-|x-x_1|/xx_1$, while in the opposite case,  $|x-x_1| \geq 2xx_1$, the function is monotonic with the minimum reached at the boundary $\mu=-1$  (see Fig. \ref{fig:gammastar}). Correspondingly, the limits of variations of $\mu$ depend on the photon energies $x,x_1$ and the
electron energy $\gamma$ and are given by  
\be \label{eq:mu_limits}
\mu_{\rm m} \leq \mu \leq \mu_+ , 
\ee
where
\beq \label{eq:mulim}
\mu_{\rm m}(x,x_1,\gamma)&=&\left\{ \begin{array}{ll}
-1       &  \mbox{if}\ |x-x_1| \geq 2\,x\,x_1,  \\
-1       & \mbox{if}\ |x-x_1| \leq 2\,x\,x_1 \ \ \mbox{and}\ \gamma \geq \gamma_*(x,x_1,-1), \nonumber \\
\mu_- &  \mbox{if}\ |x-x_1| \leq 2\,x\,x_1\ \  \mbox{and}\ \gamma \leq  \gamma_*(x,x_1,-1) ,
\end{array} \right. \\ 
\label{ted}
\mu_-(x,x_1,\gamma)&=& 1-\frac{q_+}{x\,x_1}, \\
 \mu_+(x,x_1,\gamma)&=& 1-\frac{q_-}{x\,x_1} = 1-
 \frac{(x-x_1)^2}{x\,x_1\, q_+}, \nonumber
\eeq
and
\be \label{eq:gamstar}
 \gamma_*(x,x_1,-1) = \left[x-x_1+(x+x_1)\sqrt{1+1/xx_1}\right]/2, 
\ee
\be \label{eq:Dm}
q_{\pm} =  p^2+\gamma\,(x_1-x) \pm p\,\sqrt{(\gamma+x_1-x)^2-1}.
\ee

  \begin{figure}
\centerline{\epsfig{file=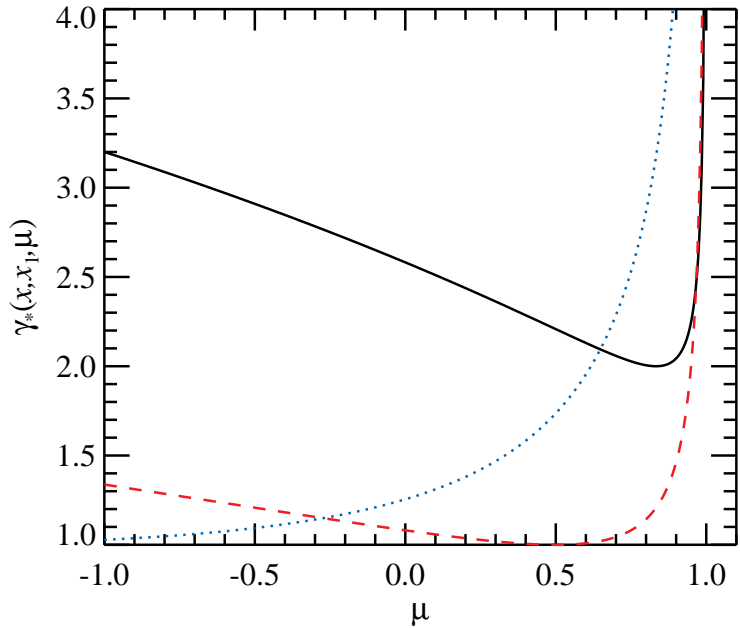,width=10cm} }
\caption{The dependence of the function $\gamma_*(x,x_1,\mu)$ on $\mu$. The energy of the incident photon 
is $x_1=2$. 
The solid curves are for $x=3$ and the dashed curves are for $x=1$ (both cases correspond to $|x-x_1|<2xx_1$), and the 
dotted curve is for $x=0.3$ (where $|x-x_1|>2xx_1$). 
}
\label{fig:gammastar}
\end{figure}

For the angle-averaged redistribution function, the lower limit on the electron energy is:
\begin{equation}\label{eq:gammaxx}
\gamma_{\star}(x,x_1) = 
\begin{cases}
 \gamma_*(x,x_1,-1)   & \text{if $ |x-x_1| \geq 2\,x\,x_1$,}   \\
\gamma_{\rm min} &  \text{if $|x-x_1| \leq 2\,x\,x_1$.}
\end{cases}	
\end{equation}
The limits of variation of the scattered photon energy $x$ as a function of  incident photon energy $x_1$ and $\gamma$ can be found by inverting   Equation (\ref{eq:gammaxx}). We obtain 
\be
x^- (x_1,\gamma) \leq x \leq  x_{\rm m} (x_1,\gamma) , 
\ee
where 
\beq \label{eq:xlim}
 x_{\rm m} (x_1,\gamma) & =&  \left\{ \begin{array}{ll}
\gamma+x_1-1   & \mbox{if} \,\, 1\leq \gamma \leq
\strut\displaystyle 1+\frac{2x_1^2}{1-2x_1}
\,\,    \mbox{and}\,\, x_1<1/2, \\
x^+ (x_1,\gamma) &  \mbox{if} \,\,
\strut\displaystyle 1+\frac{2x_1^2}{1-2x_1} \leq \gamma 
\,\,\mbox{and}\,\, x_1<1/2, \\
\gamma+x_1-1 &     \mbox{if} \,\, x_1 \geq 1/2,
\end{array} \right.  \nonumber \\
 x^{\pm} (x_1,\gamma)&=& 
 x_1 \left( \gamma \pm p \right) / \left( \gamma \mp p+2\,x_1 \right)  .
\eeq


\begin{thebibliography}{22}
\expandafter\ifx\csname natexlab\endcsname\relax\def\natexlab#1{#1}\fi

\bibitem[{{Aharonian} \& {Atoyan}(1981)}]{AA81}
{Aharonian}, F.~A. \& {Atoyan}, A.~M. 1981, \apss, 79, 321

\bibitem[{{Arutyunyan} \& {Nikogosyan}(1980)}]{an80}
{Arutyunyan}, G.~A. \& {Nikogosyan}, A.~G. 1980, Sov. Phys. -- Doklady, 25, 918

\bibitem[{{Belmont}(2009)}]{Bel09}
{Belmont}, R. 2009, \aap, 506, 589

\bibitem[{{Beloborodov}(1999)}]{B99PE}
{Beloborodov}, A.~M. 1999, \apjl, 510, L123

\bibitem[{{Belyaev} \& {Budker}(1956)}]{BB56}
{Belyaev}, S.~T. \& {Budker}, G.~I. 1956, Dokl. Adac. Nauk SSSR, 107, 807

\bibitem[{{Berestetskii} {et~al.}(1982){Berestetskii}, {Lifshitz}, \&
  {Pitaevskii}}]{LLVol4}
{Berestetskii}, V.~B., {Lifshitz}, E.~M., \& {Pitaevskii}, V.~B. 1982, {Quantum
  electrodynamics} (Oxford: Pergamon Press)

\bibitem[{{Bjornsson}(1985)}]{Bjornsson85}
{Bjornsson}, C. 1985, \mnras, 216, 241

\bibitem[{{Blumenthal} \& {Gould}(1970)}]{BG70}
{Blumenthal}, G.~R. \& {Gould}, R.~J. 1970, Rev. Mod. Phys., 42, 237

\bibitem[{{Brinkmann}(1984)}]{Bri84}
{Brinkmann}, W. 1984, JQSRT, 31, 417

\bibitem[{{Crusius-Waetzel} \& {Lesch}(1998)}]{Crusius98}
{Crusius-Waetzel}, A.~R. \& {Lesch}, H. 1998, \aap, 338, 399
 
\bibitem[{{Jones}(1968)}]{Jones68}
{Jones}, F.~C. 1968, Physical Review, 167, 1159

\bibitem[{{Kershaw}(1987)}]{K87}
{Kershaw}, D.~S. 1987, JQSRT, 38, 347

\bibitem[{{Kershaw} {et~al.}(1986){Kershaw}, {Prasad}, \& {Beason}}]{KPB86}
{Kershaw}, D.~S., {Prasad}, M.~K., \& {Beason}, J.~D. 1986, JQSRT, 36, 273

\bibitem[{{Nagirner} \& {Poutanen}(1993)}]{NP93}
{Nagirner}, D.~I. \& {Poutanen}, J. 1993, \aap, 275, 325

\bibitem[{{Nagirner} \& {Poutanen}(1994)}]{NP94}
---.  1994, 
Astrophys. \& Space Phys. Rev., 9, 1 (NP94)


\bibitem[{{Nagirner} \& {Poutanen}(2001)}]{NP01}
---. 2001, \aap, 379, 664


\bibitem[{{Malzac} {et~al.}(2001){Malzac}, {Beloborodov}, \&
  {Poutanen}}]{MBP01}
{Malzac}, J., {Beloborodov}, A.~M., \& {Poutanen}, J. 2001, \mnras, 326, 417

\bibitem[{{Pe'er} \& {Waxman}(2005)}]{PW05}
{Pe'er}, A. \& {Waxman}, E. 2005, \apj, 628, 857

\bibitem[{{Pomraning}(1973)}]{Pom73}
{Pomraning}, G.~C. 1973, {The equations of radiation hydrodynamics} (Oxford:
  Pergamon Press)

\bibitem[{{Poutanen}(1994)}]{Pou94PhD}
{Poutanen}, J. 1994, PhD thesis, University of Helsinki

\bibitem[{{Poutanen} \& {Svensson}(1996)}]{PS96}
{Poutanen}, J. \& {Svensson}, R. 1996, \apj, 470, 249

\bibitem[{{Prasad} {et~al.}(1986){Prasad}, {Kershaw}, \& {Beason}}]{PKB86}
{Prasad}, M.~K., {Kershaw}, D.~S., \& {Beason}, J.~D. 1986, Appl. Phys. Lett.,
  48, 1193

\bibitem[{{Roland} {et~al.}(1985){Roland}, {Hanisch}, {Veron}, \&
  {Fomalont}}]{Roland85}
{Roland}, J., {Hanisch}, R.~J., {Veron}, P., \& {Fomalont}, E. 1985, \aap, 148,
  323

\bibitem[{{Sazonov} \& {Sunyaev}(1998)}]{SaSu98}
{Sazonov}, S.~Y. \& {Sunyaev}, R.~A. 1998, \apj, 508, 1

\bibitem[{{Schopper} {et~al.}(1998){Schopper}, {Lesch}, \& {Birk}}]{Schopper98}
{Schopper}, R., {Lesch}, H., \& {Birk}, G.~T. 1998, \aap, 335, 26

\bibitem[{{Stern} \& {Poutanen}(2006)}]{StP06}
{Stern}, B.~E. \& {Poutanen}, J. 2006, \mnras, 372, 1217

\bibitem[{{Stern} \& {Poutanen}(2008)}]{SP08}
---. 2008, \mnras, 383, 1695

\bibitem[{{Sunyaev} \& {Zeldovich}(1972)}]{SZ72CoASP}
{Sunyaev}, R.~A. \& {Zeldovich}, Y.~B. 1972, Comments on Astrophysics and Space
  Physics, 4, 173

\bibitem[{{Vurm} \& {Poutanen}(2009)}]{VP09}
{Vurm}, I. \& {Poutanen}, J. 2009, \apj, 698, 293

\bibitem[{{Zeldovich} \& {Sunyaev}(1969)}]{ZS69}
{Zeldovich}, Y.~B. \& {Sunyaev}, R.~A. 1969, \apss, 4, 301

\end{thebibliography}

\end{document}